\documentclass[a4paper,prd,twocolumn,showpacs,superscriptaddress,floatfix,nofootinbib]{revtex4-1}
\usepackage{graphicx}
\usepackage[normalem]{ulem}
\usepackage{float}
\usepackage{amsmath}
\usepackage{amssymb}
\usepackage{booktabs}
\usepackage{hhline}
\usepackage{multirow}
\usepackage[rightcaption]{sidecap}
\usepackage{hyperref}
\usepackage{xcolor}
\usepackage{wrapfig}
\usepackage{braket}
\usepackage{lineno}
\usepackage{orcidlink}
\usepackage{comment}
\usepackage{tabularx}
\usepackage{array}
\usepackage{times}
\usepackage{mathrsfs}
\usepackage{subfigure}

\def\rhs{\rho_s}
\def\rs{r_s}
\def\a{\alpha}
\def\b{\beta}
\def\g{\gamma}
\def\d{\delta}
\def\s{\sigma} 
\def\o{\omega}
\def\O{\Omega}
\def\m{\mu}
\def\n{\nu}
\def\ph{\phi} 
\def\non{\nonumber}
\def\n{\noindent}

\begin{document}
\title{Chaotic Imprints in Gravitational Waves from Extreme-Mass-Ratio-Inspirals Embedded in Dark Matter Halo}
\author{Surajit Das\,\orcidlink{0000-0003-2994-6951}}
\email{surajitdas@mail.ustc.edu.cn, surajit.cbpbu20@gmail.com (corresponding author)}
\affiliation{%
Department of Astronomy, School of Physical Sciences, \textcolor{blue}{University of Science and Technology of China}, Hefei, Anhui 230026, China}
\affiliation{%
CAS Key Laboratory for Researches in Galaxies and Cosmology, School of Astronomy and Space Science, \textcolor{blue}{University of Science and Technology of China}, Hefei, Anhui 230026, China}
\author{Surojit Dalui\,\orcidlink{0000-0003-1003-8451}}
\affiliation{%
Department of Physics, \textcolor{blue}{Shanghai University}, Baoshan District, Shanghai 200444, China}
\author{Bum-Hoon Lee\,\orcidlink{0009-0008-3322-2087}}
\affiliation{%
Department of Physics, \textcolor{blue}{Shanghai University}, Baoshan District, Shanghai 200444, China}
\affiliation{%
Department of Physics, \textcolor{blue}{Sogang University}, Seoul 04107, Korea}
\affiliation{%
Center for Quantum Spacetime, \textcolor{blue}{Sogang University}, Seoul 04107, Korea}
\author{Yi-Fu Cai\,\orcidlink{0000-0003-0706-8465}}
\affiliation{%
Department of Astronomy, School of Physical Sciences, \textcolor{blue}{University of Science and Technology of China}, Hefei, Anhui 230026, China}
\affiliation{%
CAS Key Laboratory for Researches in Galaxies and Cosmology, School of Astronomy and Space Science, \textcolor{blue}{University of Science and Technology of China}, Hefei, Anhui 230026, China}
\begin{abstract}
We investigate imprints of chaos in gravitational waves from extreme-mass-ratio inspirals configuration, where a stellar massive object, confined in an external potential as a harmonic oscillators, orbits a supermassive Schwarzschild-like black hole embedded in a Dehnen-type dark matter halo. In our previous work (Das S, {\it et al.}, Extreme-mass-ratio inspirals embedded in dark matter halo: Existence of homoclinic orbit and horizon-induced chaos, {\it Sci.~China-Phys.~Mech.~Astron.}~{\bf 69}, 290413 (2026)), we demonstrated the system's transition from non-chaotic to chaotic dynamics by analyzing Poincaré sections, orbital evolution, and Lyapunov exponents across different energies and dark matter halo parameters. In this work, we compute the gravitational waveforms of the small celestial object along different chaotic and non-chaotic orbits by implementing the numerical kludge scheme. We further perform spectral analysis of the gravitational waveforms from such orbits. In particular, we show that when the system is in a chaotic state, the gravitational wave signals are characterized by broader frequency spectra with finite widths, irregular waveform modulations with comparatively large strain excursions for the chaotic trajectories considered here, together with modified energy-emission characteristics, distinctly differentiating them from the signals generated during the system’s non-chaotic state. Furthermore, we discuss the potential detectability of these orbits for upcoming observatories like Laser Interferometer Space Antenna (LISA), TianQin, and Taiji, emphasizing the significant potential for detecting chaotic imprints in gravitational waves to substantially enhance our understanding of chaotic dynamics in black hole physics and the dark matter environments of galactic nuclei.
\end{abstract}
\maketitle
\section{Introduction}\label{s1}
The groundbreaking detection of gravitational waves (GWs) by the LIGO, Virgo and KAGRA scientific teams represented a pivotal moment in the fields of gravitational wave astronomy \cite{LIGOScientific:2016emj,TheLIGOScientific:2017qsa,LIGOScientific:2018mvr,LIGOScientific:2020ibl,LIGOScientific:2021usb,LIGOScientific:2025rid,KAGRA:2021vkt}. This discovery initiated a brand new epoch in astronomical observation. For example, astrophysical systems such as double white dwarfs \cite{Korol:2017qcx,Huang:2020rjf}, massive binary black holes \cite{Klein:2015hvg,Wang:2019ryf}, stellar-mass binary black holes \cite{Sesana:2016ljz,Liu:2020eko}, and extreme-mass-ratio inspirals (EMRIs) \cite{Babak:2017tow,Fan:2020zhy} are predicted to emit GW signals predominantly in the millihertz (mHz) frequency band. 
Space-based GWs detectors are specifically designed to explore the mHz frequency domain \cite{LISA:2017pwj,TianQin:2020hid,Hu:2017mde}, among others \cite{Wang:2024tnk,CSST:2025ssq}. Among anticipated sources, EMRIs are particularly significant, which typically the inspirals of stellar-mass compact objects (of mass approximately $10^0M_\odot - 10^1M_\odot$) around a supermassive black hole (SMBH) (of mass approximately $10^5M_{\odot}$ - $10^{10}M_{\odot})$, frequently found in galactic nuclei. The prolonged duration of EMRIs signals that lasts thousands to millions of cycles offers a unique opportunity to perform precision tests of general relativity in the strong-field regime \cite{Amaro-Seoane:2007osp,Barack:2006pq,Yunes:2011aa,Canizares:2012is,Moore:2017lxy,Eleni:2024fgs,Destounis:2021rko,Destounis:2021mqv,Destounis:2020kss,Destounis:2023gpw,Mukherjee:2022dju,Rahman:2023sof,Strateny:2025zkd}. With the high sensitivity and long observation baseline of these future space missions, it will be possible to detect subtle dynamical effects encoded in the waveforms, thereby enhancing our understanding of gravity in the strong-field regime. \\
Despite this great scientific potential, significant challenges persist in developing accurate theoretical waveform templates for EMRI systems from observational data. Many theoretical and computational approaches have been attempted. In this work, we particularly focus on gravitational waves emitted by EMRI system from the perspective of chaotic dynamics. The following is the motivational factor. Since chaos denotes the sensitivity and unpredictability of an object's motion, it is expected to explain many significant nonlinear events of a binary system of interest in general relativity and beyond. Since the sensitivity of an object's motion can be determined by chaos, it is reasonable to think about how much significant information we can obtain from studying chaos. Chaos in gravity has attracted a lot of attention lately, numerous studies have been conducted on this subject \cite{Suzuki:1999si,Kiuchi:2004bv,Bombelli:1991eg,Sota:1995ms,Vieira:1996zf,Suzuki:1996gm,Cornish:1997hs,deMoura:1999wf,Hartl:2002ig,Han:2008zzf,Takahashi:2008zh,Apostolatos:2009vu,Lukes-Gerakopoulos:2010ipp,Lukes-Gerakopoulos:2012qpc,Lukes-Gerakopoulos:2013qva,Hashimoto:2016dfz,Hashimoto:2018fkb,Li:2018wtz,Lei:2020clg,DeFalco:2020yys,DeFalco:2021uak,Dalui:2018qqv,Dalui:2019umw,Bera:2021lgw,Das:2024iuf,Deich:2022vna,Chen:2022znf,Destounis:2023cim,Pan:2023wau,Destounis:2023khj}. Classically, substantial efforts have been devoted to understanding how the presence of horizon of black holes (BHs) can make an integrable system transit from regular to chaotic motion. Such studies cover diverse scenarios, including spinning BHs \cite{Hartl:2002ig,Han:2008zzf,Takahashi:2008zh}, magnetized BHs \cite{Li:2018wtz}, and involve test particle with varying mass, charge \cite{Lei:2020clg}, spin \cite{Suzuki:1996gm,Hartl:2002ig,Han:2008zzf}, massive particle \cite{Hashimoto:2016dfz}, massless particle \cite{Dalui:2018qqv} and as well as in modified gravity \cite{Das:2024iuf}. Recently, chaos has been investigated in an astrophysical environments with Hernquist dark matter (DM) halo \cite{Destounis:2025tjn}. \\
In a real astrophysical scenario, BHs rarely exist in isolation, instead, they inhabit dynamically rich and complex environments. For instance, SMBHs are known as driving forces of active galactic nuclei \cite{Rees:1984si,Kormendy:1995er}. Astrophysical and cosmological observations strongly suggest that most galaxies, including our Milky Way, are embedded in massive DM halos \cite{Bertone:2004pz,Bovy:2013raa,deSwart:2017heh,Wechsler:2018pic,Bertone:2018krk}. Therefore investigating gravitational interactions between DM halos and SMBHs is essential for understanding the role of DM in galactic-scale structures. Studying various DM models thereby helps address persistent theoretical challenges in gravitational physics and enriches our grasp of fundamental physics. Numerous astrophysically relevant DM halo models have been explored extensively in Refs. \cite{Dubinski:1991bm,Burkert:1995yz,Navarro:1995iw,Navarro:1996gj,Hernquist:1990be,Jaffe:1983iv,Tremaine:1993qb,Dutton:2014xda,Graham:2005xx,Urena-Lopez:2002ptf,Harko:2011xw,Begeman:1991iy} and offering insights into BHs, which are embedded in DM halos (see \cite{Cardoso:2021wlq,Cardoso:2022whc,Destounis:2022obl,Fernandes:2025osu} and references therein). \\
Among the various DM halo models, the Dehnen profile \cite{Dehnen} represents a family of spherical mass distributions that are characterized by the $(1,4,\gamma)$ model and encompasses the Hernquist profile \cite{Hernquist:1990be} ($\gamma=1$) as a special case and is widely used to describe the density distribution of elliptical galaxies and DM halos. This profile offers a highly adaptable framework for modeling both the inner and outer structural properties of DM halos in spherical galaxies and bulges. Its generalized formulation is capable of reproducing a variety of density distributions, including both cuspy and cored profiles. Several composite BH-DM halo solutions exist under the consideration of Dehnen-$(1,4,\g)$ model. For example, in Ref.~\cite{Pantig:2022whj} a BH solution with the Dehnen-$(1,4,\g)$ profile is constructed for the study of an ultrafaint dwarf galaxy. Furthermore, in Refs.~\cite{Gohain:2024eer,UktamjonUktamov:2025emm,Al-Badawi:2024asn} a combined BH-DM halo spacetimes are constructed, surrounded by $(1,4,0),~(1,4,5/2)$ and $(1,4,2)$ type DM halo profiles, respectively. On the other hand, recently Konoplya {\it et al}.~has been proposed a regular BH geometries corresponding to Einasto and Dehnen-type DM profiles with the phenomenological freedom in halo modeling \cite{Konoplya:2025ect}. Moreover, recent research has focused significantly on the impact of Dehnen-type DM halos on BH dynamics and structures, exploring consequences for star cluster survivability \cite{Graham:2005xx,shukirgaliyev2021bound}, ultra-faint dwarf galaxy evolution \cite{Pantig:2022whj}, observational shadow \cite{Luo:2025xjb} and gravitational radiation reaction \cite{Ashoorioon:2025ezk} in composite BH-DM system. Such environments naturally raise the question of how these complex surroundings modify the onset and character of chaotic dynamics because the presence of DM compositions greatly influence the horizon structure. According to current research, realistic DM halos can significantly change near-horizon dynamics, which could improve chaotic aspects in EMRI systems. Additionally, it is anticipated that chaos will leave distinct marks on gravitational wave signals, such as distorted phase coherence, expanded spectrum characteristics, or broader amplitude modulations. These could serve as observational markers of strong-field nonlinear dynamics and as probes of galactic environments near SMBHs.\\
In this work, we consider a particular BH-DM halo combined spacetime for which $\g=5/2$ \cite{Al-Badawi:2024asn}. In particular, the Dehnen-$(1,4,5/2)$ profile, exhibits a steep central cusp and provides an excellent fit to luminous elliptical galaxies \cite{Dehnen,shukirgaliyev2021bound} . In recent research, the influence of Dehnen-$(1,4,5/2)$ type DM halos on BHs has been extensively investigated from a variety of distinct perspectives \cite{Luo:2025xjb,Ashoorioon:2025ezk,Alloqulov:2025ucf,Liang:2025vux,Rani:2025esb,Xamidov:2025prl,Li:2025ver,Alloqulov:2025edn} and hence making it a natural and physically motivated testbed for our analysis. In our first work \cite{Das:2025vja}, we investigated the chaotic dynamical phenomena in EMRI-like configuration, where a stellar-mass compact object orbiting close to the horizon of a supermassive Schwarzschild BH embedded in a Dehnen-$(1,4,5/2)$ type DM halo. Therefore one can model the dynamics of the secondary inspiral as the motion of a relativistic massive particle on the background of the primary BH-DM halo combined spacetime. Our focus was on understanding the dynamical behavior of the massive particle close to the horizon in the background of such BH-DM halo combined spacetime and how the horizon influences if the particle is a part of an integrable system, and that integrable system perturbs the combined BH-DM halo background. To investigate the horizon's influence, we analyzed the dynamics of a relativistic massive particle that is subject to external potentials and is moving around a BH embedded in a DM halo environment. As a consequence, this allowed us to explore bounded motion near the BH horizon and identify how small perturbations evolve under the combined influence of BH and the surrounding DM field. Finally, we analyzed the phase-space dynamics through Poincaré sections, quantify sensitivity to initial conditions using Lyapunov exponents, and discuss how chaos emerges as a function of the halo parameters. \\
In this work as a forward study, we use the numerical kludge method in order to generate the gravitational waveforms emitted by such EMRI configuration and then particularly focus on the point that how chaos affects gravitational waves. We systematically compare the generated waveforms, the frequency and energy spectra of the gravitational waves from three typical orbits such as non-chaotic, onset-of-chaotic and chaotic orbits, which are influenced by the DM halo parameters and then analyze the imprints of chaos in gravitational waves. We discuss the clear difference of gravitational waves between a non-chaotic and a chaotic EMRIs in the background of our considered BH-DM halo combined system. Moreover, to establish the influence of DM halo parameters on chaotic dynamics and its features on the resulting gravitational waveforms, we further study the recurrence analysis and phase space reconstruction for both, the radial motion as well as the gravitational strains. One of the novel findings we obtain in this case is the possible detection of chaos through gravitational waves via characteristic strain sensitivity, and that could be detectable by the future space-based detectors, which provides a new insight into BH astrophysics and the influence of galactic settings on the dynamics of compact objects. \\
The paper is organized as follows. In Sec.~\ref{sec:background}, we mention in brief on the typical parameter spaces for which the transition between non-chaotic and chaotic dynamics are happened of EMRI configuration in the background of a Schwarzschild-like BH solution immersed in a Dehnen-$(1,4,5/2)$ type DM halo, which can be found from our previous work \cite{Das:2025vja}. In Sec.~\ref{sec:kludge}, we employs numerical kludge scheme to generate gravitational waveforms, analyzing their characteristic patterns, averaged amplitude and energy emission rate, frequency spectra and energy spectra. Sec.~\ref{sec:rp} demonstrates the signatures of chaos in gravitational waveforms through recurrence analysis. Finally in Sec.~\ref{sec:detection}, we assesses the detectability of chaotic gravitational wave signals by future space-based gravitational wave detectors, including LISA, TianQin, and Taiji. We concludes the findings of our study in Sec.~\ref{sec:conclusion}.

\section{Chaotic dynamics of extreme-mass-ratio inspirals around a Schwarzschild black hole-dark matter halo}\label{sec:background}
We review the chaotic dynamics of EMRIs (where the mass ratio is taken to be $m/M = 10^{-5}$), which involve a stellar-mass compact object in orbits around a supermassive Schwarzschild-like BH immersed in a Dehnen-$(1,4,5/2)$ type DM halo \cite{Hernquist:1990be,Dehnen,Van}. We start by considering a static spherically symmetric (SSS) BH spacetime ansatz, surrounded by a Dehnen-$(1,4,5/2)$ type DM halo profile as
    \begin{equation}
	    ds^2 = -f(r) dt^2 + \frac{dr^2}{f(r)} + r^2 d\Omega^2~,\label{eq:metric}
    \end{equation}
with $d\Omega^2 = d\theta^2 + \sin^2\theta \, d\phi^2$ representing the line element on a unit two-sphere. The solution of $f(r)$ up to leading order is given by \cite{Al-Badawi:2024asn,Das:2025vja}
    \begin{equation}
	    f(r) = 1 - \frac{2M}{r} - 32\pi \rho_s r_s^2 \sqrt{1 + \frac{r_s}{r}}~,\label{eq:model}
    \end{equation}
where $M$ is the mass of the BH and $\rhs$, $\rs$ are known as the typical density and scale radius of the central DM halo, respectively. In our earlier work \cite{Das:2025vja}, we systematically examined the transitions from a non-chaotic (regular or periodic) to chaotic dynamical behavior of a massive test particle close to the horizon under an EMRI configuration by analyzing the nonlinear dynamical systems of equations (see Eqs.~(s6), (s7), (s8) and (s9) in the Supplemental Material). This investigation has carried out by properly investigating the associated Poincar\'{e} sections, EMRI orbital evolutions, and both total, radial Lyapunov exponents as functions of energy and DM halo parameters $\rhs$ and $\rs$. This sequence of analyses elucidates the influence of both the event horizon of such a Schwarzschild-like BH embedded in a DM halo and the particular parameters of the halo ($\rhs$, $\rs$) on the emergence of chaotic motion for a massive test particle. In the current paper, we primarily concentrate on the GWs consequences arising from such chaotic orbital EMRIs, and therefore we summarize the key findings from our previous study \cite{Das:2025vja} in the following Table~\ref{tab1}. For a more detail in each and every step on the particle dynamics and transitions from non-chaotic to chaotic regime, reader refer to the Supplemental Material along with our previous work \cite{Das:2025vja} \footnote{Following the discussion on the background dynamics of the considered BH-DM halo combined spacetime in our Supplemental Material, it is worth emphasizing that the choice of the external potential as the harmonic confinement is introduced phenomenologically, following the previous works \cite{Hashimoto:2016dfz,Lei:2020clg,Wang:2026whj,Dalui:2018qqv,Bera:2021lgw,Das:2024iuf}, and serves as a minimal model that breaks integrability near the horizon in the strong-field regime of the BH-DM halo spacetime. Alternative potential forms would modify the detailed shape of the trajectories and the precise parameter spaces of the chaos threshold and could introduces an additional chaotic effects in the dynamics of a particle motion; however, the fundamental finding that the DM halo parameters $(\rho_s,r_s)$ along with energy $E$ can drive the system from regular to chaotic dynamics near the horizon is not an artifact of the potential choice. The nonlinear behavior in particle's motion near the horizon in the strong-field regime responsible for horizon-induced chaos, and the qualitative signatures - broader frequency spectra, enhanced emission rates, and characteristic recurrence patterns (which will be seen in the following sections) - are generic features of chaotic orbital motion in such strong-field environments \cite{AJ-book,Koyama:2007cj,Drasco:2003ky,Kiuchi:2007ue}.}.

\begin{table*}[htbp]
    \centering
    \renewcommand{\arraystretch}{2}
    \caption{A principal investigation is shown in the transition from ``Non-chaotic (NC)'' to ``Chaotic (C)'' dynamics, passing through the ``Onset-of-chaos (OC)'' phase, within an EMRI limit, where a stellar-mass compact object inspiraling around a supermassive Schwarzschild-like BH, which is surrounded by a Dehnen-$(1,4,5/2)$ type DM halo. The study demonstrates the influence of the energy and the dark matter halo parameters, namely $\rho_s$ and $r_s$, on the orbital dynamics of the EMRI within its galactic environment. For a comprehensive understanding of the findings summarized here, the reader is directed to our preceding work (refer to Figures~2-3 along with Figure~S6 in \cite{Das:2025vja}).}
    \label{tab1}
    \begin{tabular}{| c @{\hspace{1em}} | c @{\hspace{1em}} | c @{\hspace{1em}} | c @{\hspace{1em}}  |}
    \hline
    \textbf{~~~Case description (fixed parameters)} & \textbf{~~~Non-chaotic} & \textbf{~~~Onset-of-chaotic} & \textbf{~~~Chaotic} \\ 
    \hline
    \hline
    \textbf{Case-I:} $\rs=0.15$, $\rhs=0.02$ & $E=108$ & $E=130$ & $E=132.50$ \\
    \hline
    \textbf{Case-II:} $\rhs=0.01$, $\rs=0.25$ & $E=95$ & $E=118$ & $E=122$ \\
    \hline
    \textbf{Case-III:} $\rs=0.15$, $E=90$ & $\rhs=0.01$ & $\rhs=0.04$ & $\rhs=0.05$ \\
    \hline
    \textbf{Case-IV:} $\rhs=0.01$, $E=115$ & $\rs=0.10$ & $\rs=0.25$ & $\rs=0.27$ \\
    \hline
    \textbf{Einstein's GR:} $\rhs=0$, $\rs=0$ & $E=140$ & $E=154$ & $E=156$ \\
    \hline
    \end{tabular}
\end{table*}

\section{Kludge gravitational waveforms}\label{sec:kludge}
In the analysis of an EMRI system focused for our study, it is usual to model a supermassive BH-DM halo combined system—as the primary body at the center of halo, while a less massive stellar compact object is treated as a secondary test particle. Hence a stellar massive compact object moving in a chaotic and non-chaotic orbits around a supermassive BH surrounded by a DM halo in galaxies constitutes an EMRIs system. As the secondary approaches close to the primary, the strong gravitational field alters spacetime curvature, producing GWs. These waves extract energy and angular momentum from the test particle, causing it to gradually spiral inward. Over short timescales, the energy and angular momentum lost due to the secondary's motion are insignificant relative to the system's total energy, justifying the use of the adiabatic approximation \cite{Hughes:2005qb,Sundararajan:2007jg,Hughes:1999bq,Grossman:2011im,Zi:2023qfk}. Within the timescale considered here, we neglect the back-reaction of gravitational radiation on the orbital dynamics. The secondary trajectory is therefore evolved using the conservative equations of motion introduced in our orbital model, including the external confining potential, and the resulting trajectory is subsequently used as input to the numerical-kludge waveform construction \cite{Ruffini:1981af,Sasaki:1982ws,Babak:2006uv}. \\
When a small massive object follows a non-chaotic orbit around a supermassive BH surrounded by a DM halo, it traces a closely overlapping trajectories (as illustrated in the first column of Fig.~S6 in the supplemental material associated with our previous paper \cite{Das:2025vja}). In contrast, for chaotic orbits, variations in energy and DM halo parameters cause the particle to make them random, reducing the overlap between two initially close trajectories (see the third columns of Fig.~S6 in \cite{Das:2025vja}) and therefore, near the event horizon, chaotic orbit undergoes prolonged, high-frequency oscillations, producing more rich GW signals. These waves encode essential information about the physics near the BH’s event horizon, making them an effective tool for investigating BH properties incorporated with DM halos through both chaotic and non-chaotic orbital dynamics. Thus, such GWs generated by EMRIs establish a direct connection between two different kind of motions (chaotic and non-chaotic in our case) of the secondary and establish a fundamental properties of the BH-DM halo combined spacetime. This provides an opportunity to investigate both the behavior of compact objects and the essential features of BH in the immediate vicinity of the event horizon at the center of such halo, considered in our work. \\
The symmetric and traceless GW source quadrupole formula under the slow-motion limit can be written as \cite{Babak:2006uv,Thorne:1980ru,Poisson-Will}
    \begin{equation}
        h_{ij} = \frac{2}{D_{\mathrm{L}}} \ddot{I}_{ij}= \frac{2m}{D_{\mathrm{L}}}\big(a_i x_j + a_j x_i + 2v_i v_j\big)~,\label{5.7}
    \end{equation}
Where $D_{\mathrm{L}}$ is the luminosity distance to the detector, $v_i$ and $a_i$ are spatial velocity and acceleration of the secondary stellar-massive compact object, respectively. \\
It is important to clarify how the BH-DM halo modified spacetime enters the kludge scheme. The metric influence is entirely contained in the orbital dynamics: the orbiting particle trajectory is obtained by integrating the fully relativistic equations of motion (Eqs.~(s6)-(s9)), which depend on the DM halo-modified spacetime metric $f(r)$, given in Eq.~\eqref{eq:model}. Once the inspiral trajectory is computed, the subsequent projection onto flat space emission formula for the corresponding flat-space source orbit (as mentioned in Eq.~(s10)) and the quadrupole waveform generation formula (as given in Eq.~\eqref{5.7}) follow the standard kludge prescription \cite{Babak:2006uv,Thorne:1980ru,Poisson-Will}. The adiabatic approximation is justified because the energy and angular momentum losses due to radiation are negligible over the short simulation timescale. Higher multipole moments are not included, consistent with the leading-order quadrupole approximation. (For a brief discussion on numerical kludge waveforms, see the Supplemental Material). Our analysis will focus on the formulations given by Eq.~\eqref{5.7}. To analyze the GWs signal at a detector level, one establish a detector-aligned coordinate system $(X, Y, Z)$ sharing its origin with the original $(x, y, z)$ frame, both centered on the supermassive BH \cite{Poisson-Will}. The basis vectors of this detector frame in the original coordinates are defined as
    \begin{eqnarray}
        \mathbf{e}_X &=& [\cos \zeta, -\sin\zeta, 0]~,\label{5.8}\\
        \mathbf{e}_Y &=& [\sin \iota \sin \zeta, -\cos \iota \cos \zeta, -\sin \iota]~,\label{5.9}\\
        \mathbf{e}_Z &=& [\sin \iota \sin \zeta, -\sin \iota \cos \zeta, \cos \iota]~,\label{5.10}
    \end{eqnarray}
where $\zeta$ denotes the pericenter longitude within the orbital plane while $\iota$ represents the inclination angle between the orbital plane and the $(X-Y)$ plane. The GW polarizations can then be derived by projecting the metric perturbations (Eq.~\eqref{5.7}) onto the detector frame, as the following.
    \begin{eqnarray}
        h_{+} &=& \frac{1}{2} (\mathbf{e}_X^i \mathbf{e}_X^j - \mathbf{e}_Y^i \mathbf{e}_Y^j) h_{ij}~,\label{5.11}\\
        h_{\times} &=& \frac{1}{2} (\mathbf{e}_X^i \mathbf{e}_Y^j + \mathbf{e}_Y^i \mathbf{e}_X^j) h_{ij}~.\label{5.12}
    \end{eqnarray}
Hence using Eqs.~\eqref{5.8}, \eqref{5.9} in Eqs.~\eqref{5.11}, \eqref{5.12} the polarizations can be formulated in the detector frame as specific combinations of the $h_{ij}$ elements and can be expressed as the following forms \cite{Babak:2006uv,Yang:2024lmj}.
    \begin{eqnarray}
        h_{+} &=& \frac{1}{2}\bigg[\cos ^2\zeta \left(h_{xx}-h_{yy} \cos ^2\iota\right)+\sin ^2\zeta \left(h_{yy}-h_{xx} \cos ^2\iota\right)\nonumber\\
        && -h_{xy} \sin 2 \zeta \left(\cos ^2\iota +1\right) - h_{zz} \sin ^2\iota \nonumber\\
        && + \sin 2\iota\left(h_{xz} \sin \zeta + h_{yz} \cos \zeta \right) \bigg]~,\label{5.13}\\
        h_{\times} &=& \cos \iota \Big[ \frac{1}{2} h_{xx} \sin 2\zeta + h_{xy} \cos 2\zeta - \frac{1}{2} h_{yy} \sin 2\zeta \Big] \nonumber \\
        && + \sin \iota [h_{yz} \sin \zeta - h_{xz} \cos \zeta] ~.\label{5.14}
    \end{eqnarray}
This approach, while approximate and applicable only for the secondary body's small orbital velocities, leads to an adiabatic progression of EMRIs. In the subsequent sections that follow, we conduct a comprehensive investigation of GWs utilizing the methodology outlined in this section.

\begin{figure*}[htbp]
    \centering
    \scalebox{1.0}{\includegraphics[width=0.9\textwidth, height=0.9\textheight]{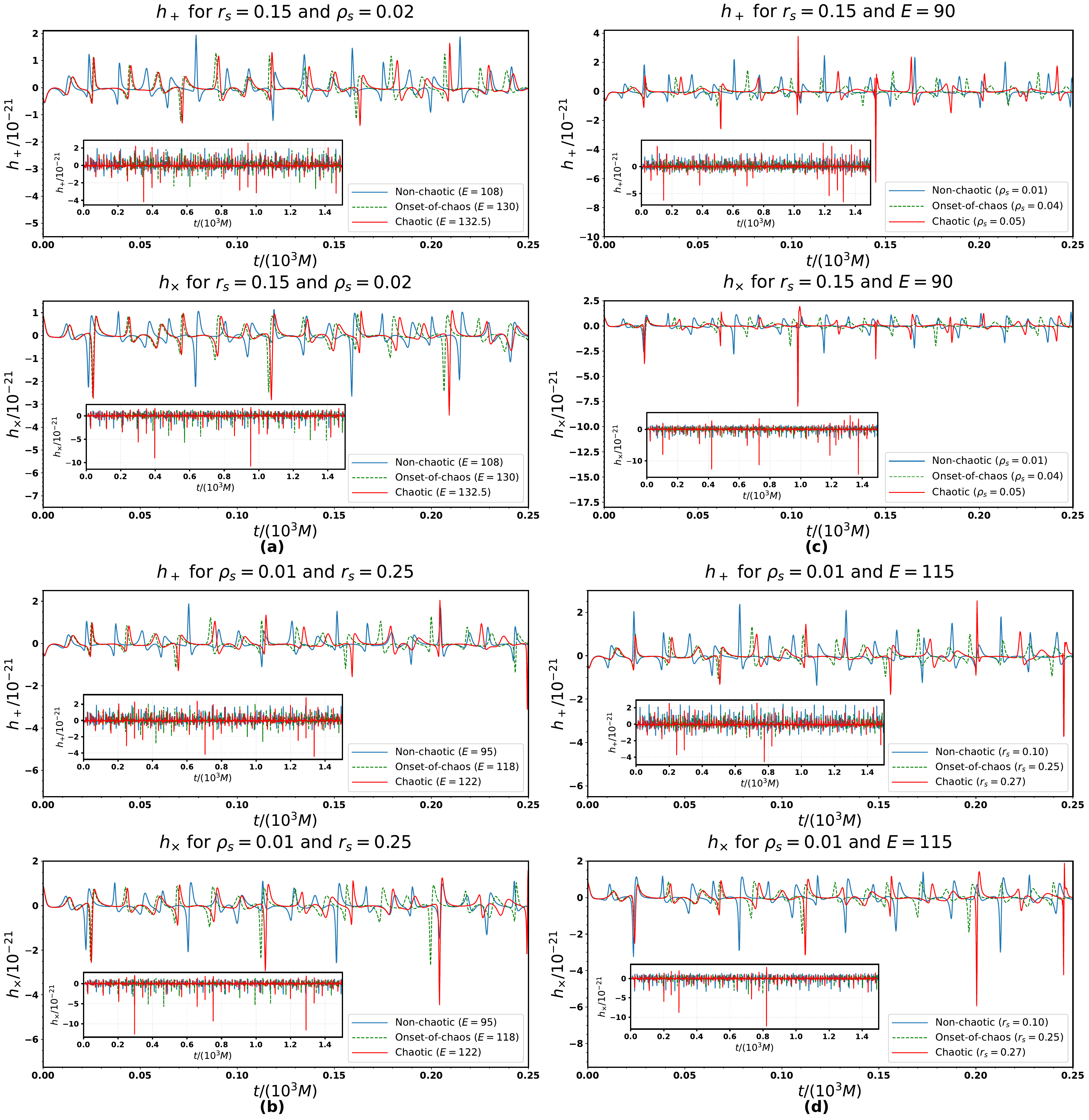}}
    \caption{The gravitational waveforms of the $+$ and $\times$ modes corresponding to the non-chaotic, onset-of-chaos and chaotic orbits for four different cases (a,b,c and d) (see Table \ref{tab1}). The time range is taken from 0 to 0.25 in unit of $t/(10^3 M)$, whereas the rectangular box inserted in each figures display of an extended view of the corresponding waveforms over a more extended time range (from 0 to 1.5 in unit of $t/(10^3 M)$. We emphasize that the ordering of the instantaneous strain amplitudes should not be interpreted as an ordering of the degree of chaos; the NC, OC, and C classifications refer to the underlying phase-space dynamics of the corresponding trajectories.}\label{fgw}
\end{figure*}

\subsection{Analysis of the generated gravitational waveforms}\label{sec:modes}
To investigate how the central density $\rho_s$ and scale radius $r_s$ of the halo affect the waveforms, we study a secondary stellar-massive compact object of mass $m = 10\,M_\odot$ orbiting a supermassive Schwarzschild-like BH embedded in a Dehnen-$(1,4,5/2)$ type DM halo of mass $M = 10^6\,M_\odot$. The system is assumed to be at a luminosity distance of $D_\mathrm{L} = 2\,\mathrm{Gpc}$. This setup allows us to analyze GWs generated by both chaotic and non-chaotic orbital trajectories under the influence of a Dehnen-$(1,4,5/2)$ type DM halo. For simplicity, we set the inclination angle $\iota = \pi/4$ and pericenter longitude $\zeta = \pi/4$ to calculate the polarizations of GWs. \\
Fig.~\ref{fgw} shows the gravitational waveforms from an EMRI system, where a secondary object orbits a supermassive Schwarzschild-like BH embedded in a Dehnen-$(1,4,5/2)$ type DM halo. The analysis of the generated waveforms for the onset-of-chaos orbits (i.e., the first appearance of the broken tori in the associated Poincar$\Acute{e}$ sections in \cite{Das:2025vja}), situated between non-chaotic and chaotic orbits, is to illustrate the transition from orderly motion to irregular or chaotic dynamics (see our first work \cite{Das:2025vja} for further clarification). Thus the waveforms have the hidden gravitational influences of a Dehnen-$(1,4,5/2)$ type DM halo on the formation of the respective chaotic and non-chaotic orbits. On the other hand, Fig.~\ref{fgw} also shows that the overall amplitudes remain nearly constant for non-chaotic orbits (as one can visualize the waveforms from non-chaotic orbits in the insets of each four figures). However, as chaos begins to influence the dynamics (evident from the first appearance of broken tori in the associated Poincaré sections as the ``onset-of-chaos"), the amplitudes of each polarization mode start fluctuating. Eventually, for chaotic orbits, they exhibit irregular and non-constant behavior. \\
We emphasize that the gravitational-wave amplitude should not be interpreted as a monotonic measure of chaoticity. Within the quadrupole approximation, the strain is determined by the second time derivative of the source quadrupole moment and therefore depends on the detailed position, velocity, and acceleration along each trajectory. In contrast, the classifications non-chaotic, onset-of-chaos, and chaotic characterize the phase-space structure and stability of the orbital motion. Consequently, the transition from regular to chaotic dynamics does not require a monotonic ordering of the waveform amplitudes. In particular, the onset-of-chaos state corresponds to the first distortion or breaking of regular invariant structures in the Poincaré section; it does not imply that the instantaneous quadrupole derivatives, and hence the waveform amplitude, must take values intermediate between those of the selected non-chaotic and fully chaotic trajectories. Therefore, the smaller amplitude of the onset-of-chaos waveform observed in some panels of Fig.~\ref{fgw} is not inconsistent with the dynamical transition toward chaos. \\
This distinction is particularly transparent in Cases III and IV. As summarized in Table \ref{tab1}, the conserved orbital energy is fixed at E=90 in Case III and E=115 in Case IV while the halo parameters are varied across the non-chaotic, onset-of-chaos, and chaotic regimes. Nevertheless, the corresponding waveforms exhibit different amplitudes and temporal structures. The differences therefore cannot be attributed simply to an increase in the conserved orbital energy; rather, they originate from the different orbital trajectories and the associated time dependence of the quadrupole moment. \\
To understand it more clearly, we have also presented an enlarged view of the gravitational waveforms emitted by such EMRI system for non-chaotic, onset-of-chaos, and chaotic orbits, focusing on two different cases (see our Supplemental Material), which reveal that the waveforms associated with the non-chaotic orbits exhibit an overall periodic patterns, despite variations in the amplitude of individual modes. In contrast as chaos begins to influence the dynamics, the periodicity start to degrade and finally, in the chaotic regime, the waveforms entirely lose their periodic structure, displaying irregular fluctuating pattern (see Figs.~S1,~S2 in the Supplemental Material). These findings are in agreement with the Poincar\'e maps analyzed in our previous work \cite{Das:2025vja} and reveals how different DM halo parameters affect on the formation of chaotic dynamics in such EMRIs, considered here. \\
We have also explored the averaged amplitude and energy-emission characteristics of the gravitational waves associated with different orbital regimes. For the particular chaotic trajectories considered here, the irregular orbital evolution can produce comparatively large excursions in the quantities entering the quadrupole moment, resulting in larger strain excursions and stronger variations of the radiative observables at particular times, as illustrated in Fig.~S3 (see the Supplemental Material). The enhanced GW amplitude observed in chaotic orbits should be understood in terms of the time variability of the source's quadrupole moment rather than the total orbital energy, which is fixed within each case. According to the quadrupole formula (Eq.~\eqref{5.7}), the strain amplitude is proportional to the acceleration and velocity of the secondary. Chaotic trajectories, by exploring a larger region of phase space, undergo more frequent and rapid close approaches to the black hole, where the gravitational field is strongest. This results in larger time derivatives of the quadrupole moment $\dot{I}_{ij}$ and $\ddot{I}_{ij}$, and consequently enhanced GW amplitude and energy emission rate. We stress, however, that this behavior is trajectory dependent and should not be interpreted as a universal or monotonic increase of gravitational-wave amplitude with chaoticity. In particular, the onset-of-chaos trajectory need not have an amplitude intermediate between the selected non-chaotic and fully chaotic trajectories. \\
In conclusion, this subsection demonstrates that merely examining the waveforms of different orbital trajectories (chaotic, onset-of-chaos and non-chaotic) through visual inspection is inadequate to determine whether the original orbit was non-chaotic or chaotic. While there appear to be slight distinctions among non-chaotic, onset-of-chaos, and chaotic orbits, these differences may not be easily distinguishable. Therefore, in the next subsection, we analyze the frequency and energy spectrum of the gravitational waveforms for a more precise assessment.

\begin{figure*}[!t]
    \centering
    \includegraphics[width=0.9\linewidth, height=0.6\textheight]{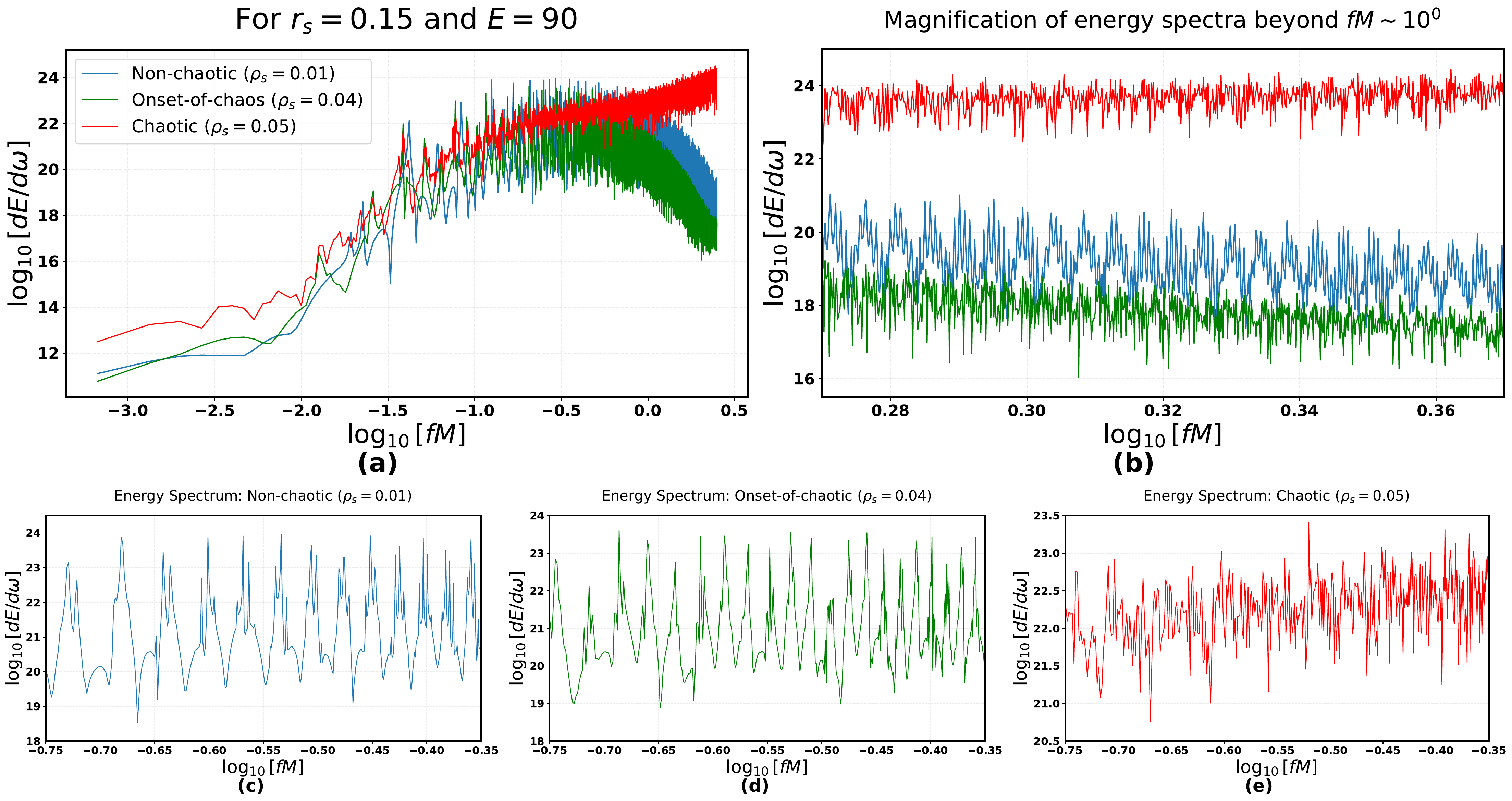}
    \caption{{\it (a):} GWs energy spectra corresponding to different orbits for Case-III (fixed $\rs = 0.15$, $E = 90$ with various $\rhs$). For frequencies $fM \geq 10^{0}$, both the non-chaotic and onset of chaos orbits exhibit a decay in their respective energy spectra. But this decay feature is notably absent in the case of chaotic orbit. {\it (b):} The zoomed-in view beyond $fM\sim10^{-0.25}$ of Fig.~2(a) is represented here, where the blue curve corresponds to a non-chaotic orbit ($\rhs = 0.01$), the green curve indicates the onset of chaos ($\rhs = 0.04$), and the red curve represents a fully chaotic orbit ($\rhs = 0.05$). The bottom panel ((c), (d) and (e)) display magnified views of the energy spectra for non-chaotic, onset-of-chaos, and chaotic orbits within the frequency $fM\sim 10^0$, respectively. In (c), it reveals numerous sharp peaks at specific characteristic frequencies, a signature of regular motion, same as in Fig.~2(b). In contrast, (e) shows broader peaks with additional overlapping spikes, a consequence of the chaotic nature of the orbit, same as in Fig.~2(b).} \label{es_combined} 
\end{figure*}

\subsection{Analysis of the spectrum of gravitational waves}\label{sec:spectra}
To further investigate the imprints of chaos in GWs, we analyze the frequency spectra of the gravitational waveforms associated with different type of orbits, building upon our earlier orbital computations (for more details on the orbital dynamics and computations, see \cite{Das:2025vja}). This analysis involves applying discrete Fourier transforms to the time-domain gravitational waveforms (as shown in Fig.~\ref{fgw}). The resulting frequency spectra, presented in Fig.~S4, display the absolute values of $\hat{h}_+(f)$ and $\hat{h}_{\times}(f)$, representing the Fourier transforms of the plus and cross polarization modes, respectively (see the Supplemental Material). \\
Our investigation reveals a fundamental distinction: gravitational waveforms from non-chaotic orbits exhibit discrete characteristic frequencies, whereas from chaotic orbits display a broader range of frequencies and contain many small spikes in the entire region compared with those in the non-chaotic and onset-of-chaos cases, hence the spectrum from chaotic systems appearing as a quasi-continuous with finite bandwidth. This typical spectral behavior in the Fourier analysis indicates that the signal is not periodic/regular, rather than chaotic, which is a distinctive and peculiar characteristic of chaotic systems \cite{AJ-book}. This is one of the most important and interesting results of our paper, which is truly consistent with two previous findings on GWs energy spectra producing from the dynamics of a spinning particle around a Schwarzschild and Kerr BH by Maeda \textit{et al.} \cite{Suzuki:1999si} and Keuchi \textit{et al.} \cite{Kiuchi:2004bv}, respectively. Notably, while Refs.~\cite{Suzuki:1999si,Kiuchi:2004bv} studied energy spectra from a spinning secondaries in Schwarzschild and Kerr spacetime, our research focuses on a spinless secondary around a supermassive Schwarzschild-like BH surrounded by a Dehnen-$(1,4,5/2)$ type DM halo, yet both approaches consistently demonstrate distinguishable spectral features between chaotic and non-chaotic orbital dynamics. \\
We further compute the energy spectra of GWs associated with non-chaotic, onset-of-chaos and chaotic orbits, as these are expected to become crucial observational parameters. To study the effect of the DM halo parameters $\rhs,\rs$ on GWs spectrum in our context, we particularly consider two cases as follows. Fig.~\ref{es_combined}(a) presents the energy spectra of the gravitational waveforms corresponding to each different orbits for Case-III (fixed $E=90,~\rs=0.15$ with different values of $\rhs$), while Fig.~S5 is the same for Case-IV (fixed $E=115,~\rhs=0.01$ with different values of $\rs$) (see the Supplemental Material for Fig.~S5). It is clearly visible from both of the figures that the non-chaotic orbit exhibits numerous distinct characteristic peaks at specific characteristic frequencies. Moreover for regular orbital motion, one typically observes several fundamental frequencies along with their harmonics. Therefore our analysis indicate the presence of regular motion in this case. On the other hand the observed spectrum from chaotic orbit exhibits a distinct behavior compared to the non-chaotic orbit. The signal from chaotic orbit is accompanied by a numerous small, irregular spikes, which is a distinctive hallmark of chaotic dynamics. These fluctuations cause the spectral peaks to broaden significantly compared to those observed in non-chaotic system. \\
However, beyond a characteristic frequency $fM\sim 10^{-0.25}$, the energy spectra associated with non-chaotic and onset-of-chaos orbits appear to be flat and similar to white noise, suggesting a significant stochastic contribution from non-chaotic and onset-of-chaos motions. This interpretation, however, is misleading on an overall view. Upon closer inspection of the frequency range above $fM\sim 10^{-0.25}$, particularly when examining the energy spectra of the individual orbits — the distinction between regular and chaotic dynamics becomes evident (see Fig.~\ref{es_combined}(b) for Case-III and Fig.~S5(b) for Case-IV). Figs.~\ref{es_combined}(b) and S5(b) clearly show that non-chaotic orbits display multiple sharp peaks (regular) at well-defined characteristic frequencies. In contrast, chaotic orbits exhibit a greater number of irregular spikes compared to non-chaotic orbits, with these spikes often overlapping and forming a quasi-continuous spectrum. This particular behavior is entirely distinct from that of a non-chaotic systems, which is a key findings of this work in the context of a galaxy environments. \\
Figs.~\ref{es_combined}(b) and S5(b) also suggest a significant changes in the GWs energy spectra beyond the frequency scale $fM\sim10^0$, particularly for the chaotic regime for both of the Cases-III and IV (highlighted with red color). For the chaotic trajectory, sensitivity to initial conditions and the resulting irregular orbital evolution generate a more complex time dependence of the quadrupole moment. This redistributes spectral power over a broader range of frequencies and can maintain appreciable high-frequency components compared with the selected non-chaotic and onset-of-chaos trajectories. On the other hand, since energy is emitted in discrete, narrow frequency bands, the non-chaotic (highlighted with blue color) and onset-of-chaos (highlighted with green color) orbits show a decline in the GWs energy spectra beyond this typical characteristic frequency contrasting with the chaotic case. The spectrum likely follows an exponential or steep power-law decay, typical of ordered systems \cite{Koyama:2007cj}. \\
For a more clarity of our results, we have also examined each spectra for different orbits by magnifying its frequency range within $fM\sim 10^0$, which allows for a clearer interpretation of the behavior occurring just before the frequencies reach $fM\sim 10^0$. Figs.~\ref{es_combined}(c), (d) and (e) display zoomed-in views of the energy spectra for non-chaotic, onset-of-chaos, and chaotic orbits, respectively for Case-III, while Figs.~S5(c), S5(d) and S5(e) display the same for Case-IV, respectively. Particularly, the GW energy spectra from non-chaotic orbit show many sharp peaks at certain characteristic frequencies due to its regular periodic motion, same as the case beyond $fM\sim 10^{-0.25}$. It is noteworthy to mention here that the appearance of such discrete characteristic frequencies from a regular or non-chaotic motions has been analytically derived by Drasco {\it{et al.}} for rotating BHs in Ref.~\cite{Drasco:2003ky}, which is in well agreement with our findings. In contrast to the relatively clean (discrete) spectra of non-chaotic cases, the spectra of chaotic cases display numerous irregular spikes. However, the peaks are not distinctly sharp; instead, they appear broadened due to the presence of numerous overlapping spikes, same as beyond $fM\sim 10^{-0.25}$. Such noisy spectral features are characteristic signatures of GWs emanating from chaotic orbital dynamics in EMRIs \cite{Kiuchi:2007ue}. Therefore we may conclude here that the distinct spectral signatures between non-chaotic and chaotic motion for the entire frequency range $fM\sim(10^{-3.0}-10^{0.5})$ provide a potential tool for constraining particle motion as well as DM halo parameters. \\
Additionally, let us mention that that for our considered BH-DM halo spacetime, the characteristic frequencies of GWs produced by EMRIs for three different non-chaotic, onset-of-chaos and chaotic orbits fall within the detectable range. These frequencies are particularly well-suited for detection by the future space-based GWs observatories, which we will discuss in the following section. Although spectral analysis provides a clear distinction between regular and chaotic dynamical systems and its imprints on GWs, it would be advantageous to validate our findings using a more sophisticated approach and interpret our conclusions strongly as stated in the above sections. Such a method should be capable of distinctly characterizing chaotic and regular dynamics and its signatures in GWs. In what follows, before proceeding with further discussions on the possible detection of chaotic imprints in GWs, let us briefly examine the signatures of chaos in Gws through orbital and waveform recurrence analysis, inspired by the recent works of Zelenka \textit{et al.} \cite{Zelenka:2019nyp,Zelenka:2024fjg}.

\begin{figure*}[!t]
    \centering
    \includegraphics[width=0.9\linewidth]{Figures/Figure3.png}
    \caption{{\it (a):} Recurrence plots of the gravitational waveforms $h_+$, generated using the kludge scheme illustrating three distinct orbits for Case-III, with fixed parameters $\rs = 0.15$ and $E = 90$, are presented. {\it Left:} It displays $h_+$ mode for the non-chaotic (regular) orbit corresponding to $\rhs = 0.01$. For this regular motion, the embedding parameters are $d = 8$, $\tau = 14$, and $\epsilon = 1.7489$ with a recurrence rate $\mathcal{RR} = 0.06$. {\it Middle:} This shows the $h_+$ mode of orbit at the onset of chaos for $\rhs = 0.04$. Here, the embedding dimension is $d = 7$, the delay time $\tau = 16$, the threshold $\epsilon = 1.7835$, and the recurrence rate $\mathcal{RR} = 0.08$. {\it Right:} Finally, it depicts the $h_+$ waveform of fully chaotic orbit for $\rhs = 0.05$. The parameters for this chaotic motion are $d = 9$, $\tau = 22$, $\epsilon = 1.9769$, and $\mathcal{RR} = 0.15$. We note that the plus polarization associated with non-chaotic orbits generally produces extended diagonal lines. In contrast, chaotic orbits generate a more intricate, square-shaped pattern in the corresponding plus polarization of the waveforms as the central density $\rhs$ increases. \\
    {\it (b):} Recurrence plots of the gravitational waveforms $h_\times$, generated using the kludge scheme illustrating three distinct orbits for Case-III, with fixed parameters $\rs = 0.15$ and $E = 90$, are presented. {\it Left:} Displays $h_\times$ mode for the non-chaotic (regular) orbit corresponding to $\rhs = 0.01$. For this regular motion, the embedding parameters are $d = 8$, $\tau = 13$, and $\epsilon = 1.6365$ with a recurrence rate $\mathcal{RR} = 0.06$. {\it Middle:} It shows the $h_\times$ mode of orbit at the onset of chaos for $\rhs = 0.04$. Here, the embedding dimension is $d = 7$, the delay time $\tau = 18$, the threshold $\epsilon = 1.7704$, and the recurrence rate $\mathcal{RR} = 0.08$. {\it Right:} Finally, this figure depicts the $h_\times$ waveform of fully chaotic orbit for $\rhs = 0.05$. The parameters for this orbit are $d = 10$, $\tau = 16$, $\epsilon = 1.6388$, and $\mathcal{RR} = 0.15$. We note that similar to the plus mode, the cross polarization associated with non-chaotic orbits generally produces extended diagonal lines. In contrast, chaotic orbits generate a more intricate, square-shaped pattern in the corresponding cross polarization of the GWs as the central density $\rhs$ increases.}
    \label{frp_plus} 
\end{figure*}

\section{Signature of chaos through recurrence analysis of orbits and gravitational waveforms}\label{sec:rp}
Recurrence analysis is a powerful tool for investigating the dynamical properties of a nonlinear system from time series data. The core idea involves reconstructing the system's phase space and analyzing the recurrences of its states, which are visualized through a Recurrence Plot (RP). It serves as a tool for examining the dynamics of both regular and chaotic systems. This technique enables the estimation of dynamical invariants, including the second-order Rényi entropy and correlation dimension, while also revealing features such as unstable periodic orbits and chaotic orbits. Additionally, it provides a means to differentiate between linear and nonlinear dynamical behaviors. For comprehensive details on recurrence analysis and various metrics used to quantify recurrence plots for a thorough time series examination, refer to the following Refs.~\cite{Marwan,Yu,Lukes-Gerakopoulos:2017jub}. In what follows, we give a brief discussion on the theoretical foundations of recurrence analysis in our Supplemental Material. \\
An additional benefit of recurrence analysis is its applicability to time series data, which is particularly useful when examining GWs strain signals. Therefore, in this work, we evaluate the effectiveness of recurrence analysis by comparing its results with those obtained from the conventional Poincaré section and Lyapunov exponent methods (see our previous work \cite{Das:2025vja} for details). \\
For the recurrence analysis, we reconstruct the phase space from the time series using Takens's time-delay embedding method \cite{Takens}. According to Takens's theorem, the reconstructed time series is defined as
    \begin{eqnarray}
        \mathbf{Y}_i = \left(\mathbf{X}_i, \mathbf{X}_{i+\tau}, ..., \mathbf{X}_{i+\tau\left(d-1\right)}\right)~,\label{eq:Takens}
    \end{eqnarray}
where $\mathbf{X}_i$ is a time series of dimension $n$ and length $l \geq i \in \mathbb{N}$. Here $\tau\in\mathbb{N}$ is the time delay and $d\in\mathbb{N}$ is the embedding dimension. The recurrence threshold $\varepsilon$ is determined by tuned to fit for a fixed recurrence rate $\mathcal{RR}$, following the standard procedures \cite{Zelenka:2019nyp,Marwan,Yu}. We have utilized the Euclidean norm \cite{Marwan, Kopacek:2010yr} for the numerical computation and in our recurrence analysis, the time series is standardized to achieve zero mean and unit standard deviation. To reconstruct the phase space trajectory, we employ the embedding method. The embedding dimension $d$ is computed for each time series using the False Nearest Neighbors (FNN) algorithm \cite{Kennel}, while the embedding time delay $\tau$ is identified as the first minimum of the time-delayed mutual information, as described detailed in the Supplemental Material. We have computed these parameters separately for each orbits (see the Supplemental Material for a brief discussion on Takens's embedding and FNN methods). \\
First we begin by performing recurrence analysis on the radial coordinate $r$ along the orbit at constant time interval $\Delta t$ by generating a time series for two of our cases i.e., for fixed $E=90~\rs=0.15$ with different values of the central density $\rhs$ (Case-III) and for fixed $E=115~\rhs=0.01$ with different values of the scale radius $\rs$ (Case-IV) . This particular time series is used to construct the radial recurrence plots and their corresponding reconstructed phase space (showing in $3$ dimensions) along each orbits. We set a fixed recurrence rate $(\mathcal{RR})$ to $0.06$ for non-chaotic orbit, $0.08$ for onset-of-chaos orbit, whereas for chaotic orbit, a higher value of $\mathcal{RR}=0.15$ proves more suitable for both of the Cases-III and IV. These values were chosen following common practice in recurrence analysis \cite{Zelenka:2019nyp,Marwan,Yu}. For regular (non-chaotic/periodic) dynamics, a low $\mathcal{RR}$ (around $0.05–0.08$) is typical to avoid saturation, while for chaotic dynamics a slightly higher value ($0.10–0.20$) is often used to capture sufficient recurrences without over-embedding. We chose $\mathcal{RR}=0.06,~0.08$, and $0.15$ as representative values that clearly distinguish the three regimes i.e., non-chaotic, onset-of-chaos and chaotic, respectively and we also verified that small variations $(\pm 0.01)$ do not change the qualitative patterns. We also note that the choice is adaptive in the sense that we keep $\mathcal{RR}$ fixed for each regime across different cases to allow fair comparison. \\
The computed recurrence plots of orbit along radial direction are displayed in Figs.~S6 and S8 for Case-III and IV, respectively in the Supplemental Material. A visual examination of these plots reveal distinct patterns corresponding to non-chaotic and chaotic dynamical trajectories. In the non-chaotic case (Figs.~S6(a) and S8(a)), the recurrence plots exhibit a simple diagonal structure, which is a characteristic signature of a regular (or periodic) motion. However, this clear pattern becomes little bit disrupted when weak chaos introduced via the onset-of-chaos orbits (Figs.~S6(b) and S8(b)). Finally for $\rhs=0.05$ and $\rs=0.25$, we typically find a recurrence plots of strongly chaotic dynamics (Figs.~S6(c) and S8(c)). \\
On the other hand, the reconstructed phase space trajectories are also shown in Figs.~S7 and S9 corresponding to different orbits for both of the considered cases in the Supplemental Material. It is clearly visible in Figs.~S7(c) and S9(c) that chaos actually appears for DM halo parameters $\rhs=0.05$ and $\rs=0.27$, respectively, compared to the non-chaotic and onset-of-chaos orbits (Figs.~S7(a), S7(b) for Case-III and Figs.~S9(a), S9(b) for Case-IV). \\
Now, we apply the same recurrence analysis technique to the simulated time-domain gravitational waveforms associated with these particular orbits, namely non-chaotic, onset-of-chaos and chaotic orbits. These waveforms are numerically generated using the kludge method as outlined in Sec.~\ref{sec:kludge}. Specifically, we analyze the time series of $h_+$ and $h_\times$, as discussed in Subsec.~\ref{sec:modes}. Similar to the previous analysis, we reconstruct the phase space trajectory using the delay embedding method and fix the recurrence rate to determine the threshold. The resulting recurrence plots for the two modes of gravitational waveforms are presented in Figs.~\ref{frp_plus}(a), (b) for Case-III, respectively and Figs.~S10, S11 for Case-IV, respectively. Specifically for Case-III, it is clearly visible that in Figs.~\ref{frp_plus}(a), (b), the left panel exhibit diagonal lines, a signature of ordered motion, while in the right panel, they are characterized by square-shaped patterns, which are indicative of chaotic behavior. The similar patterns are also observed by looking the individual plots for Case-IV in Figs.~S10(a), S11(a), which exhibit diagonal structure for non-chaotic waveforms, while they are broken for the chaotic polarizations of GWs in Figs.~S10(c), S11(c). Therefore our analysis on these plots clearly indicate the presence of chaos and exhibit a qualitative resemblance to the recurrence plots of the corresponding orbital trajectories. Thus, we conclude that gravitational waveforms emitted by EMRIs in a galactic settings, in which a supermassive Schawarzschild-like BH immeresed in a Dehnen-$(1,4,5/2)$ type DM halo, retain the signature of chaotic dynamics, which can also be clearly identified easily (and cross-checked) through the powerful methods of recurrence analysis.  In conclusion, our analysis reveals a strong connection between the dynamical behavior of a secondary’s motion around a supermassive Schwarzschild-like BH with DM halo and their corresponding gravitational waveforms. This link is evident not only in the spectrum analysis of the GWs but also in the recurrence analysis of the orbits and their associated gravitational waveforms. These findings suggest a possible realistic signature of chaos in the emitted GWs by an EMRI system in a galactic environments.

\section{Scopes of detectability by the future-based detection antennas}\label{sec:detection}
\begin{figure*}[!t]
    \centering
    \includegraphics[width=0.9\linewidth, height=0.7\textheight]{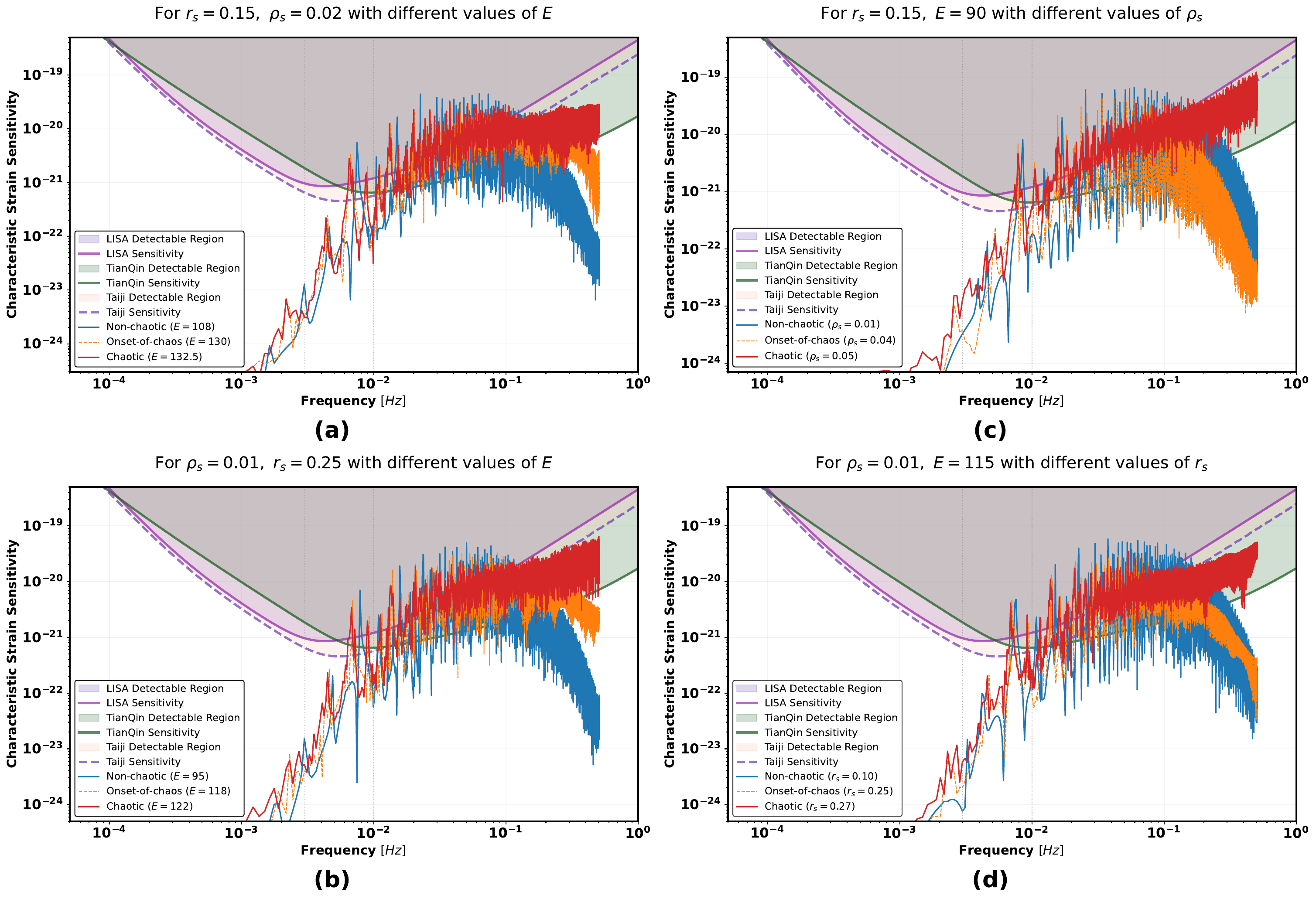}
    \caption{Comparison of the characteristic strain $h_c(f)$ of the gravitational waveforms generated using kludge scheme corresponding to the non-chaotic, onset-of-chaos and chaotic orbits for our four different cases with the sensitivity curves of LISA, TianQin and Taiji. In all of the figures ((a)-(d)), we plot the dimensionless characteristic strain sensitivity $\sqrt{fS_n^L},~\sqrt{fS_N},$ and $\sqrt{fS_n^T}$ for LISA, TianQin and Taiji, respectively.}
    \label{fdetect}
\end{figure*}

Our numerical analysis demonstrate that the typical characteristic frequencies of GWs emitted by a secondary object orbiting around a supermassive Schwarzschild-like BH embedded in a Dehnen-$(1,4,5/2)$ type DM halo generally fall within the observable range of space-based gravitational wave observatories, which is associated with three typical orbital motions namely, non-chaotic, onset-of-chaos and chaotic orbits (refer to the earlier discussions in Sec.~\ref{sec:spectra}). This analysis allows us to evaluate the detectability of the produced GWs arising from such non-chaotic, onset-of-chaos, and chaotic orbits using upcoming space-based gravitational wave detectors like LISA, TianQin, and Taiji. In the Supplemental Materiale, we provide a concise overview of the sensitivity curves for these detectors, which will be used for our analysis. \\
Let us now examine the detectability challenges associated with these gravitational wave detectors. To assess the detectability of GWs emitted from EMRI in a supermassive Schwarzschild-like BH embedded in a Dehnen-$(1,4,5/2)$ type DM halo associated with non-chaotic, onset-of-chaos, and chaotic orbits, we calculate the characteristic strain sensitivity using the frequency spectra derived from such typical orbits. The characteristic strain is computed using the following definition.
    \begin{equation}
        h_c(f) = 2|f| \sqrt{|\hat{h}_+(f)|^2 + |\hat{h}_{\times}(f)|^2}~.\label{eq:strain}
    \end{equation}
Subsequently, we analyze the characteristic strain derived for three distinct orbital classifications-non-chaotic, onset-of-chaos, and chaotic-across our four different cases (see Table~\ref{tab1}). These results are then contrasted with the sensitivity curves of the gravitational wave observatories, including LISA, TianQin, and Taiji. \\
From Fig.~\ref{fdetect}, it is evident that certain portions of the strains, which are associated with varying central densities and radii of the DM halo, or differing energy values, are above the sensitivity thresholds of LISA, TianQin, and Taiji. This suggests that upcoming space-based gravitational wave detectors could potentially observe GW signals originating from EMRIs in a supermassive Schwarzschild-like BH with a DM halo associated with the regular and chaotic orbital dynamics. Particularly for TianQin, acceleration noise dominates at lower frequencies ($f < 10^{-3}~\mathrm{Hz}$), while laser frequency noise and shot noise become significant at higher frequencies ($f > 10^{-1}~\mathrm{Hz}$). Consequently, around $10^{-2}~\mathrm{Hz}$, these noise contributions balance out, resulting in TianQin's optimal strain sensitivity \cite{TianQin:2020hid,Li:2024rnk}. Consequently, TianQin performs better than LISA and Taiji at frequencies above $10^{-2}~\mathrm{Hz}$, due to its lower optical metrology noise. Conversely, in the lower frequency range ($f < 10^{-3}~\mathrm{Hz}$), TianQin exhibits higher noise due to its shorter arm lengths, though its acceleration noise performance remains better. Meanwhile, Taiji demonstrates slightly better sensitivity than LISA in this regime because of its marginally longer arms with respect to LISA. On the other hand, in the intermediate frequency band, $f \sim (10^{-3} - 10^{-2})~\mathrm{Hz}$, LISA and Taiji exhibit comparable sensitivity, while TianQin is less effective here due to its higher transfer frequency. The optimal frequency for GWs detector corresponds to the frequency where it achieves its best sensitivity (i.e., minimal strain noise). \\
Moreover considering all of our four different cases, we observe that TianQin can detect the typical signal, coming from chaotic orbital dynamical motion at most level, whereas LISA and Taiji are only capable of detecting a certain portions of it.
Additionally, it is worth noting that beyond $f \sim 10^{-1}~\mathrm{Hz}$, we observe a sharp decline in the characteristic strain for non-chaotic and onset-of-chaotic motions. However, no such decline is observed for fully developed chaotic motion, due to the presence of high non-linearity in the orbital dynamics. \\
We also note that our detectability assessment uses the instrumental sensitivity only, without including the unresolved galactic binary confusion noise, which can raise the effective noise floor, especially for LISA. This confusion noise is non-stationary and diminishes over the mission lifetime as sources are resolved. Its inclusion would reduce the signal-to-noise ratio for all orbits, considered in this work; however, chaotic signals, which exhibit broader frequency content, higher amplitudes and non-declined strain, are likely to be more resilient than the narrower-band, declined non-chaotic signals. Thus, while the exact detectability thresholds would shift, the relative advantage of chaotic orbits and the qualitative conclusions of our analysis remain robust. A more precise study including confusion noise using the new LISA design \cite{Cornish:2017vip} is left for future work.

\section{Discussions and conclusion}\label{sec:conclusion}
Let us briefly discuss and summarise our results regarding what we have achieved in this paper.   

\begin{itemize}

\item Our previous work \cite{Das:2025vja} involved a detailed analysis of the chaotic dynamics exhibited by a massive test particle inspirals around a supermassive Schwarzschild-like BH surrounded by a Dehnen-$(1,4,5/2)$ type DM halo, within an EMRI configuration. We began with a rigorous derivation and examination of the dynamical equations governing the motion of a massive test particle in the vicinity of the BH's event horizon. We introduced and justified the adoption of Painlevé-Gullstrand coordinates, emphasizing their effectiveness in dealing with numerical instabilities usually encountered in the near-horizon regime. Furthermore, we discussed the strategic use of an external potentials (harmonic confinement in this case) to maintain stable particle trajectories, which is essential for investigating the chaotic transitions induced by gravitational effects in the near-horizon region.

\item Subsequently, in \cite{Das:2025vja}, we performed an extensive numerical study of the system's chaotic dynamics. This involved computing and analyzing Poincaré sections, tracking the orbital evolution of two close initial geodesics, and calculating Lyapunov exponents. Our simulations demonstrated that chaotic behavior becomes markedly more pronounced with increasing particle energy and with variations in the DM halo parameters, specifically the core density $\rho_s$ and the scale radius $r_s$. The emergence of chaos was clearly identified through the disintegration of regular toroidal structures in the Poincaré sections. These qualitative observations were quantitatively validated by positive Lyapunov exponents, which were verified to be consistent with the universal chaos bounds established by the BH's surface gravity.

\item The present work is primarily dedicated to investigating the GWs emitted by an EMRI system involving a supermassive BH with a DM halo. Building upon our previous findings that the system transitions to chaos under specific parameter space \cite{Das:2025vja}, we specifically analyze the horizon's influence during both regular (non-chaotic) and chaotic dynamical states. We observe that this behavioral transition of the system imprints itself on the GWs emitted from the near-horizon region, modifying their morphology and frequency spectrum in distinctive ways. To this end, we detailed the extraction of gravitational waveforms using the numerical kludge scheme, explicitly contrasting the signals originating from chaotic orbits with those from non-chaotic orbits. Our results demonstrate that the chaotic trajectories considered here produce gravitational waveforms characterized by loss of regular periodicity, broader and more quasi-continuous frequency content, irregular strain excursions, modified radiative behavior, and distinctive recurrence patterns. We emphasize that the absolute strain amplitude is trajectory dependent and is not, by itself, a monotonic diagnostic of chaoticity. The more robust signatures of the transition are instead the degradation of waveform periodicity, the redistribution of spectral power over a broader frequency range, and the qualitative changes observed in the recurrence structure.

\item A key novel contribution of this study is a comprehensive detectability analysis. We rigorously compared our numerically generated chaos-imprinted gravitational wave signatures with the projected sensitivity curves of upcoming space-based gravitational wave observatories, namely LISA, TianQin, and Taiji. Our findings indicate that these detectors possess the requisite sensitivity to observe and analyze GW signals emanating from chaotic EMRI systems. This analysis underscores the astrophysical feasibility and significance of detecting the imprints of chaos within gravitational wave data, forecasting substantial advancements for the field of gravitational wave astronomy.

\end{itemize}
\noindent
Looking forward, our findings provide a foundation for future exploration of chaotic dynamics in other astrophysically significant BH systems. This includes astrophysical environments such as accretion disks, stellar clusters, matter spikes, and disks, which can significantly alter the orbital dynamics. Furthermore, since our current analysis deals with a spinless secondary, a natural extension for subsequent work would be to include the effects of secondary's spin. In the present study, we employ the adiabatic approximation, which neglects the back-reaction of gravitational radiation on the orbital dynamics. This approach remains valid for simulations encompassing a few of orbital periods, as is the situation, considered in this work. Examining the influence of gravitational radiation on the long-term behavior of chaotic orbits is a promising direction for further investigation. Additionally, our model does not account for multipole contributions beyond the quadrupole order. For the development of more precise gravitational waveform templates, it is essential to incorporate the higher-order multipole moments into the GWs expansion. Finally, with the eventual development of such accurate waveform templates, we will be positioned to study how next-generation gravitational wave detectors could potentially constrain or probe the influence of DM on chaotic orbital behavior. We look forward to addressing these challenges in subsequent studies.

\section*{Supplementary Information} 
\noindent
Readers are encouraged to consult the Supplemental Material, where detailed explanations, interpretations of other plots are briefly presented with some additional References.
\\
\\
\textbf{Data Availability:} 
The analytical computations supporting the findings of this study are presented within the article and the Supplemental Material. The numerical code will be made available upon reasonable request from the corresponding author.
\\
\\
\textbf{Conflict of interest:}
The authors declare that they have no conflict of interest.

\section*{Acknowledgements}{The authors would like to thank the anonymous Reviewers for their constructive and insightful reports, which have significantly improved this paper. We thank Kyriakos Destounis for his insightful comments and for bringing key references to our attention. Surajit Das is grateful to Rickmoy Samanta and Surhud More for numerous discussions. Surajit Das further acknowledges the support of USTC Fellowship Level A--CAS-ANSO Scholarship 2024 for PhD candidates (formerly the ANSO Scholarship for Young Talents). Surojit Dalui thanks the Department of Physics, Shanghai University, for providing postdoctoral funds and support from the Super Postdoctoral Fund (Grant No. 2023307) during the period of this work. Bum-Hoon Lee thanks APCTP, KIAS and USTC for the hospitality during his visit, where a part of this project was done. This work was supported in part by the National Key R \& D Program of China (Grant No. 2021YFC2203100), the National Natural Science Foundation of China (Grant Nos. 12433002, 12261131497, and 125B1023), the Chinese Academy of Sciences Young Interdisciplinary Innovation Team (Grant No. JCTD-2022-20), the 111 Project (Grant No. B23042), the China Scholarship Council Innovation Talent Funds, the USTC Fellowship for International Cooperation, and the USTC Research Funds of the Double First-Class Initiative. Bum-Hoon Lee was supported by the National Research Foundation of Korea (NRF) (Grant Nos. RS-2020-NR049598, and RS-2026-25473640) and Overseas Visiting Fellow Program of Shanghai University. Numerical computations were performed on the computer facilities “LINDA $\&$ JUDY" in the Particle Cosmology group (COSPA) at USTC.}

\newpage
\onecolumngrid
\renewcommand{\theequation}{s\arabic{equation}}

\renewcommand{\thefigure}{S\arabic{figure}}

\renewcommand{\thetable}{S\arabic{table}}

\hypersetup{
	colorlinks   = true, 
	urlcolor     = blue, 
	linkcolor    = blue, 
	citecolor    = blue    
}


\hypersetup{
	colorlinks   = true, 
	urlcolor     = blue, 
	linkcolor    = blue, 
	citecolor    = blue 
}
\maxdeadcycles=1000 

\def\rhs{\rho_s}
\def\rs{r_s}
\def\a{\alpha}
\def\b{\beta}
\def\g{\gamma}
\def\d{\delta}
\def\s{\sigma} 
\def\o{\omega}
\def\O{\Omega}
\def\m{\mu}
\def\n{\nu}
\def\ph{\phi} 
\def\non{\nonumber}
\def\n{\noindent}

\definecolor{lightblue}{RGB}{173,216,230}
\definecolor{lightgreen}{RGB}{144,238,144}
\definecolor{magenta}{RGB}{255,0,255}
\definecolor{myolive}{RGB}{128,128,0}
\definecolor{masteredyellow}{RGB}{255,255,102}
\definecolor{mymaroon}{RGB}{128,0,0}
\newcommand{\hb}[1]{\textcolor{orange}{\textit{\textbf{HB:} #1}}}

\setcounter{figure}{0}
\setcounter{equation}{0}

\section*{Supplemental Material for ``\textit{Chaotic Imprints in Gravitational Waves from Extreme-Mass-Ratio-Inspirals Embedded in Dark Matter Halo}"}

\section*{Background dynamics of an extreme-mass-ratio inspirals around a Schwarzschild-like black hole embedded in dark matter halo}\label{S:background}
\noindent
Here we mention in short about the background spacetime and the procedures, we have implemented to find the typical parameter space from a non-chaotic to chaotic transitions, investigated in our previous work \cite{Das:2025vja}. For a more detail in each and every step on the particle dynamics and transitions from non-chaotic to chaotic regime, reader refer to our previous work \cite{Das:2025vja}. \\
The double power-law density distributions of DM halos and elliptical galaxies is given by \cite{Van,Hernquist:1990be}
    \begin{equation}  
        \rho = \rho_s \left( \frac{r}{r_s} \right)^{-\gamma} \left[ \left( \frac{r}{r_s} \right)^{\alpha} + 1 \right]^{\frac{\gamma - \beta}{\alpha}}~,\label{2ca}  
    \end{equation} 
where $\a,\b,\g$ are free parameters and $\rhs$, $\rs$ are known as the typical density and scale radius of the central DM halo, respectively. Under the consideration of a background spacetime metric ansatz (Eq.~(1)) with the DM halo density profile (Eq.~\eqref{2ca}), one can find a Schwarzschild-like BH solution embedded in a Dehnen-$(1,4,5/2)$ type DM halo, which is given in Eq.~(2). In the current study, we will conduct our entire analysis based on this specific solution (Eq.~(2)). For more details on the solution of Schwarzschild-like BH embedded in a Dehnen-$(1,4,5/2)$ type DM halo, see \cite{Al-Badawi:2024asn,Das:2025vja} and the references therein. \\
Now we review the dynamics of a massive probe particle in a BH-DM halo combined spacetime. We are specifically interested in examining how the DM halo parameters ($\rhs,\rs)$ influence a massive test particle situated near the event horizon of a Schwarzschild-like BH embedded in a galactic DM halo. To facilitate this investigation, we adopt a more suitable coordinate system, the Painlev\'e-Gullstrand (PG) coordinates \cite{PG}. Under the Painlev\'e-Gullstrand transformation, Eq.~(1) transforms to the following metric, which is non-singular at the horizon.
    \begin{equation}
        ds^2 = -f(r)dt^2 + 2\sqrt{1-f(r)}dt dr + dr^2 + r^2 d\Omega^2.\label{eq:PG}
    \end{equation}
Next, we consider by defining the Lagrangian, subject to an external potential $V(x^\alpha)$ of a massive test particle of mass $m$ as given by
    \begin{equation}
        \mathcal{L} = -m \sqrt{ -g_{\mu\nu} \dot{x}^\mu \dot{x}^\nu }-V(x^\mu)~,\label{eq:Lag1}
    \end{equation}
where $g_{\mu\nu}$ denotes the metric tensor, given by Eq.~\eqref{eq:PG}. Throughout this analysis, we employ the static gauge condition, meaning the over dot denotes a derivative with respect to the proper time. Now, regarding the external potential, in a realistic astrophysical setting this external potential $V(x^\mu)$ may be interpreted as an effective perturbation arising, for example, from the gravitational influence of nearby stars or from perturbations of the surrounding DM halo. Such effects can break the exact integrability of the idealized background motion. We introduce $V(x^\mu)$ phenomenologically in order to keep the particle in a bounded near-horizon region and to probe the resulting horizon-induced nonlinear dynamics in a controlled manner. Similar effective external potentials have also been employed in earlier studies in chaotic dynamics \cite{Hashimoto:2016dfz,Lei:2020clg,Wang:2026whj}, where the origin of such potentials can be considered as the form of electromagnetic, scalar or higher spin potentials. Due to the existence of spherical symmetry, we are considering the particle motion in the equatorial plane, defined by $\theta=\frac{\pi}{2}$ with $p_{\theta}=0$. Next we take the general external potential $V(x^\mu)$ appearing in Eq.~\eqref{eq:Lag1} in a separable form, $V(x^\mu)\big|_{\theta=\pi/2}=a(r)+b(\phi)$, where $a(r)$ and $b(\phi)$ represent, the radial and azimuthal contributions to the same effective external perturbation, respectively. Therefore, the two potentials introduced in Eq.~\eqref{eq:Lag1} should be understood as a decomposition of $V(x^\mu)$ in the reduced phase-space description, rather than as two unrelated external interactions. For the numerical analysis, we adopt the simplest phenomenological choice, i.e. the harmonic potentials which provides a controlled confinement of the particle near the horizon and allows us to isolate the resulting nonlinear dynamics.\\
From the preceding Lagrangian, the generalized momenta corresponding to the $r$ and $\phi$ coordinates are given by:
    \begin{eqnarray}
        p_r & = & \frac{m\left( \dot{r} + \sqrt{1-f(r)} \right)}{\sqrt{ f(r) - 2\sqrt{1-f(r)}\,\dot{r} - \dot{r}^2 - r^2\dot{\phi}^2 }}~, \nonumber \\
        p_\phi & = & \frac{m r^2 \dot{\phi}}{\sqrt{ f(r) - 2\sqrt{1-f(r)}\,\dot{r} - \dot{r}^2 - r^2\dot{\phi}^2 }}~.\label{eq:momenta}
    \end{eqnarray}
Therefore the conserved energy for the massive test particle can be written as:
    \begin{align}
        E & = p_r \dot{r} + p_\phi \dot{\phi} - \mathcal{L} \nonumber \\
        & = -\sqrt{1-f(r)}\, p_r \pm \sqrt{ p^2_r + \frac{p^2_\phi}{r^2} + m^2 } + \left( a(r) + b(\phi) \right)~,\label{3.4}
    \end{align}
Given our focus on dynamics extremely close to the horizon, our analysis specifically concerns outgoing particle (i.e., considering the positive sign in the above expression). Accordingly, one can derive the following set of dynamical equations of motion:
    \begin{eqnarray}
        \frac{dr}{dt} & = & \frac{\partial E}{\partial p_r} = -\sqrt{1-f(r)} + \frac{p_r}{\sqrt{p^2_r + \frac{p^2_\phi}{r^2} + m^2}}~,\label{3.8} \\
        \frac{dp_r}{dt} & = & -\frac{\partial E}{\partial r} = -\frac{\partial_r f(r)}{2\sqrt{1-f(r)}} p_r + \frac{p_\phi^2 / r^3}{\sqrt{p_r^2 + \frac{p_\phi^2}{r^2} + m^2}} \non\\
        &-& \partial_r a(r)~,\label{3.9} \\
        \frac{d\phi}{dt} & = & \frac{\partial E}{\partial p_\phi} = \frac{p_\phi / r^2}{\sqrt{p_r^2 + \frac{p_\phi^2}{r^2} + m^2}}~,\label{3.10} \\
        \frac{dp_\phi}{dt} & = & -\frac{\partial E}{\partial \phi} = -\partial_\phi b(\phi)~.\label{3.11}
    \end{eqnarray}
As we have mentioned earlier that it is instructive to consider a phenomenological choice that describes the near-horizon geometry of the BH-DM halo combined spacetime with external potentials and here we choose this as a harmonic oscillator, as the simplest case, since a particle motion near the event horizon of a Schwarzschild BH-DM halo spacetime does not refer to any concrete form of such external potentials. Hence we have considered $a(r)=\frac{1}{2}K_r (r-r_c)^2;~b(\phi)=\frac{1}{2}K_\phi r_H^2 \phi^2$. Here $K_r$ and $K_{\phi}$ are two strength constants of $a(r)$ and $b(\phi)$, respectively and $r_c$ denotes the center position of $a(r)$. It is reasonable to mention that this framework follows established approaches (see \cite{Hashimoto:2016dfz,Lei:2020clg,Wang:2026whj} for massive particle and \cite{Dalui:2018qqv,Bera:2021lgw,Das:2024iuf} for massless cases), while alternative potential choices would modify the trajectory dynamics. Importantly, these harmonic potentials represent an externally imposed constraint rather than an inherent feature of such Schwarzschild BH-DM halo spacetime. At sufficiently large distances from the horizon, the system remains integrable and regular, exhibiting periodic orbital behavior with preserved invariant tori. However, close to the horizon introduces strong nonlinear gravitational effects that disrupt these regular tori (in the sense of a Kolmogorov-Arnold-Moser tori) and thus the breakdown of integrability facilitates the detection of the horizon's distinctive influence.\\
The dynamical equations, mentioned above form the basis for the numerical investigations on chaotic dynamics. The rationale for employing the Painlev\'e-Gullstrand coordinate transformation, together with the specification of the Lagrangian for a massive probe particle that is influenced by an external potentials, is provided in our earlier work \cite{Das:2025vja}, where a comprehensive treatment of these topics can be found.

\section*{Brief discussion on numerical kludge gravitational waveforms}\label{S:kludge}
\noindent
The numerical kludge waveform approach involves combining a fully relativistic particle's motion with an approximate model for GWs emission \cite{Babak:2006uv}. To implement such method in our case, we first obtain the secondary's trajectory, which is generated in ``phase space" defined in the equatorial plane with constant energy and angular momentum in PG coordinates, as discussed in Sec.~\ref{S:background}. Next, the dynamics of a secondary object around a Schwarzschild BH, immeresed in a Dehnen-$(1,4,5/2)$ type DM halo, is numerically integrated to compute the coordinates of the secondary as a function of time\footnote{Since we have investigated chaos in the dynamics of a massive test particle and chaotic phenomena is probed by the near-horizon effect, we adopt the PG coordinates due to its regularity across the horizon (see the related text of our previous paper \cite{Das:2025vja}, along with the associate supplemental material for details), avoiding the coordinate singularities present in the Schwarzschild coordinates. While GWs emission is typically analyzed in the asymptotic wave zone, where both coordinate approaches converge to the same Minkowski limit, the choice of PG coordinates remains physically consistent. This approach ensures precise representation of the chaotic behavior close to the horizon, considering numerical complexity, while remaining consistent with conventional wave extraction techniques in the distant field region.} Finally, the kludge gravitational waveform is produced by constructing an ``equivalent'' trajectory in flat space. This is achieved by projecting the respective coordinate onto a spherical polar coordinate grid. Subsequently, the corresponding coordinate system is defined, as the Cartesian coordinates in flat space, as described below:
    \begin{eqnarray}
        &x(t) = r(t) \sin \theta(t) \cos \phi(t)~,\non\\
        &y(t) = r(t) \sin \theta(t) \sin \phi(t)~,\non\\
        &z(t) = r(t) \cos \theta(t)~.\label{5.1}
    \end{eqnarray}
Before proceeding further, it is important to note that as discussed by Babak {\it{et al.}} in their Ref.~\cite{Babak:2006uv}, the use of a derived flat-space trajectory serves solely as an input for the phenomenological waveform-generation modeling. Employing such an approximation results in GW amplitudes that lack precision; however, the derived GW frequencies remain accurate. This specific kludge approach has demonstrated strong performance, achieving approximately $95\%$ accuracy when benchmarked against Teukolsky-based waveforms for an EMRIs \cite{Babak:2006uv}. Consequently, after determining the trajectory of a particle within a flat spacetime geometry—where the particle orbits a supermassive BH situated within a Dehnen$(1,4,5/2)$ DM halo—it becomes feasible to utilize these kludge waveforms. Their acceptable quality enables them to serve a function in the analysis of space-borne gravitational-wave data, particularly in the search for an EMRIs where the central BH is influenced by a galactic DM halo.\\
In the weak-field regime, expressing the spacetime metric as $g_{\mu\nu} = \eta_{\mu\nu} + h_{\mu\nu}$, where $\eta_{\mu\nu}$ represents the Minkowski metric and $h_{\mu\nu}$ denotes small metric perturbations (i.e., $|h_{\mu\nu}|<<1$). The trace-reversed metric perturbation is defined through the relation $\bar h^{\mu\nu} \equiv h^{\mu\nu} - \frac{1}{2}\eta^{\mu\nu} h$, with $h = \eta^{\mu\nu}h_{\mu\nu}$ being the trace of the perturbation. By imposing the Lorentz gauge condition $\partial_\a\bar{h}^{\mu\alpha} = 0$, we obtain the linearized Einstein field equations in the form \cite{Poisson-Will}:
    \begin{eqnarray}
        \Box \bar{h}^{\mu \nu} = -16\pi \mathcal{T}^{\mu \nu}~\label{5.2},
    \end{eqnarray}
where $\Box$ represents the standard flat-space d'Alembertian operator, and $\mathcal{T}^{\mu\nu}$ is the effective energy momentum tensor, obeys the conservation relation $\partial_{\nu}\mathcal{T}^{\mu\nu}=0$. For gravitational wave detection at large distances, we specifically require the transverse-traceless components of $\bar{h}^{jk}$, necessitating an appropriate projection operation. In our coordinate system in which a supermassive BH centered DM halo, designate $(t,\mathbf{x})$ as the observer's position and $(t_p,\mathbf{x}_p)$ as the particle's position. The solution to the wave equation Eq.\eqref{5.2} takes the familiar retarded potential form as
    \begin{equation}
        \bar{h}^{ij}(t,\mathbf{x})
        = 4 \int \frac{\mathcal{T}^{ij}(t-|\mathbf{x} - \mathbf{x'}|, \mathbf{x'})}
        {|\mathbf{x} - \mathbf{x'}|} \; d^{3}x'~.\label{5.3}
    \end{equation}
Building upon Ref.~\cite{Press:1977ps}, Press developed an expression valid for fast motion, extended sources by systematically applying the conservation relation ($\partial_{\nu}\mathcal{T}^{\mu\nu}=0$) to Eq.~\eqref{5.3}. However in the slow-motion approximation, Eq.~\eqref{5.3} reduces to the standard quadrupole formula 
    \begin{eqnarray}
        \bar{h}^{ij}(t, \mathbf{x})
        &= \frac{2}{r} \left[ \ddot{I}^{ij}(t') \right]_{t'=t-r}~,\label{5.4}
        \end{eqnarray}
where overdots denote time derivatives. The source symmetric and traceless mass quadrupole moment, appearing in the above expression is defined as
    \begin{eqnarray}
        I^{ij} &=& \int x^{i}x^{j} T^{00} d^3x~.\label{5.5}
    \end{eqnarray}
Here $T^{00}$ is the $tt$-component of the energy-momentum tensor for the small secondary of mass $m$ with trajectory $Z^i(t)$, defined as \cite{Thorne:1980ru}
    \begin{eqnarray}
        T^{00}(t,x^i)=m~\delta^3\left(x^i-Z^i(t)\right)~.
    \end{eqnarray}
One can also include the next order terms (the mass octupole and current quadrupole moments of the source) in the Press formula. However, the presence of delta-function in the above expression of energy-momentum tensor simplifies the evaluation of the source quadrupole moment and under the slow-motion limit, we obtain the symmetric and traceless GW source quadrupole formula as mentioned in Eq.~(3).

\section*{Enlarged view of kludge waveforms for Case-III and Case-IV}
\begin{widetext}
\begin{figure}[H]
    \centering

    \subfigure[]{
        \includegraphics[width=0.30\textwidth,height=0.28\textwidth]{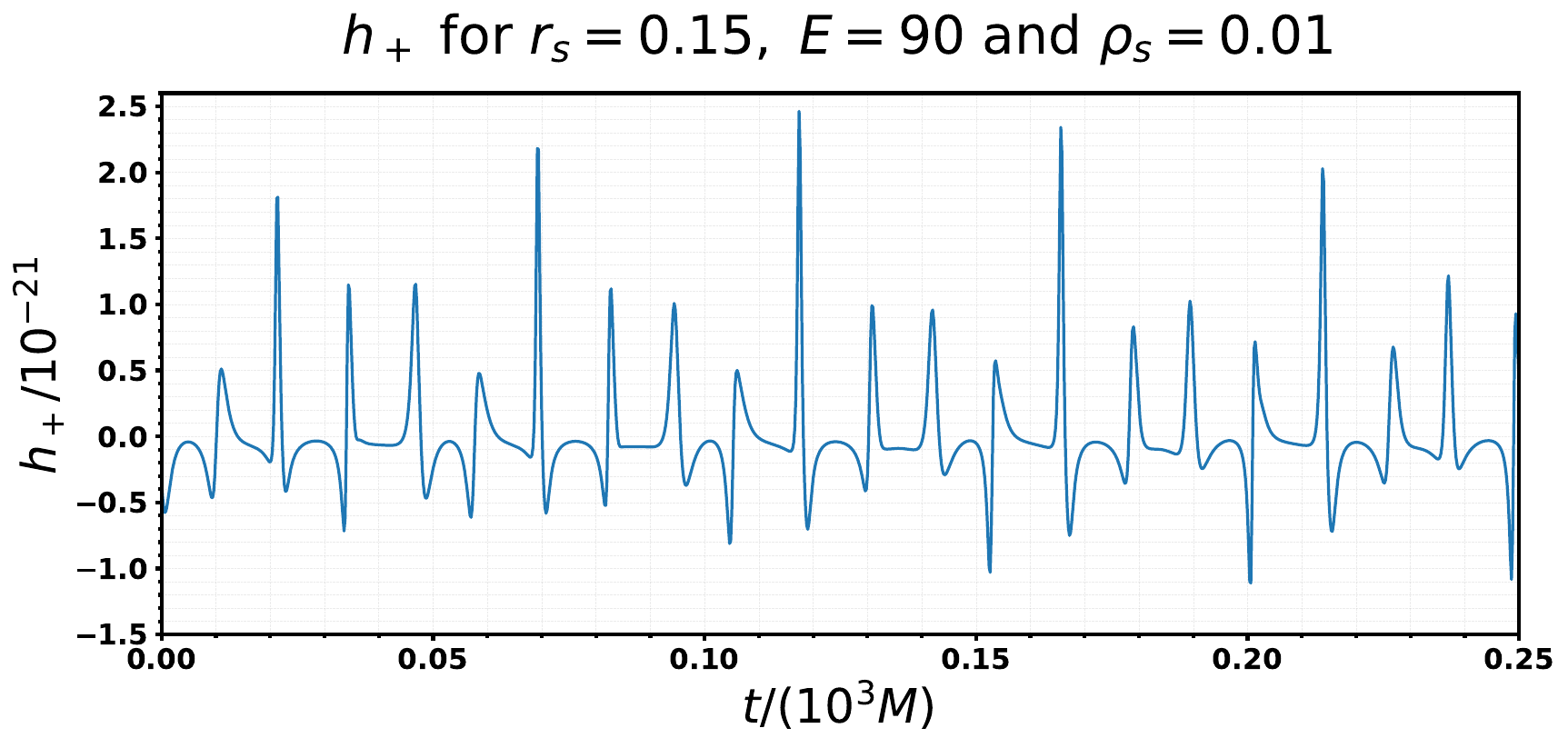}
        \label{gw_plus_p}
    }
    \hfill
    \subfigure[]{
        \includegraphics[width=0.30\textwidth,height=0.28\textwidth]{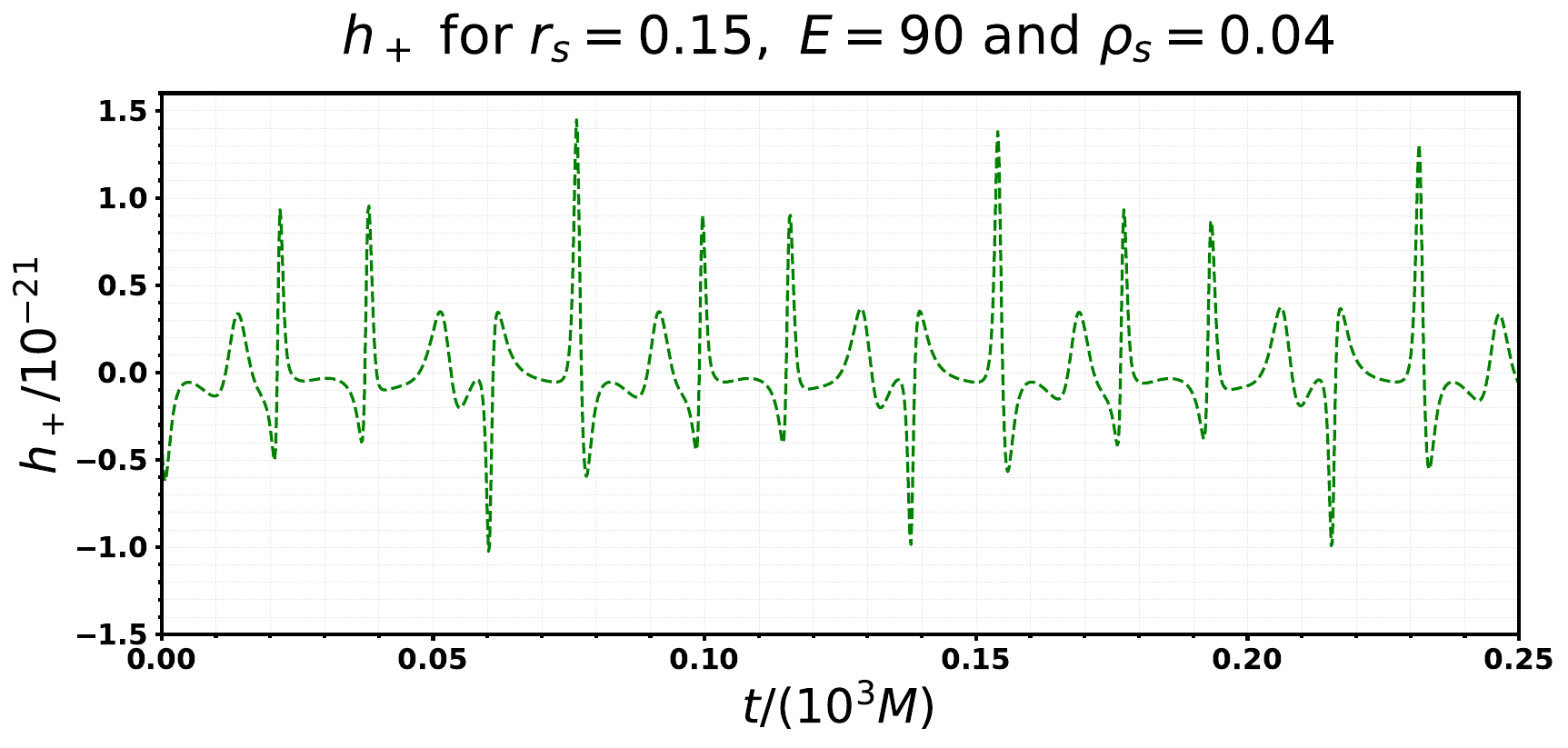}
        \label{gw_plus_oc}
    }
    \hfill
    \subfigure[]{
        \includegraphics[width=0.30\textwidth,height=0.28\textwidth]{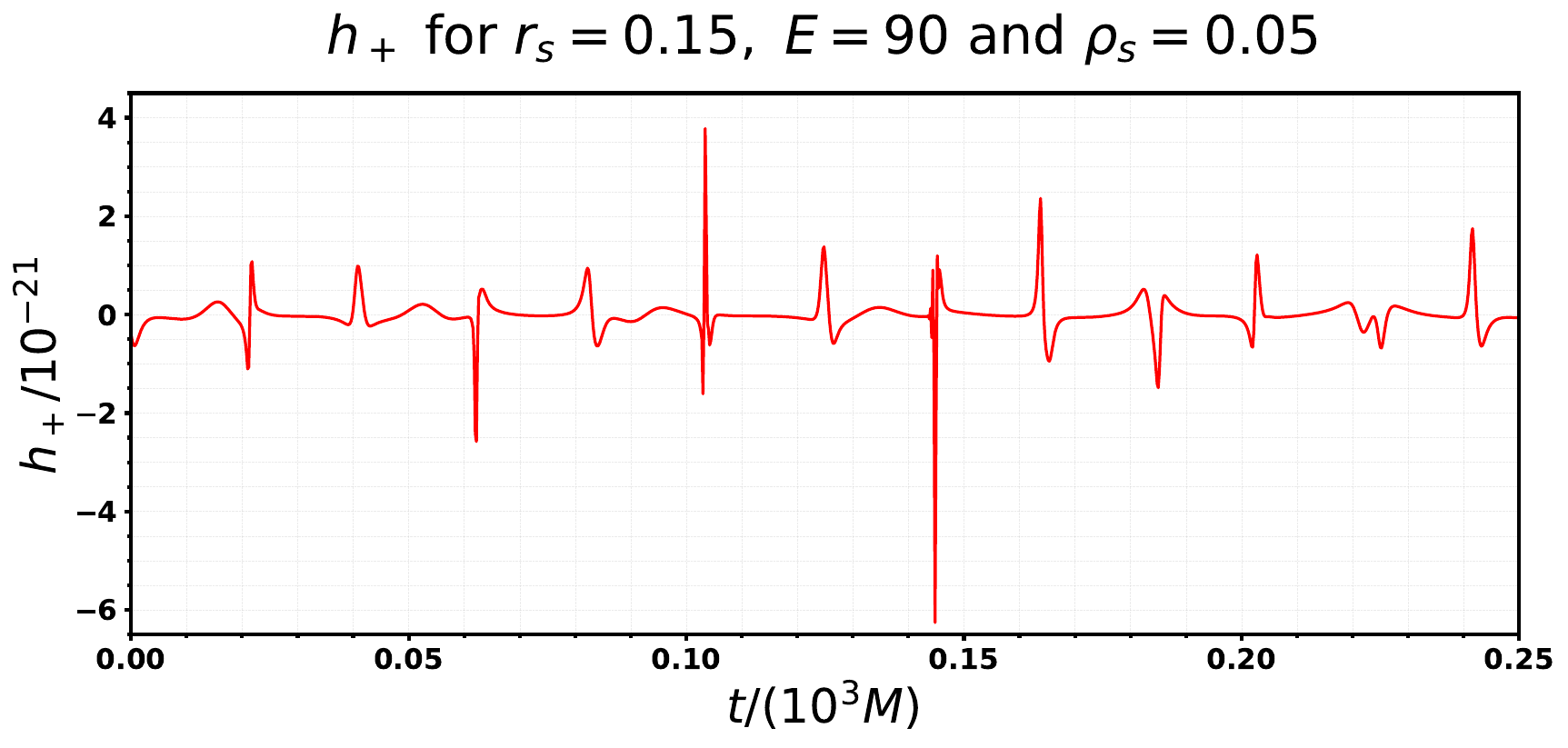}
        \label{gw_plus_c}
    }

    \vspace{0.15cm}

    \subfigure[]{
        \includegraphics[width=0.30\textwidth,height=0.28\textwidth]{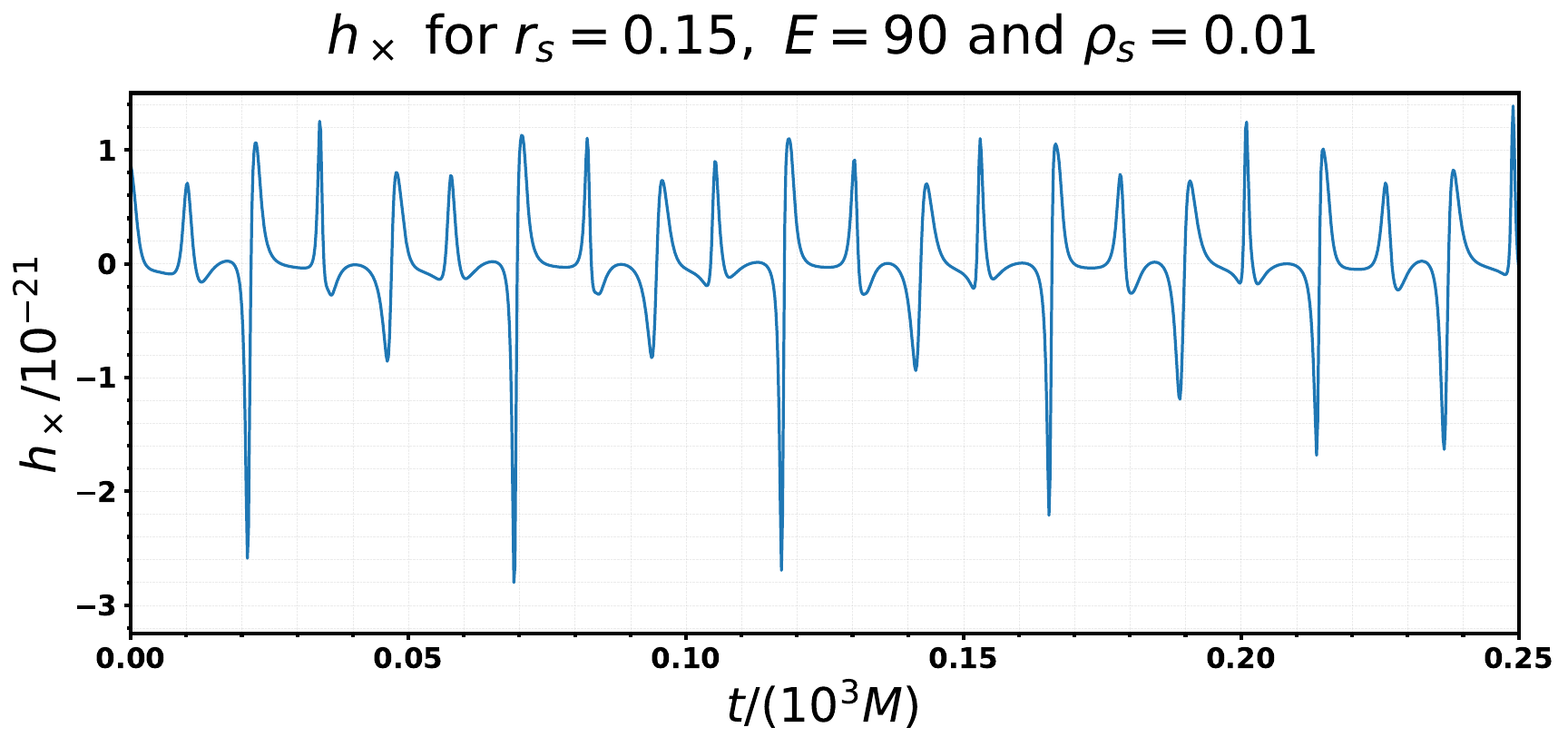}
        \label{gw_cross_p}
    }
    \hfill
    \subfigure[]{
        \includegraphics[width=0.30\textwidth,height=0.28\textwidth]{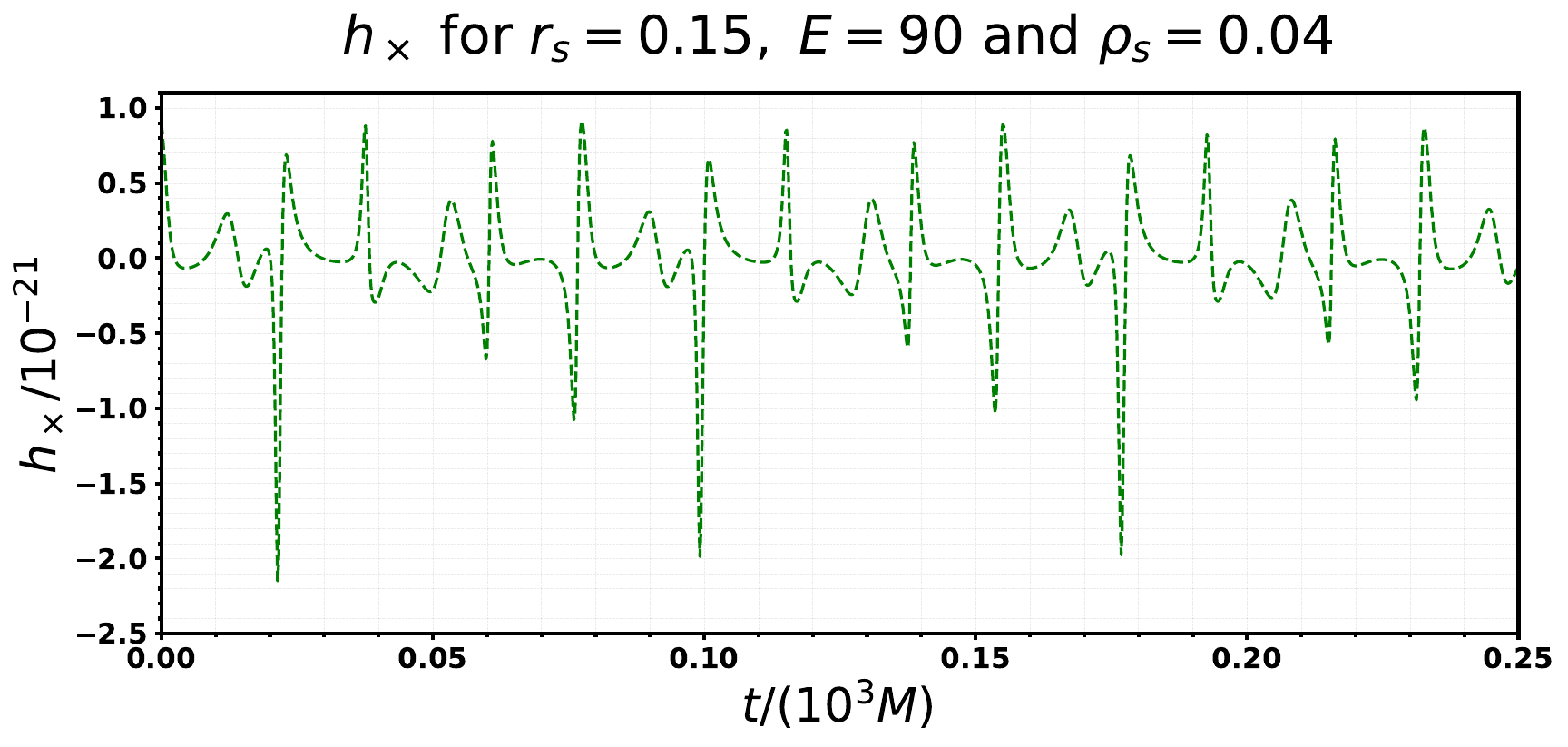}
        \label{gw_cross_oc}
    }
    \hfill
    \subfigure[]{
        \includegraphics[width=0.30\textwidth,height=0.28\textwidth]{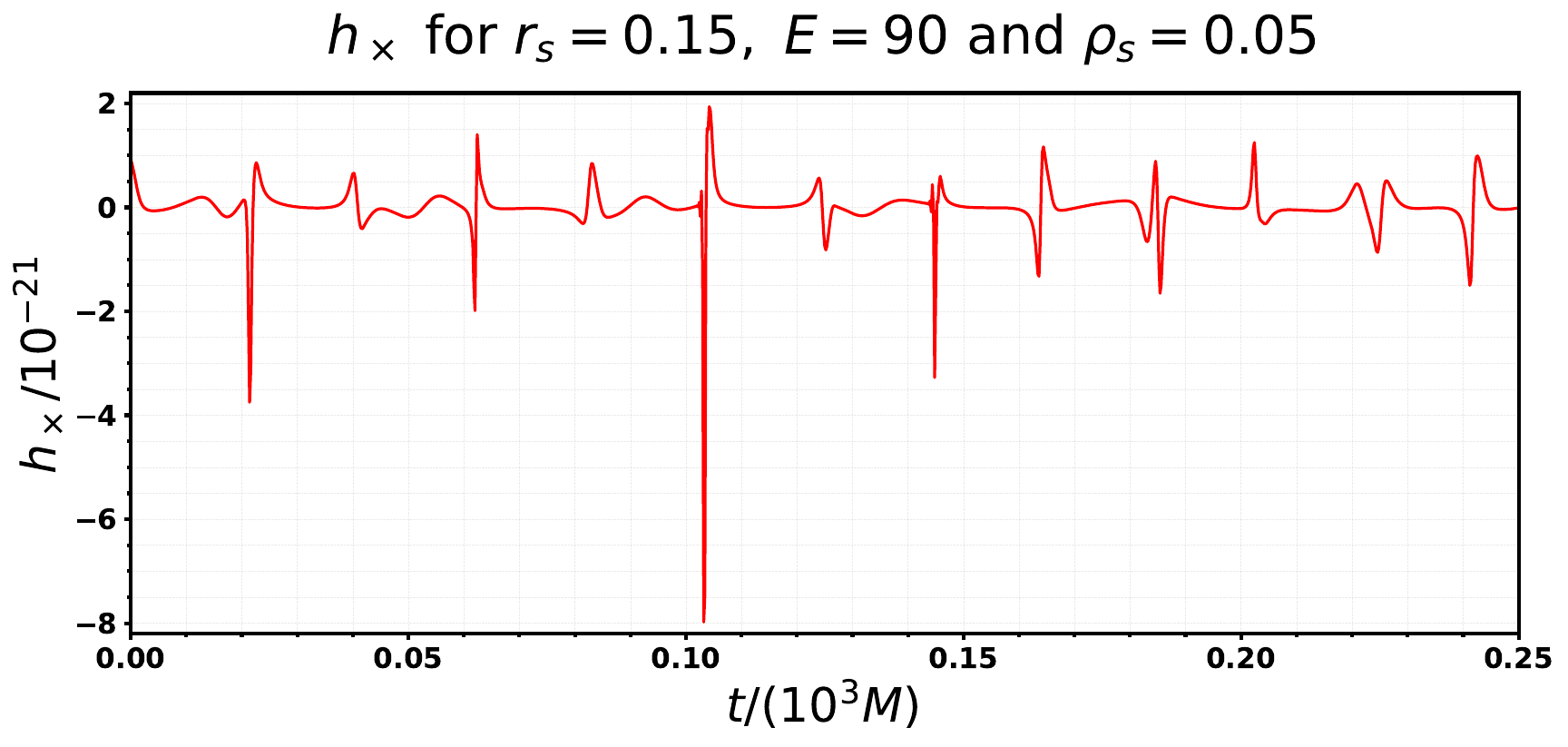}
        \label{gw_cross_c}
    }

    \caption{Gravitational waveforms of the $+$ and $\times$ modes corresponding to the non-chaotic, onset-of-chaos and chaotic orbits for fixed $\rs=0.15,~E=90$ with variations in density parameter $\rhs$ as our considered Case-III. Here the time range is shown from 0 to 0.25 in unit of $t/(10^3 M)$.}
    \label{fgw_individual}

    \hrulefill
\end{figure}
\end{widetext}

\begin{widetext}
\begin{figure}[H]
    \centering

    \makebox[\textwidth][c]{%
        \subfigure[]{
            \includegraphics[width=0.30\textwidth,height=0.28\textwidth]{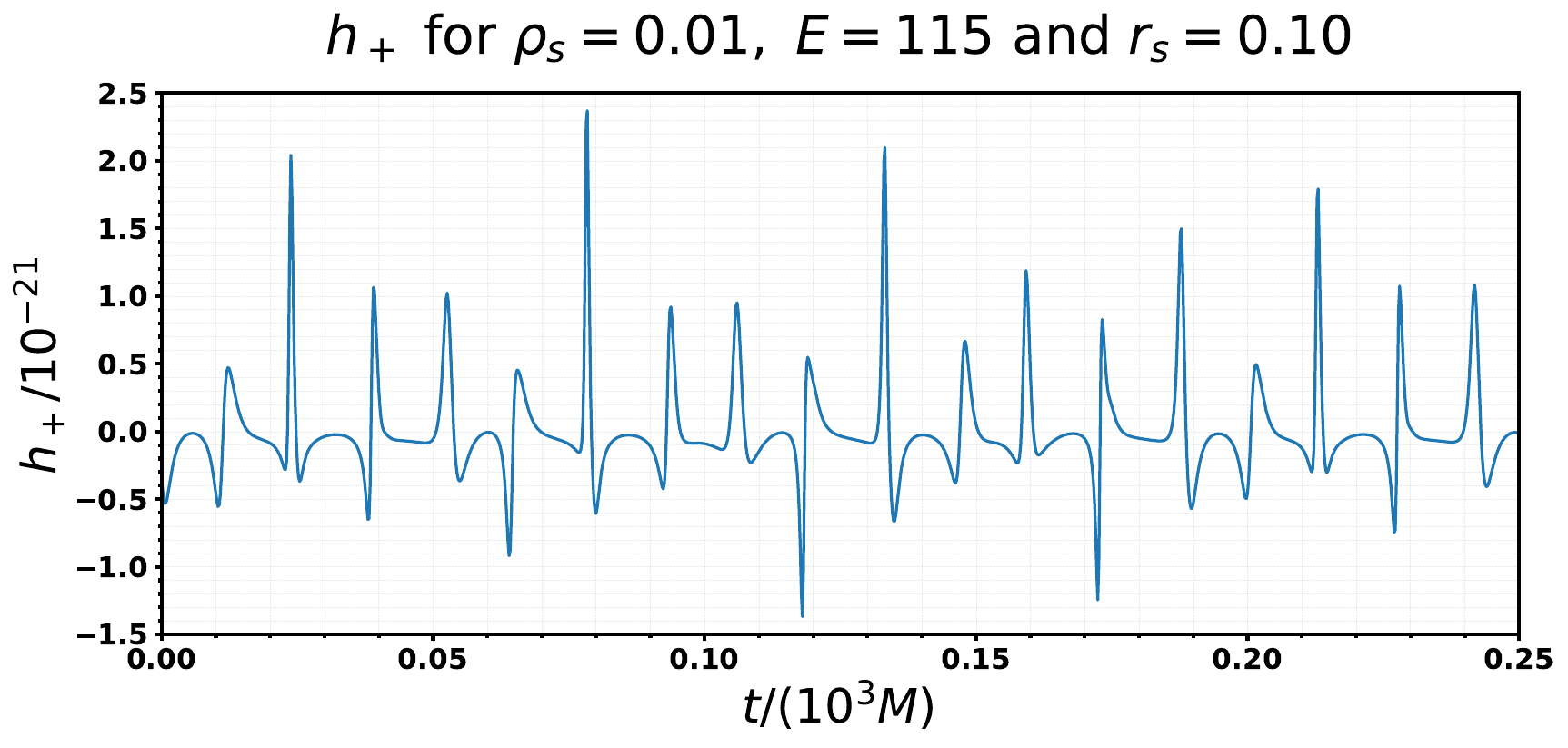}
            \label{gw_plus_p1}
        }\hspace{0.015\textwidth}%
        \subfigure[]{
            \includegraphics[width=0.30\textwidth,height=0.28\textwidth]{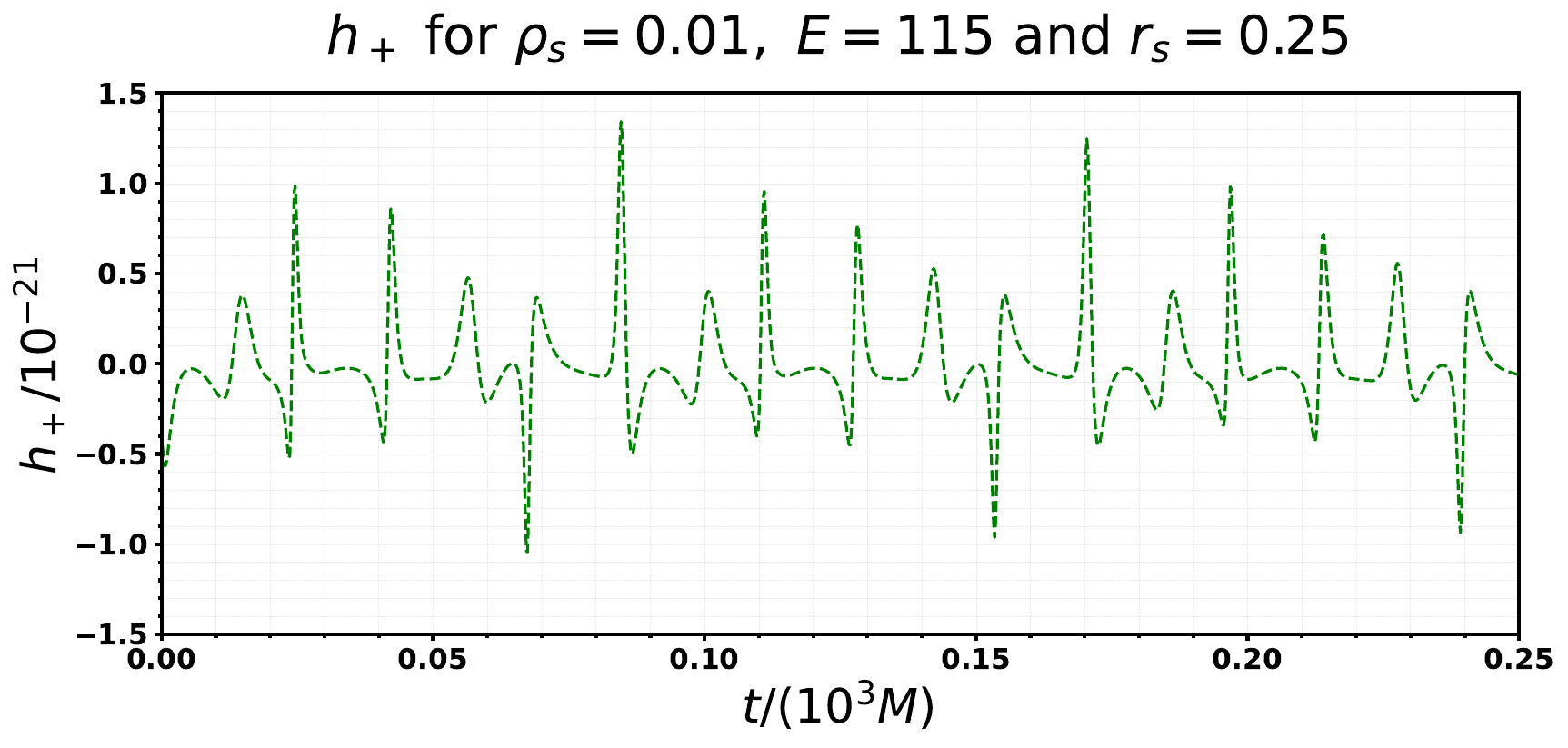}
            \label{gw_plus_oc1}
        }\hspace{0.015\textwidth}%
        \subfigure[]{
            \includegraphics[width=0.30\textwidth,height=0.28\textwidth]{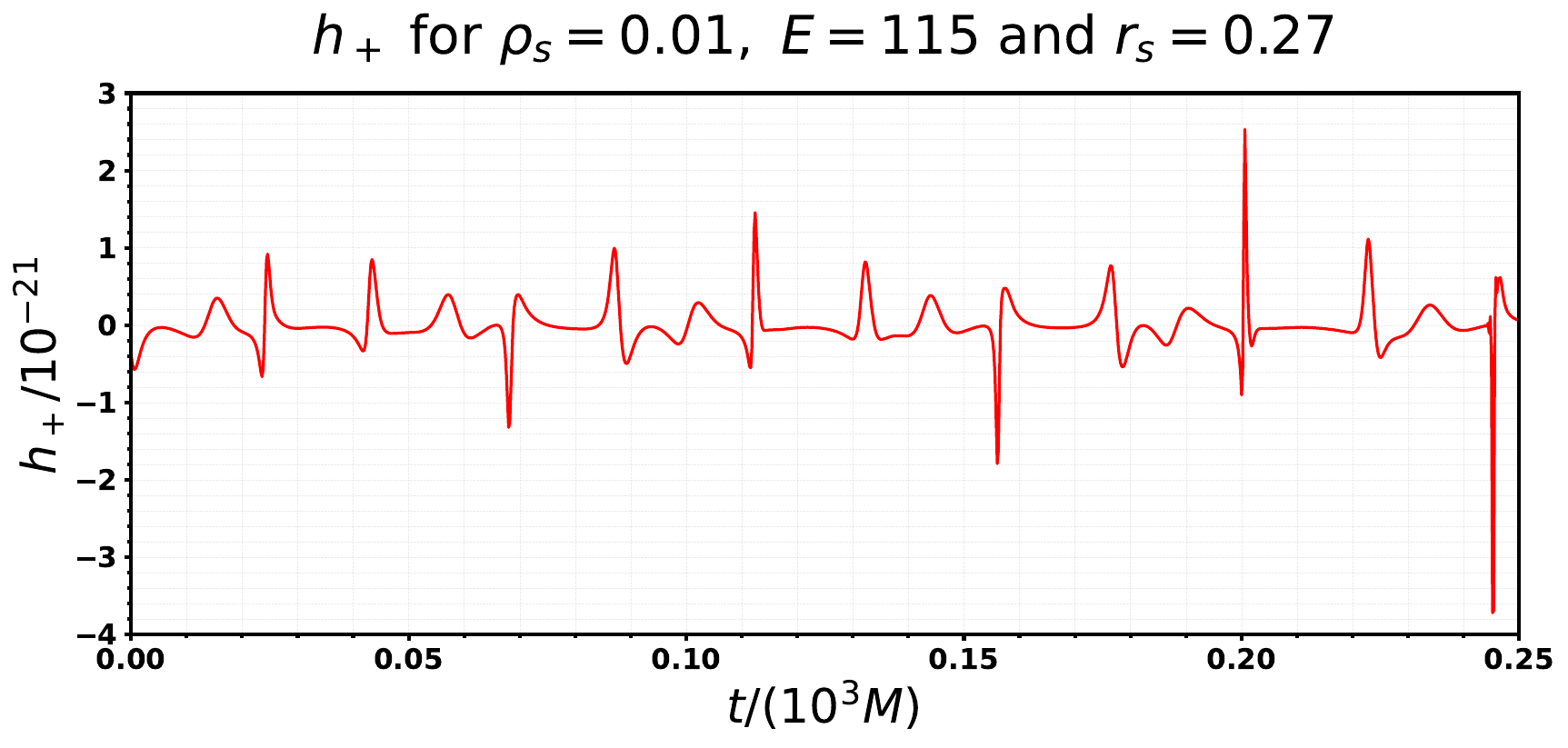}
            \label{gw_plus_c1}
        }%
    }

    \vspace{0.15cm}

    \makebox[\textwidth][c]{%
        \subfigure[]{
            \includegraphics[width=0.30\textwidth,height=0.28\textwidth]{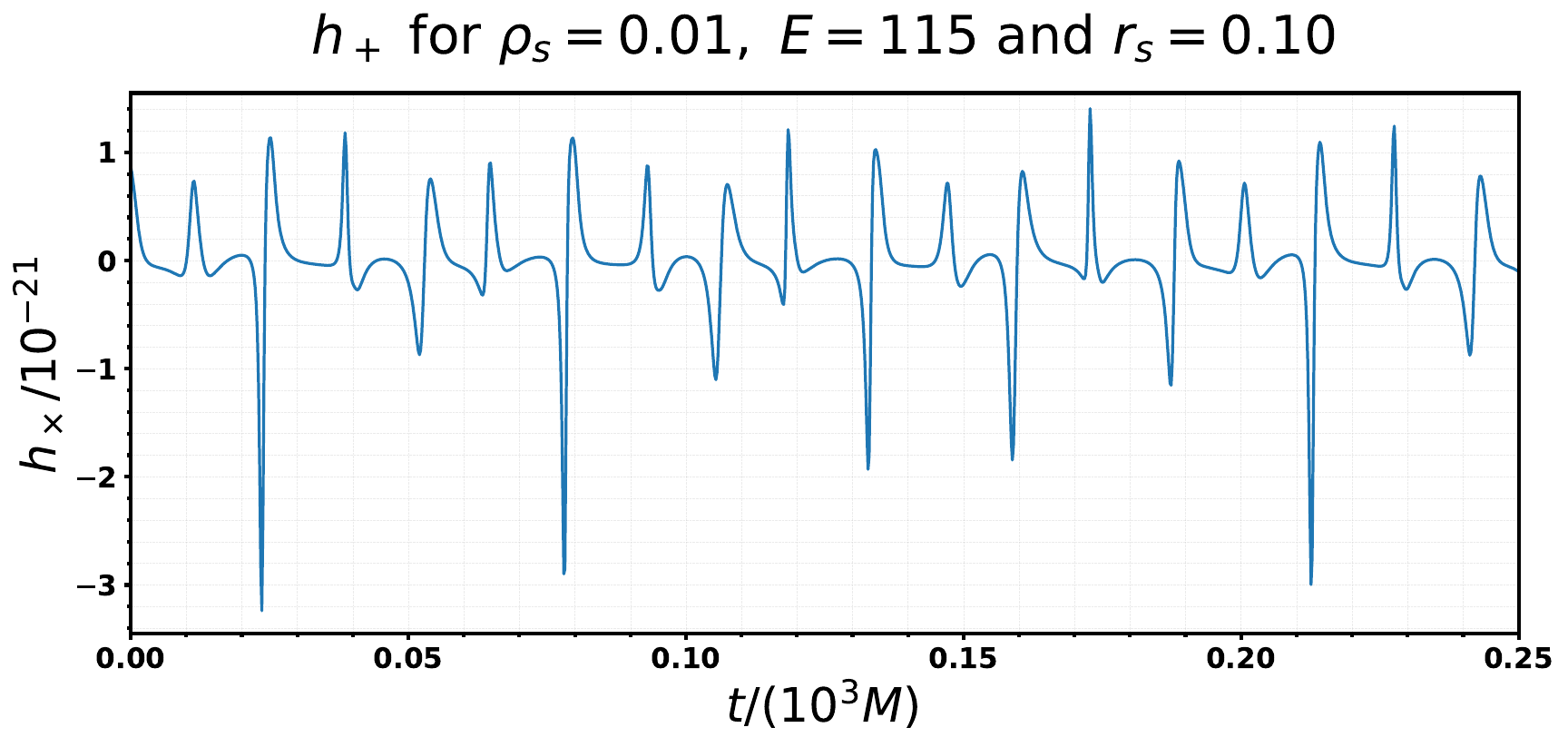}
            \label{gw_cross_p1}
        }\hspace{0.015\textwidth}%
        \subfigure[]{
            \includegraphics[width=0.30\textwidth,height=0.28\textwidth]{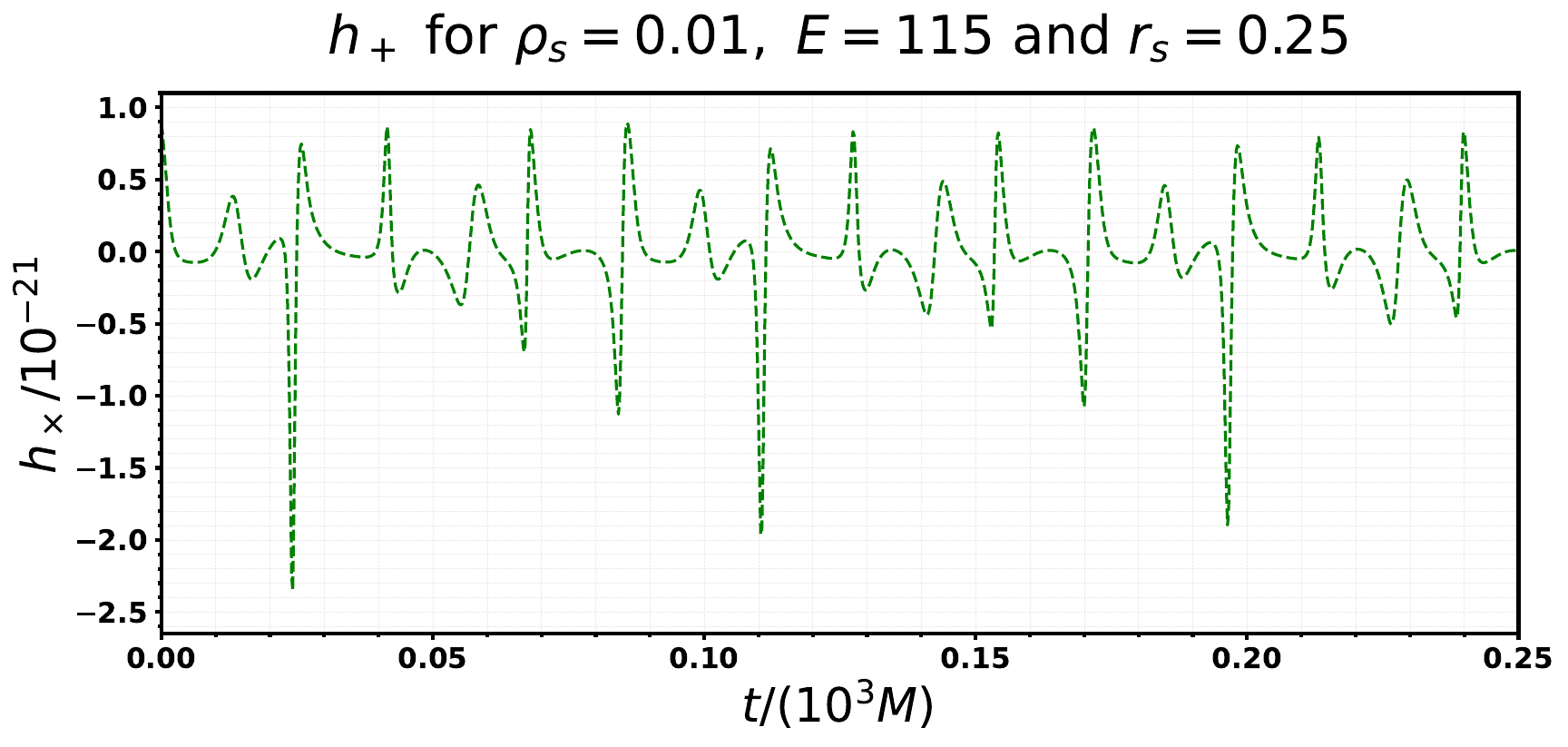}
            \label{gw_cross_oc1}
        }\hspace{0.015\textwidth}%
        \subfigure[]{
            \includegraphics[width=0.30\textwidth,height=0.28\textwidth]{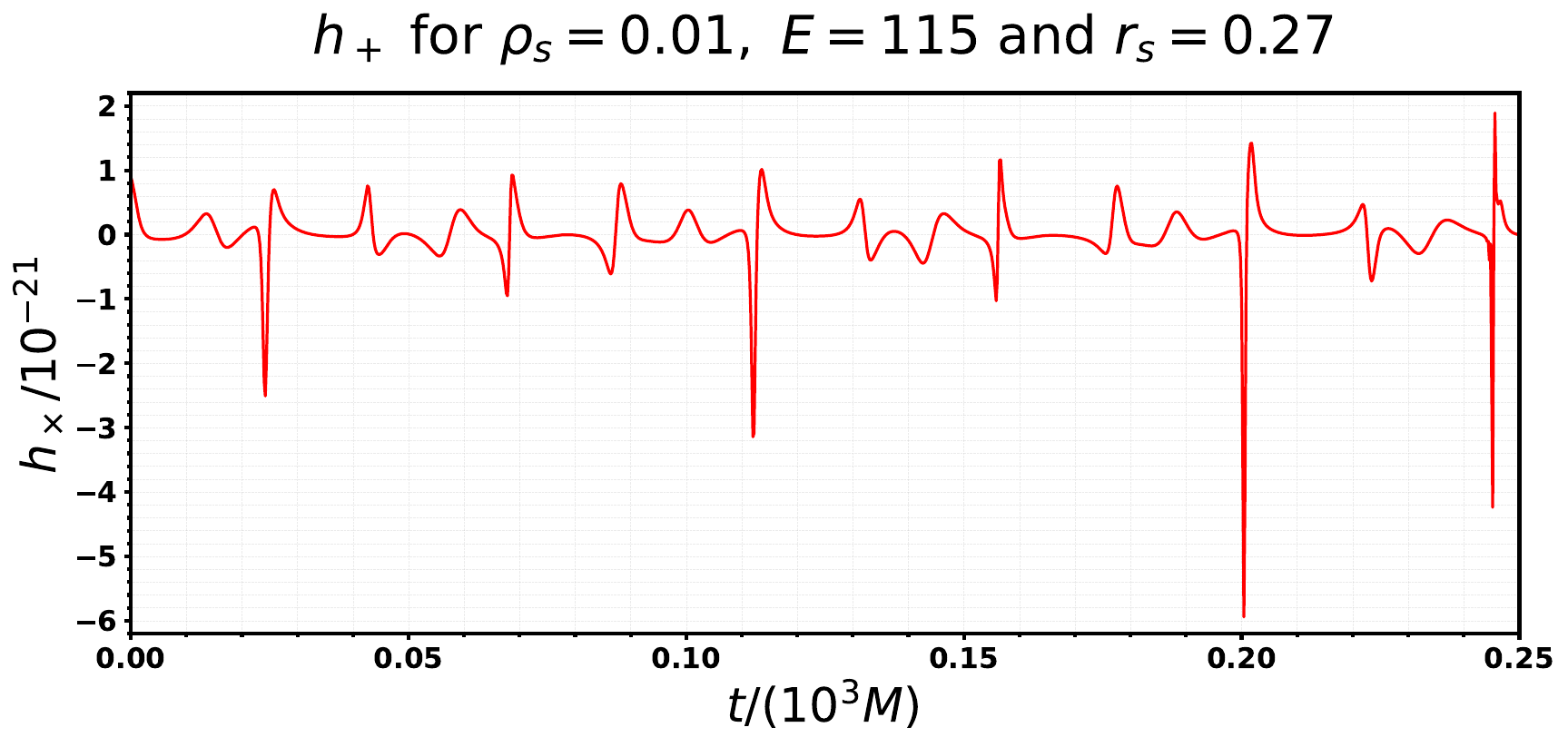}
            \label{gw_cross_c1}
        }%
    }

    \begin{minipage}{\textwidth}
    \caption{Gravitational waveforms of the $+$ and $\times$ modes corresponding to the non-chaotic, onset-of-chaos and chaotic orbits for fixed $\rhs=0.01,~E=115$ with variations in DM halo's scale radius $\rs$ as our considered Case-IV. Here the time range is shown from 0 to 0.25 in unit of $t/(10^3 M)$.}\label{fgw_individual1}
    \hrulefill
    \end{minipage}
    \end{figure}
\end{widetext}

\noindent
Here we present an enlarged view of the gravitational waveforms for non-chaotic, onset-of-chaos, and chaotic orbits, focusing on two different cases in our context: Case-III is on varying central density parameter $\rho_s$ with fixed energy $E=90$ and scale radius $r_s=0.15$ (see Fig.~\ref{fgw_individual}). Another Case-IV, which is on varying the scale radius $r_s$ with fixed energy $E=115$ and central density $\rho_s=0.01$ (see Fig.~\ref{fgw_individual1}). The waveforms associated with the non-chaotic orbits exhibit an overall periodic patterns, despite variations in the amplitude of individual modes (Figs.~\ref{fgw_individual}(a), \ref{fgw_individual}(d) for $\rhs=0.01$ and Figs.~\ref{fgw_individual1}(a), \ref{fgw_individual1}(d) for $\rs=0.10$, respectively). This behavior is consistent with the associated Poincar\'e sections for these type of non-chaotic orbits, as two very close initial non-chaotic trajectories are mostly overlapping to one another (see our previous work \cite{Das:2025vja} for details). In contrast, as chaos begins to influence the dynamics, the periodicity observed in Figs.~\ref{fgw_individual}(b), \ref{fgw_individual}(e) for $\rhs=0.04$ and Figs.~\ref{fgw_individual1}(b), \ref{fgw_individual1}(e) for $\rs=0.25$ start to degrade. Finally, in the chaotic regime (for $\rho_s=0.05$ and $\rs=0.27$, respectively), the waveforms entirely lose their periodic structure, displaying irregular fluctuating pattern. These findings are in agreement with the Poincar\'e maps analyzed in our previous work \cite{Das:2025vja} and reveals how different DM halo parameters affect on the formation of chaotic dynamics in such EMRIs, considered here.

\section*{Averaged amplitude and energy-emission rate}
\begin{widetext}
    \begin{figure}[H]
	\centering
	\begin{center} 
	$\begin{array}{ccc}
	\subfigure[]              
    {\includegraphics[width=1.02\linewidth,height=1.2\linewidth]{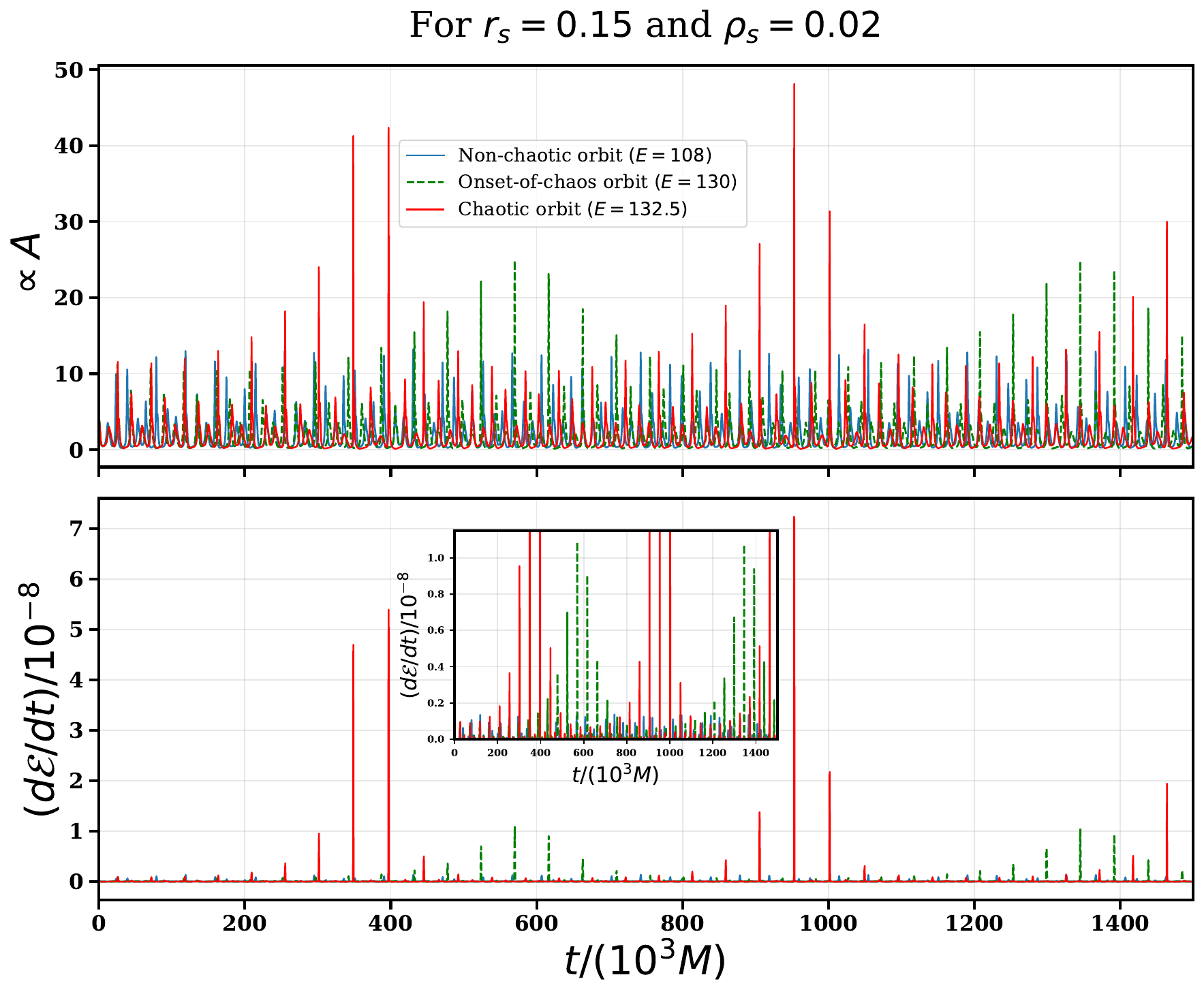}\label{ampee_r_rho}}
	\subfigure[]           
    {\includegraphics[width=1.02\linewidth,height=1.2\linewidth]{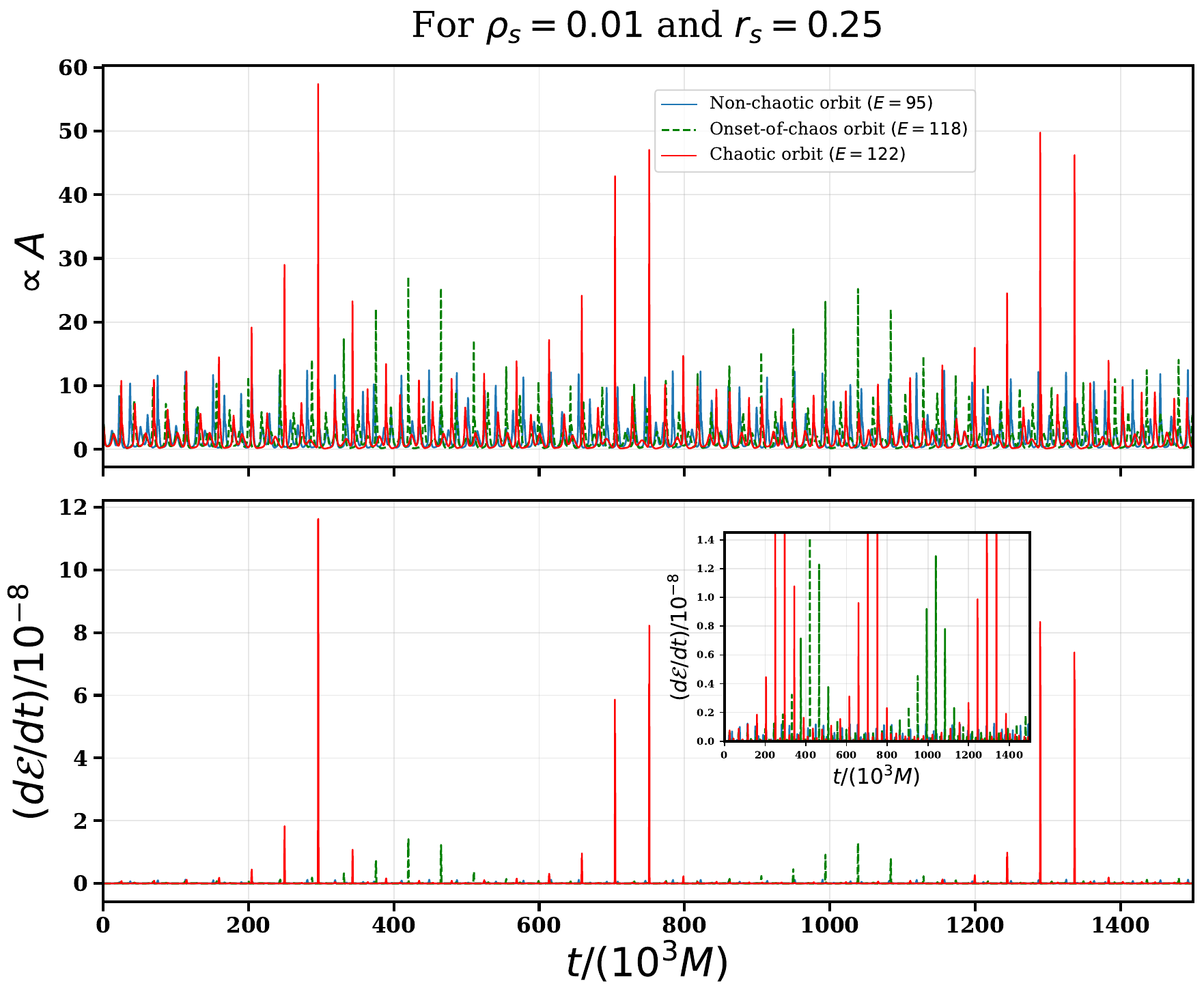}\label{ampee_rho_r}}\\
	\subfigure[] 
    {\includegraphics[width=1.02\linewidth,height=1.2\linewidth]{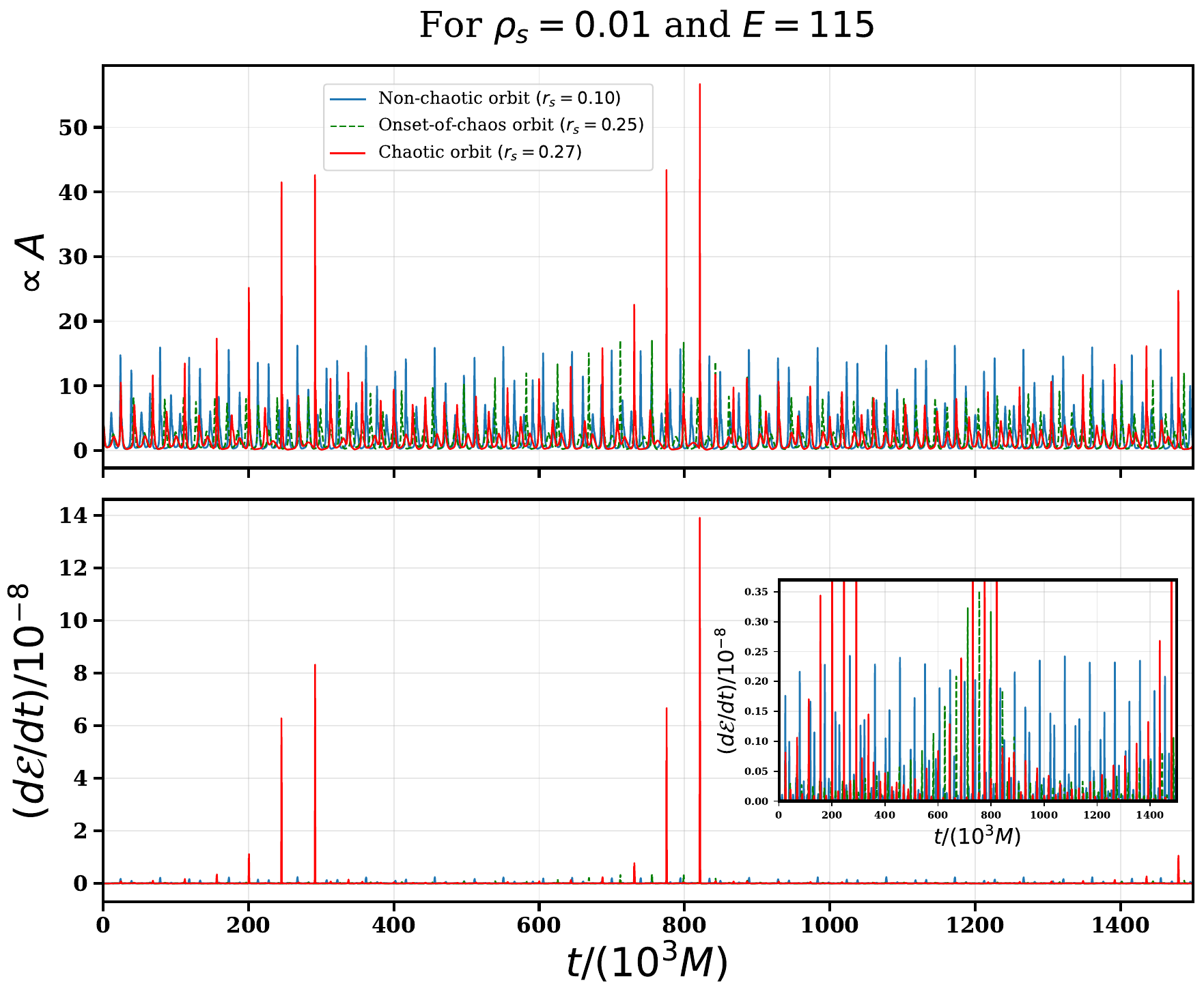}\label{ampee_rho_E}}
	\subfigure[] 
    {\includegraphics[width=1.02\linewidth,height=1.2\linewidth]{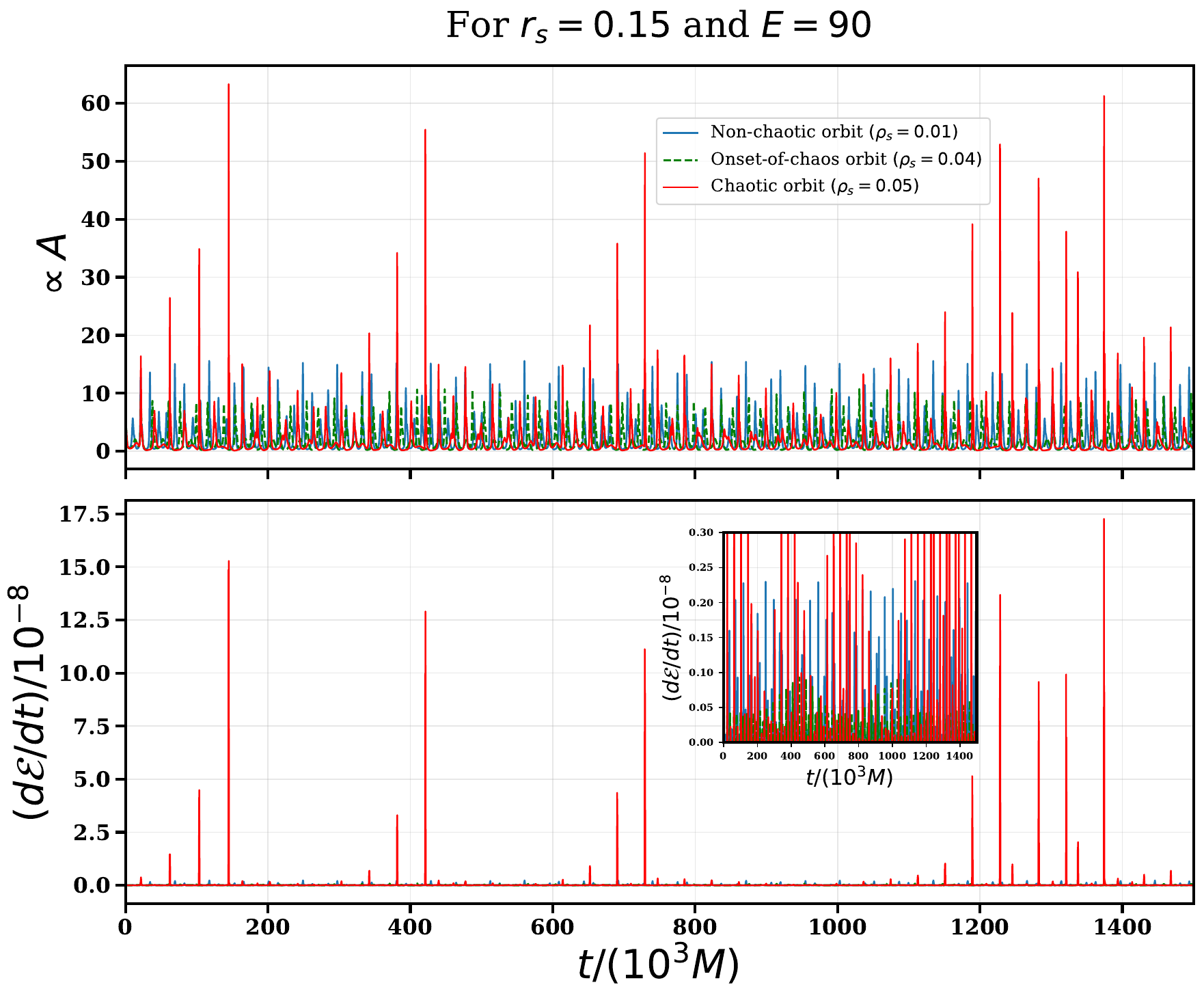}\label{ampee_r_E}}
    \end{array}$
    \end{center}
    \begin{minipage}{\textwidth}
    \caption{The averaged GWs amplitude across all directions and the energy emission rate of the GWs corresponding to the non-chaotic, onset-of-chaos and chaotic orbits for all of the cases, mentioned in Table 1.}\label{fampee}
    \end{minipage}
    \end{figure}
\end{widetext}

\noindent
To explore how chaotic and non-chaotic orbits evolve over time, we study the averaged amplitude and energy emission rate of their respective GWs emission. As stated by the quadrupole formula under kludge scheme, the averaged gravitational wave amplitude across all directions can be expressed as \cite{Landau}
    \begin{equation}
        A\propto \sqrt{\ddot{I}_{ij} \ddot{I}^{ij}}~,\label{amplitude}
    \end{equation}
and the power radiated by the GWs can be described as \cite{Landau}
    \begin{equation}
        \frac{d\mathcal{E}}{dt}=-\frac{{\dddot{I}}_{ij}^2}{45}~.\label{energyloss}
    \end{equation}
Here the negative sign in the above expression implies a loss of energy by radiation. It is important to emphasize that the amount (numerical value) of energy lost is extremely small, so that its impact on dynamical motion remains entirely negligible, even for celestial objects observed over long timescales \cite{Landau}. \\
In Fig.~\ref{fampee}, we represent the averaged GWs amplitude across all directions and the energy emission rate of the associated GWs corresponding to non-chaotic, onset-of-chaos and chaotic orbits for four different cases, as mentioned in Table 1. As evident from the given expressions in Eqs.~\eqref{amplitude}, \eqref{energyloss}, GWs amplitude increases when the system's mass quadrupole distribution undergoes rapid changes. Thus, it can be inferred that the averaged GWs amplitude across all directions from a chaotic orbit would likely exceed those from a non-chaotic or onset-of-choas orbits. This type of observation has been shown in \cite{Suzuki:1999si} for a spinning test particle orbiting around a Schwarzschild BH. On the other hand, it is anticipated that such chaotic motion will also amplify the energy emission rate, since the energy emission rate is strongly correlated with the temporal changes in the system's mass quadrupole distribution. 

In this context, it is worth to mention that the energy emission rate for a particle moving within the Hénon-Heiles potential is studied in Ref.~\cite{Kokubun:1996kg}. On the other hand in Ref.~\cite{Cornish:1997hs}, Cornish {\it{et al.}} investigated trajectories of a spinless particle in the Majumdar-Papapetrou spacetime containing two black holes and their imprints of gravitational amplitude on several orbits. All of these references particularly revealed that the chaotic dynamics leads to an increase in both amplitude and energy emission rate, which are clearly consistent with our findings presented in Fig.~\ref{fampee}, when a comparison of GWs amplitude and energy emission rate from non-chaotic and chaotic orbits are taken into account.

\section*{Frequency and energy spectrum of the generated gravitational waveforms}
\begin{widetext}
    \begin{figure}[H]
	\centering
	\begin{center} 
	$\begin{array}{ccc}
	\subfigure[]              
    {\includegraphics[width=1.02\linewidth,height=1.2\linewidth]{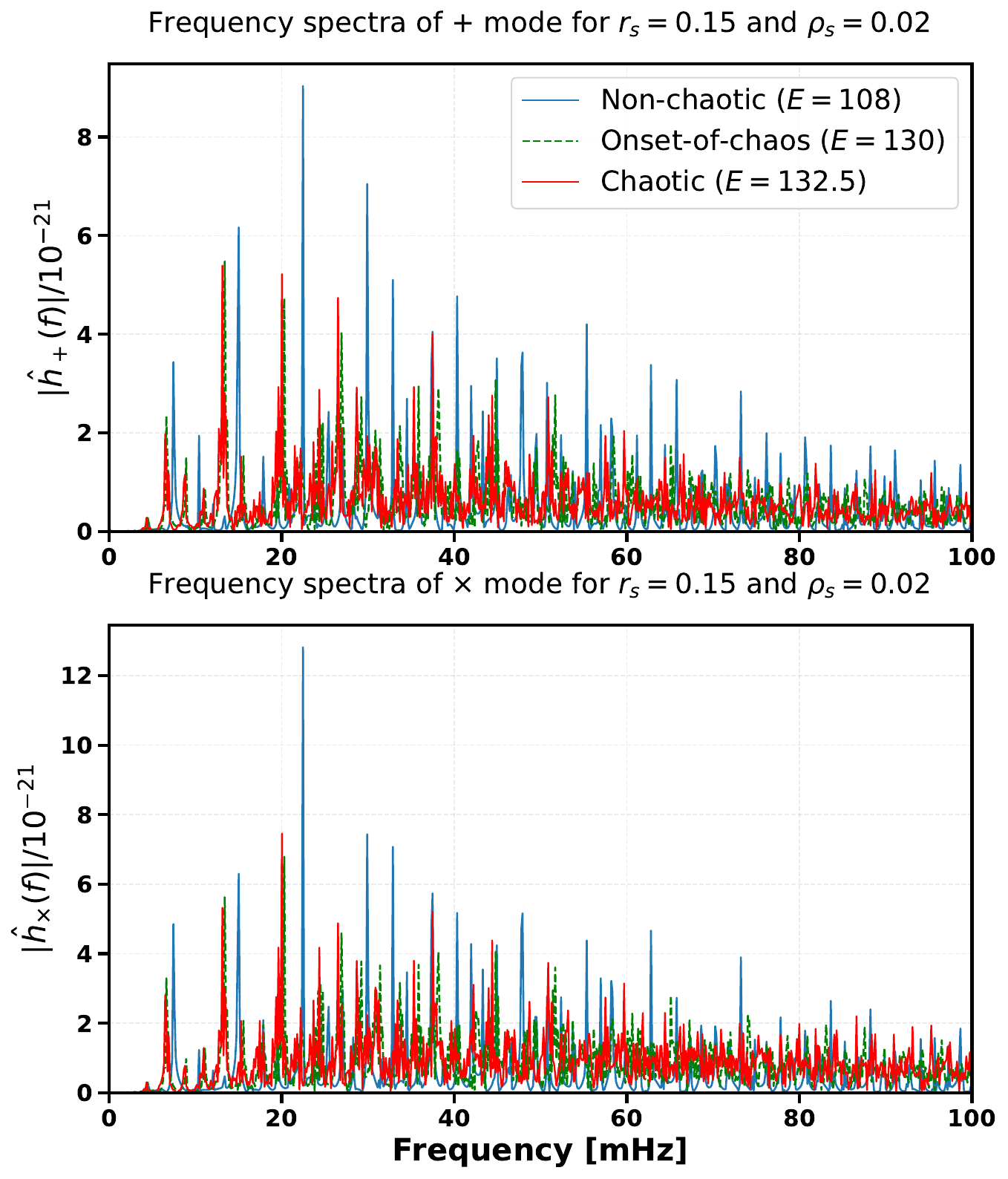}\label{fq_r_rho}}
	\subfigure[]           
    {\includegraphics[width=1.02\linewidth,height=1.2\linewidth]{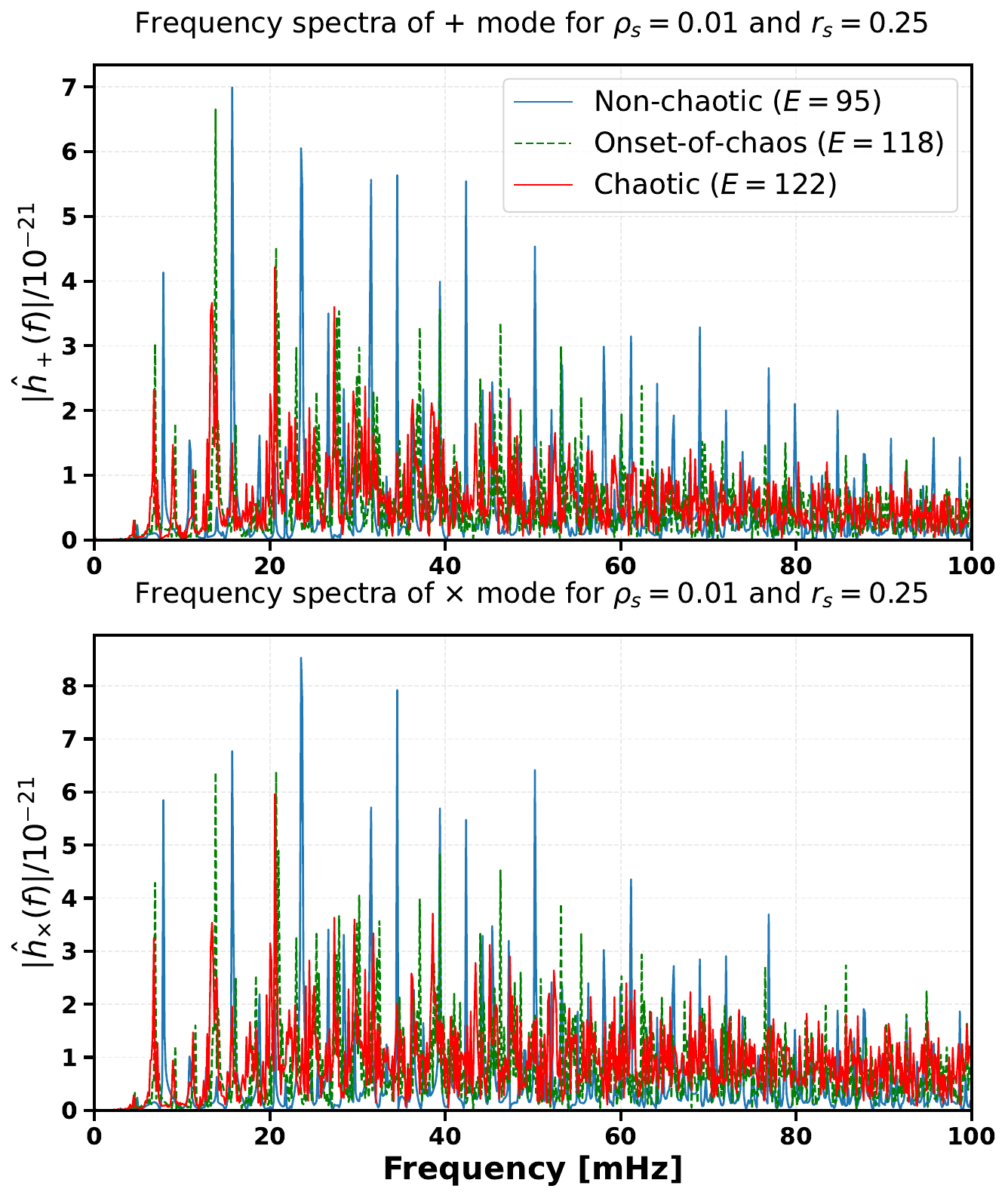}\label{fq_rho_r}}\\
	\subfigure[] 
    {\includegraphics[width=1.02\linewidth,height=1.2\linewidth]
    {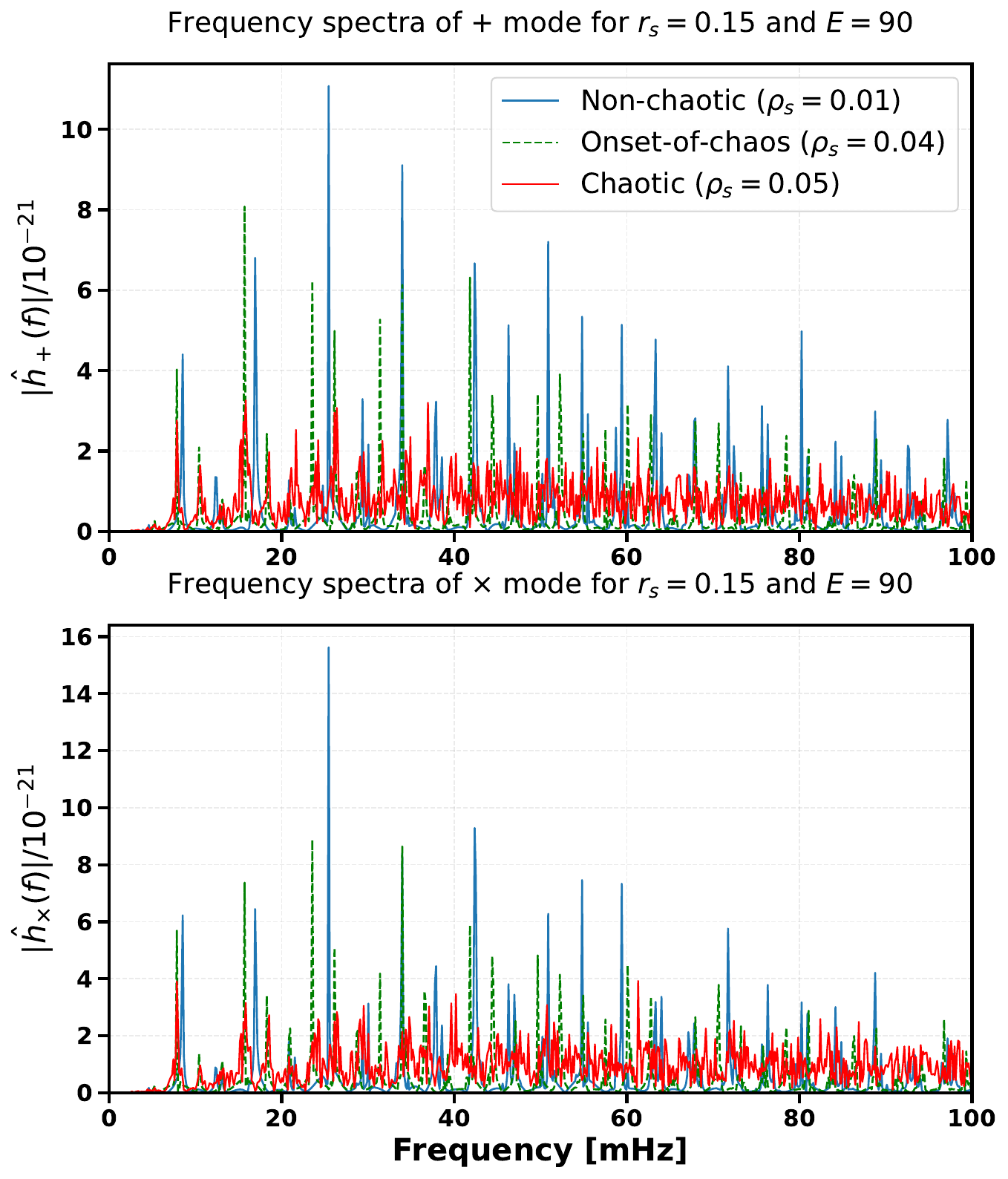}\label{fq_r_E}}
	\subfigure[] 
    {\includegraphics[width=1.02\linewidth,height=1.2\linewidth]{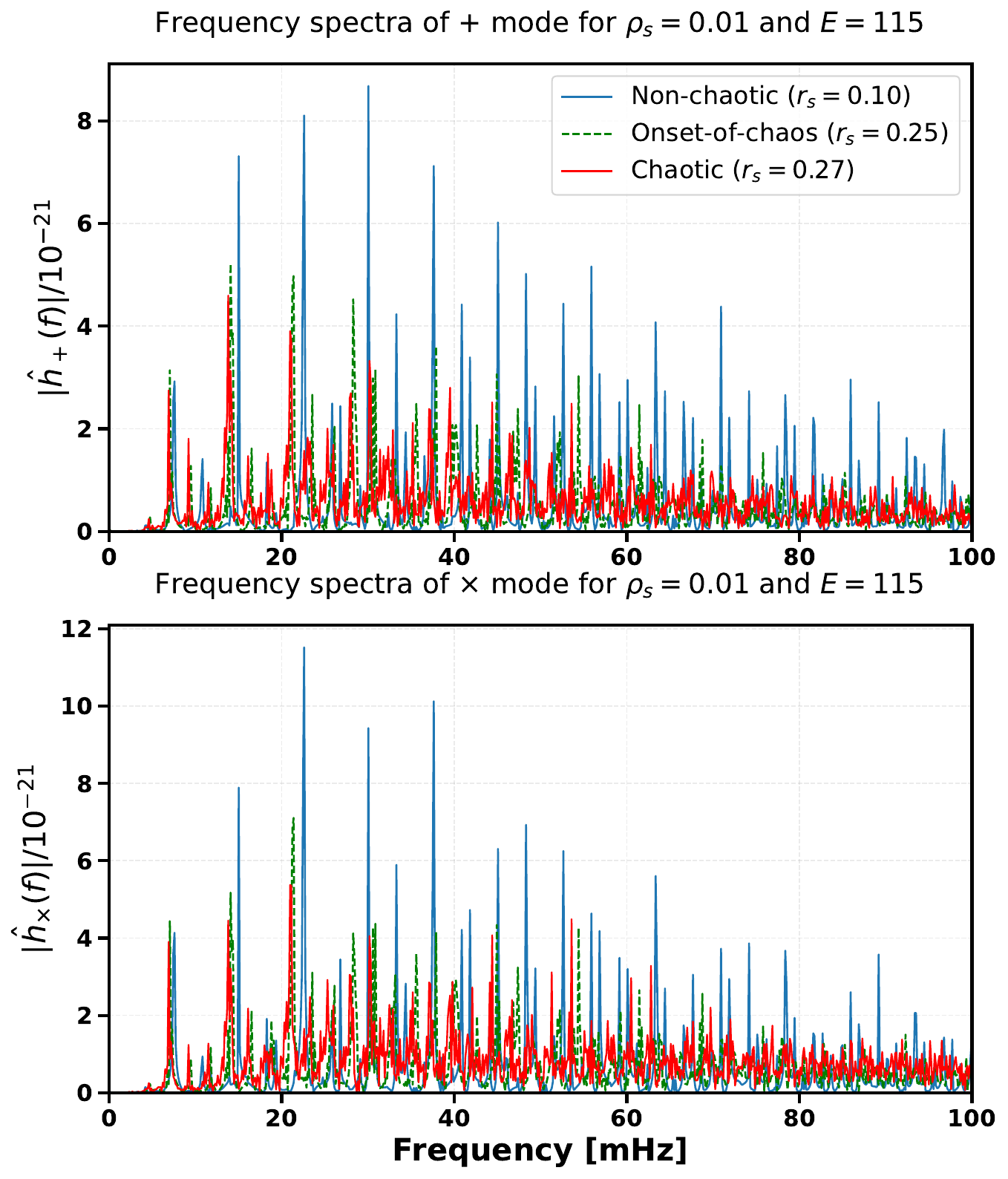}\label{fq_rho_E}}
    \end{array}$
    \end{center}
    \begin{minipage}{\textwidth}
    \caption{Absolute values of Fourier transforms of the time-domain gravitational waveforms corresponding to the Non-chaotic, onset-of-chaos and chaotic orbits for all of the cases, mentioned in Table 1.}\label{freq}
    \end{minipage}
    \end{figure}
\end{widetext}

\begin{widetext}
    \begin{figure}[H]
	\centering
	\begin{center} 
	$\begin{array}{ccc}
	\subfigure[]              
    {\includegraphics[width=1.5\linewidth,height=0.71\linewidth]
    {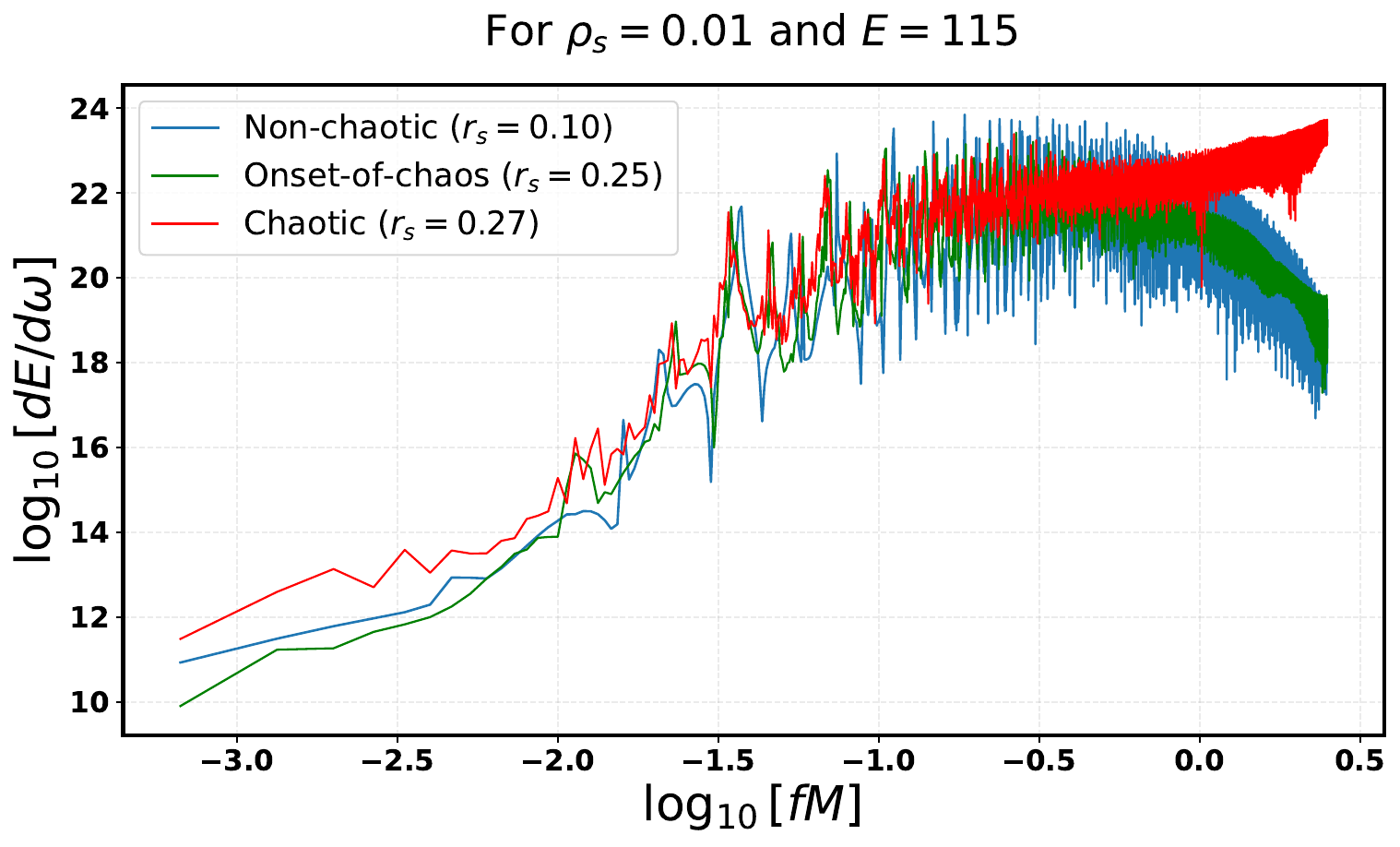}\label{es_combined1}}\\
    \subfigure[]
    {\includegraphics[width=1.5\linewidth,height=0.71\linewidth]{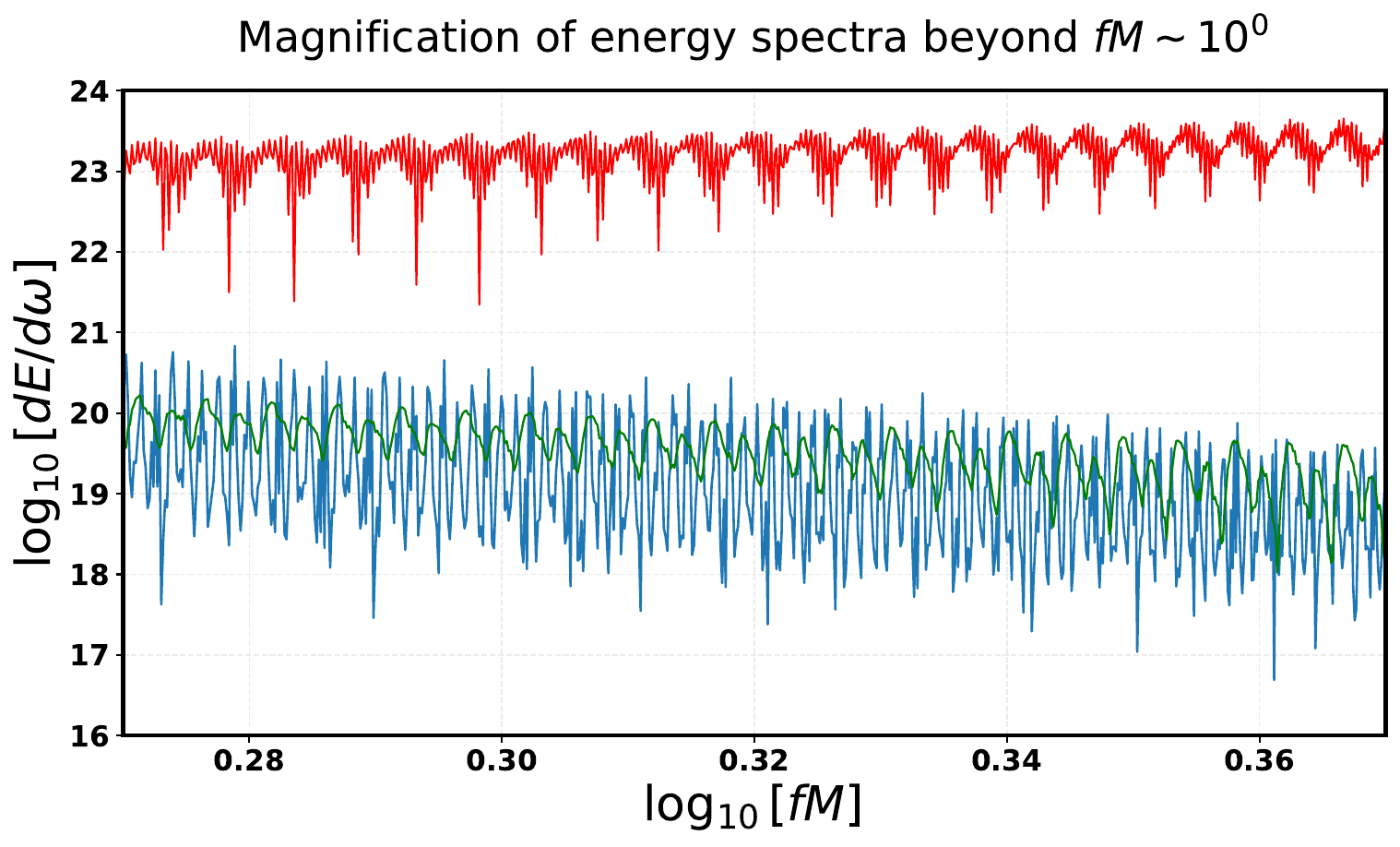}\label{es_combined_zoom1}}\\
	\subfigure[]           
    {\includegraphics[width=0.68\linewidth,height=0.65\linewidth]
    {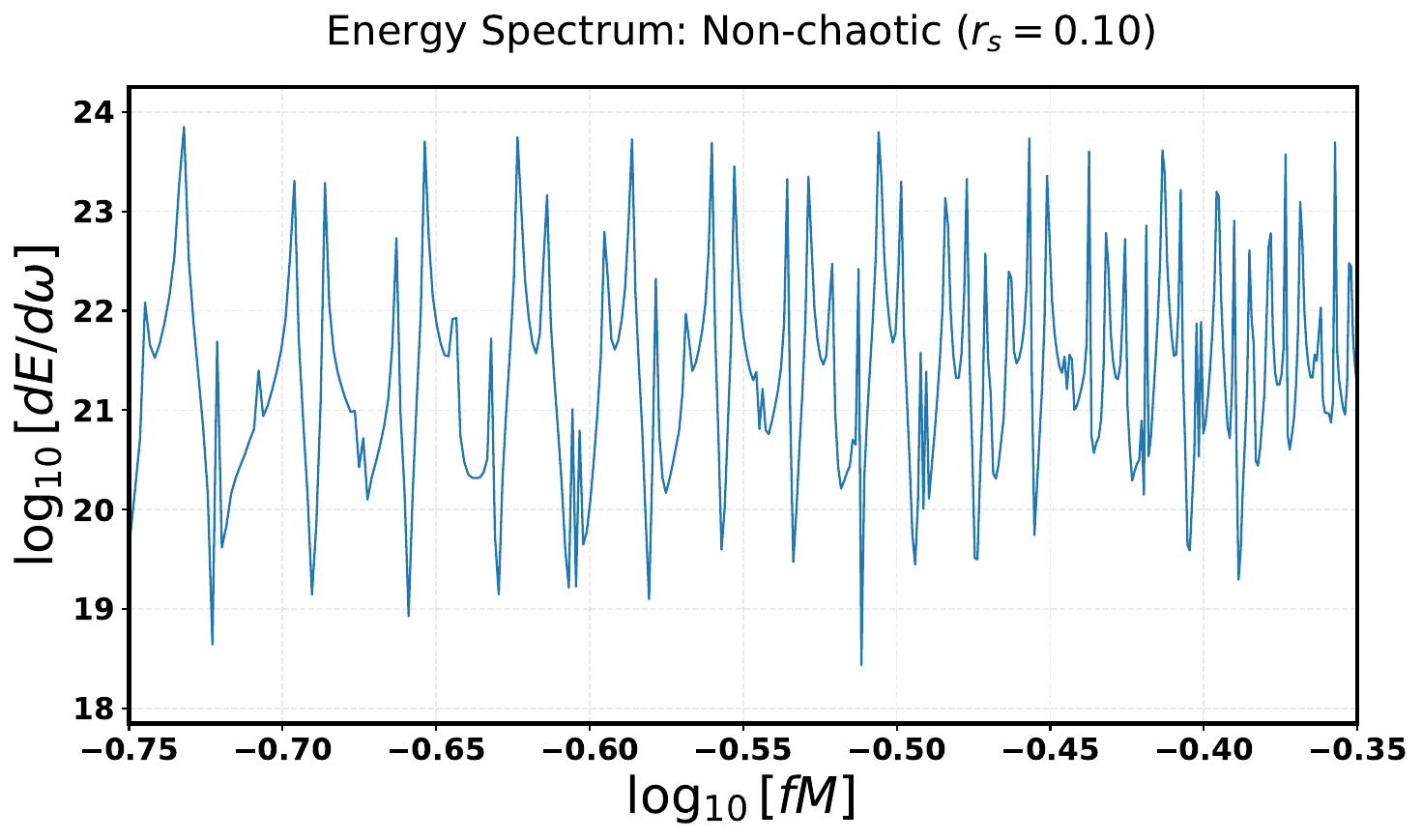}\label{es_p1}}
	\subfigure[] 
    {\includegraphics[width=0.68\linewidth,height=0.65\linewidth]{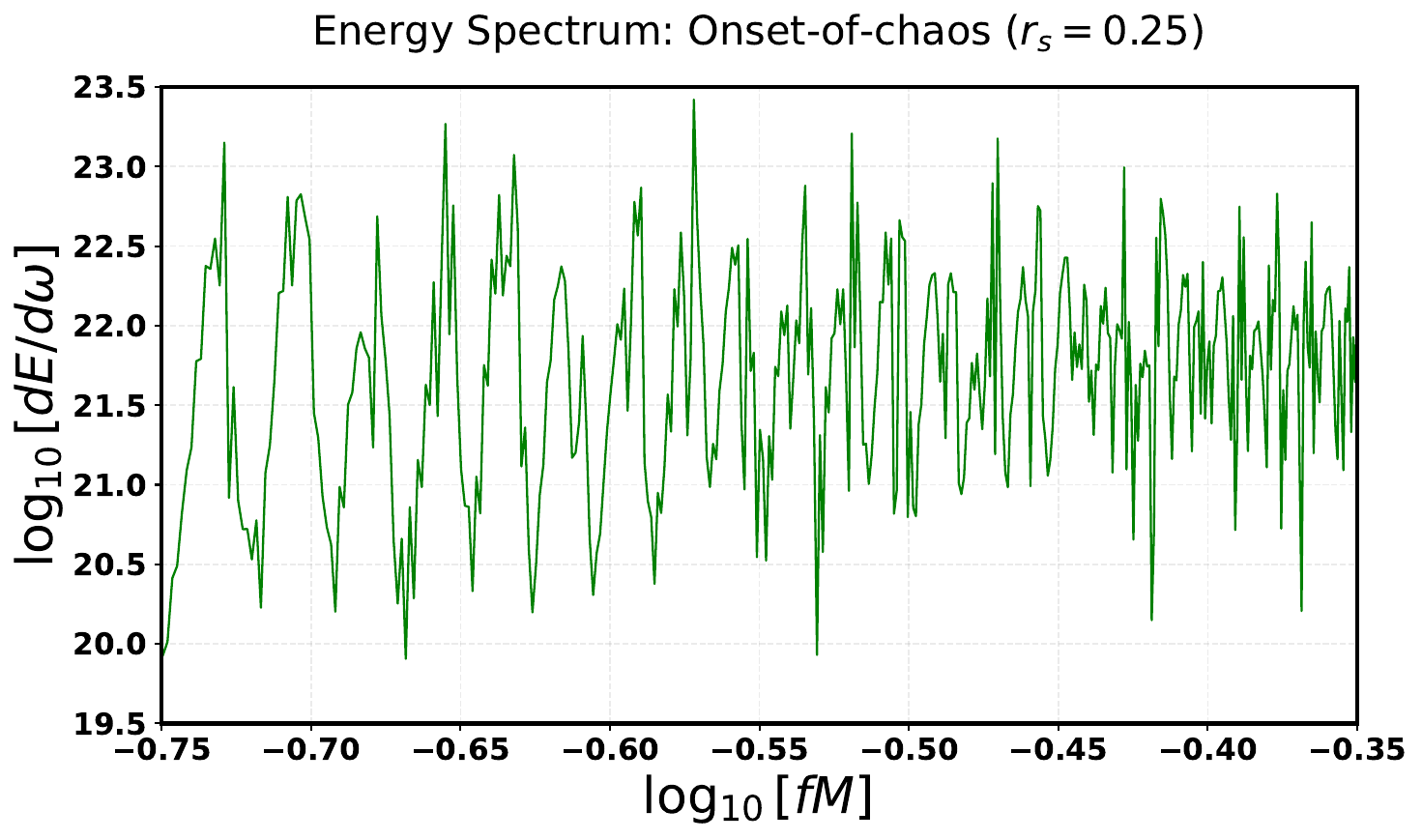}\label{es_oc1}} 
    \subfigure[]
    {\includegraphics[width=0.68\linewidth,height=0.65\linewidth]
    {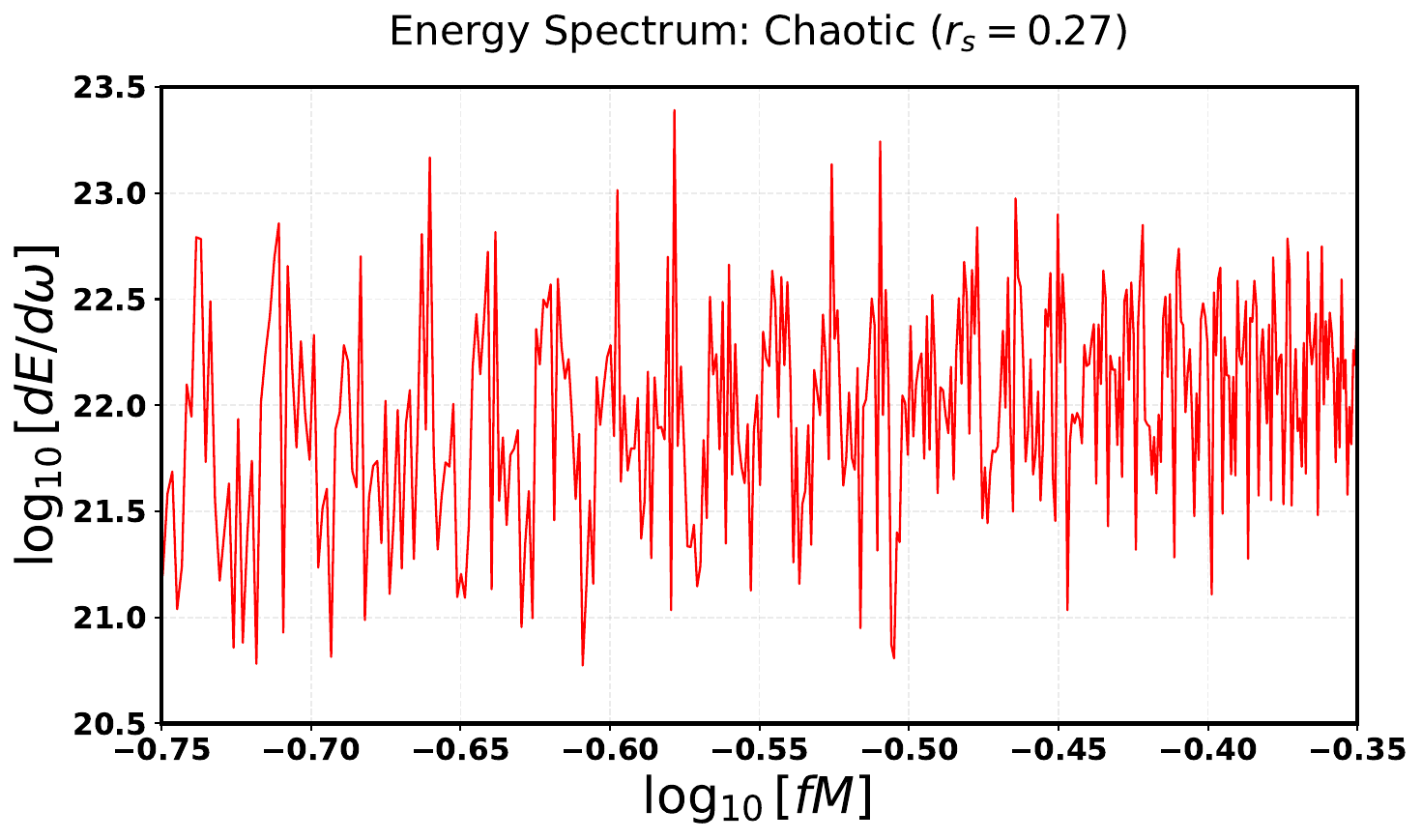}\label{es_c1}} 
    \end{array}$
    \end{center}
    \begin{minipage}{\textwidth}
    \caption{Fig.~\ref{fes1}(a) presents the GWs energy spectra corresponding to different orbits for Case-IV (fixed $\rhs = 0.01$, $E = 115$ with various $\rs$). For frequencies $fM \geq 10^{0}$, both the non-chaotic and onset of chaos orbits exhibit a decay in their respective energy spectra. But this decay feature is notably absent in the case of chaotic orbit.\\
    However, the zoomed-in view beyond $fM\sim10^{-0.25}$ is represented in Fig.~\ref{fes1}(b), where the blue curve corresponds to a non-chaotic orbit ($\rs = 0.10$), the green curve indicates the onset of chaos ($\rs = 0.25$), and the red curve represents a chaotic orbit ($\rs = 0.27$).\\
    Figs.~\ref{fes1}(c), \ref{fes1}(d), and \ref{fes1}(e) display magnified views of the energy spectra for non-chaotic, onset-of-chaos, and chaotic orbits within the frequency $fM\sim 10^0$, respectively. Fig.~\ref{fes1}(c) reveals numerous sharp peaks at specific characteristic frequencies, a signature of regular motion, same as in Fig.~\ref{fes1}(b). In contrast, Fig.~\ref{fes1}(e) shows broader peaks with additional overlapping spikes, a consequence of the chaotic nature of the orbit, same as in Fig.~\ref{fes1}(b).}\label{fes1}
    \end{minipage}
    \end{figure}
\end{widetext}

\noindent
The frequency spectra of the plus $\hat{h}_{+}$ and cross polarizations $\hat{h}_{\times}$ are presented in Fig.~\eqref{freq}, which involves applying discrete Fourier transforms to the time-domain gravitational waveforms (as shown in Fig.~1). \\
On the other hand, the energy spectra of the generated gravitational waveforms for Case-IV is presented in Fig.~\ref{fes1}.

\section*{Brief discussion on recurrence analysis}
\noindent
Here we give a brief discussion on the theoretical foundations of recurrence analysis and then study the computational implementation of this approach in our context.

\newpage

\newpage

\newpage

\subsection*{A. Phase Space Reconstruction via Takens's Theorem}\label{sec:Takens} 
\noindent
The phase space of a dynamical system is reconstructed from a time series $\mathbf{X}_i$ of dimension $n$ and length $l \geq i \in \mathbb{N}$ using the time-delay embedding method proposed by Takens \cite{Takens}. According to Takens's theorem, the reconstructed time series is defined as
\begin{equation}
        \mathbf{Y}_i = \left(\mathbf{X}_i, \mathbf{X}_{i+\tau}, ..., \mathbf{X}_{i+\tau\left(d-1\right)}\right)\nonumber~,
    \end{equation}
where $\tau\in\mathbb{N}$ is the time delay and $d\in\mathbb{N}$ is the embedding dimension. Here the reconstructed time series has a dimension of $(n_\cdot d)$ and also has reduced length $l - \tau\left(d-1\right) = L \geq i\in\mathbb{N}$. The trivial case $d=1$ corresponds to no reconstruction. The theorem guarantees that for sufficiently large $d$ and suitable $\tau$, the reconstructed attractor is topologically equivalent to the original system's attractor. It is worthy to mention that this method establishes a diffeomorphism between the original and reconstructed phase spaces under certain conditions \cite{Takens}.
    \begin{itemize}
        \item{\underline{Selection of time delay ($\tau$):}}
        The time delay $\tau$ is generally chosen as the first local minimum of the mutual information (MI) between the time series and its delayed version. The MI for a delay $\tau$ is computed by the following definition.
    \begin{equation}
        I(\tau)=\sum_{i,j} p_{i,j}(\tau)\log\frac{p_{i,j}(\tau)}{p_i p_j}\nonumber
    \end{equation}
    where $p_{i,j}(\tau)$ is the joint probability of observing $\mathbf{X}_i$ in bin $i$ and $\mathbf{X}_{i+\tau}$ in bin $j$, with $p_i,~p_j$ are marginal probabilities. The first minimum of $I(\tau)$ indicates the delay where the time series provides maximal independent information.
    \item{\underline{Selection of embedding dimension ($d$):}}
        The embedding dimension $d$ is determined using the False Nearest Neighbors (FNN) algorithm \cite{Kennel}. A false nearest neighbor is a point that appears close in dimension $d$ but becomes distant when embedded in $d+1$ dimensions, indicating insufficient unfolding of the attractor. A neighbor is considered false if the following condition holds:
    \begin{equation}
        \frac{\|\mathbf{X}_i^{(d+1)} - \mathbf{X}_j^{(d+1)}\|}{\|\mathbf{X}_i^{(d)} - \mathbf{X}_j^{(d)}\|} > f\nonumber
    \end{equation}
        The smallest $d$ where the fraction of FNN drops below a threshold (e.g., $5\%$ in our analysis) is selected as the optimal embedding dimension. Ofcourse an optimal $d$ minimizes the fraction of FNNs (ideally reducing it to zero).
    \end{itemize}

\subsection*{B. Recurrence Plot Construction}\label{sec:recurrence}
\noindent
A Recurrence Plot (RP) visualizes the recurrences of states in the reconstructed phase space. The RP matrix $\mathcal{R}$ is defined as the following \cite{Marwan,Yu}.
    \begin{equation}
        \mathcal{R}_{ij} = \left\{ \begin{array}{lr}
        \mathbf{\Theta}\left(\varepsilon - \left|\left|\mathbf{X}_i-\mathbf{X}_j\right|\right|\right) & i \neq j \\
        0 & i = j \end{array} \right.\nonumber~,
    \end{equation}
where $\mathbf{\Theta}$ is the Heaviside function and $\varepsilon$ is known as the recurrence threshold. The free parameter $\epsilon$ is chosen adaptively to achieve a fixed Recurrence Rate (RR), defined as the fraction of recurrent points in the RP as the following.
    \begin{equation}
        \mathcal{RR} = \frac{1}{l^2}\sum_{i,j=1}^{N}\mathcal{R}_{ij} \:~.\label{eq:rr}
    \end{equation}    
Here a specific value of $\mathcal{RR}$ is chosen and then the threshold $\epsilon$ is tuned to fit the given $\mathcal{RR}$. By construction, the recurrence matrix is symmetric. Therefore, $\mathcal{R}_{ij} = 1$ indicates a recurrence at times $i$ and $j$, meaning $\mathbf{X}_i$ and $\mathbf{X}_j$ lie within a specified small threshold distance.\\
Visualizing this matrix as a graph with axes $i$ and $j$ representing times $t_i$ and $t_j$, where a dot marks the nonzero entries, yields a Recurrence Plot (RP). \\
RPs can be visually inspected to identify fundamental dynamical features, particularly to distinguish linear from nonlinear behavior. For periodic or quasiperiodic time series, a delay $\Delta i$ exists where recurrences appear for sufficiently large $\varepsilon$, satisfying $\forall i\: \mathcal{R}_{i, i+\Delta i}=1$. This results in diagonal lines parallel to the main diagonal, offset by $\Delta i$ either in the vertical or in the horizontal direction. \\
On the other hand in chaotic systems, this regular structure emerges only during irregular (chaotic) phases.
The corresponding RP region $\mathcal{I}\times \mathcal{I}$ exhibits diagonal-parallel lines, forming a regular-looking square along the main diagonal. If chaotic behavior occurs in intervals $\mathcal{I}_1$ and $\mathcal{I}_2$, the region $\left(\mathcal{I}_1\times \mathcal{I}_2\right) \bigcup \left(\mathcal{I}_2\times \mathcal{I}_1\right)$ may either display a similar structure or remain completely empty based on the characteristics of chaotic particle's trajectory.
Therefore it creates the characteristic square-like patterns in RPs \cite{Zelenka:2019nyp,Zelenka:2024fjg}.

\section*{Radial orbital recurrence plots and reconstructed phase space trajectories}
\begin{widetext}
    \begin{figure}[H]
	\centering
	\begin{center} 
	$\begin{array}{ccc}
	\subfigure[]              
    {\includegraphics[width=0.68\linewidth,height=0.65\linewidth]
    {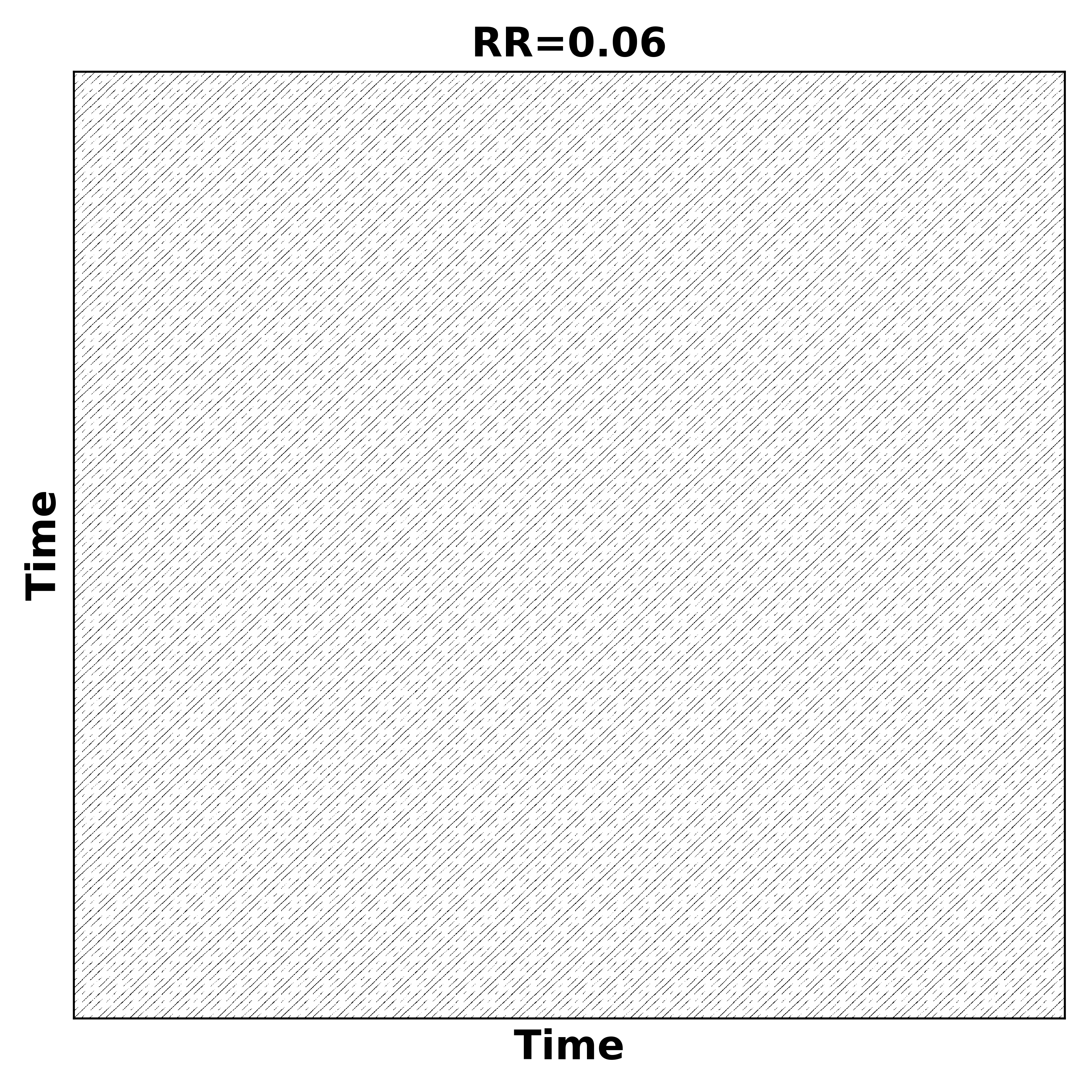}\label{frp_r_p}}
	\subfigure[]           
    {\includegraphics[width=0.68\linewidth,height=0.65\linewidth]
    {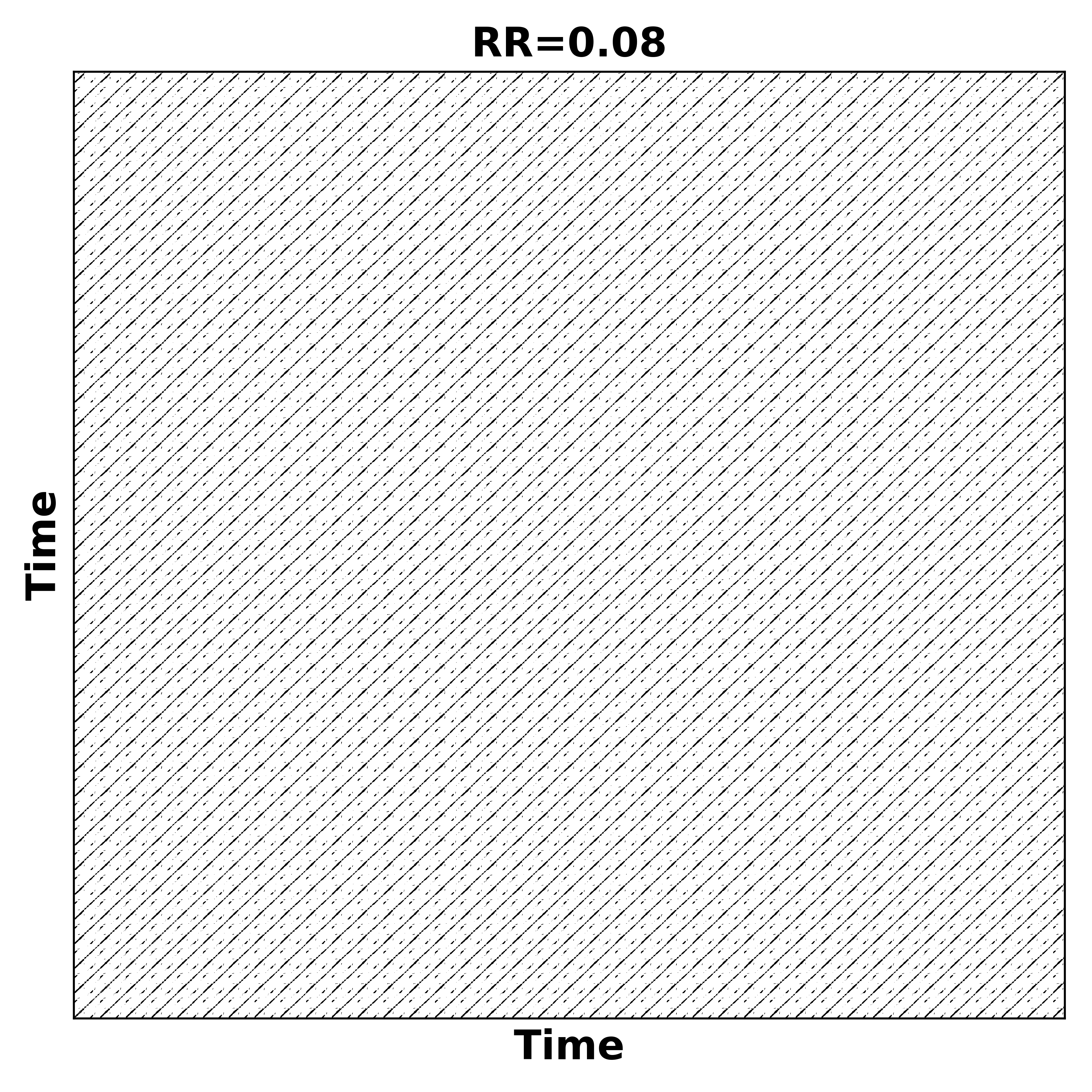}\label{frp_r_oc}}
	\subfigure[] 
    {\includegraphics[width=0.68\linewidth,height=0.65\linewidth]
    {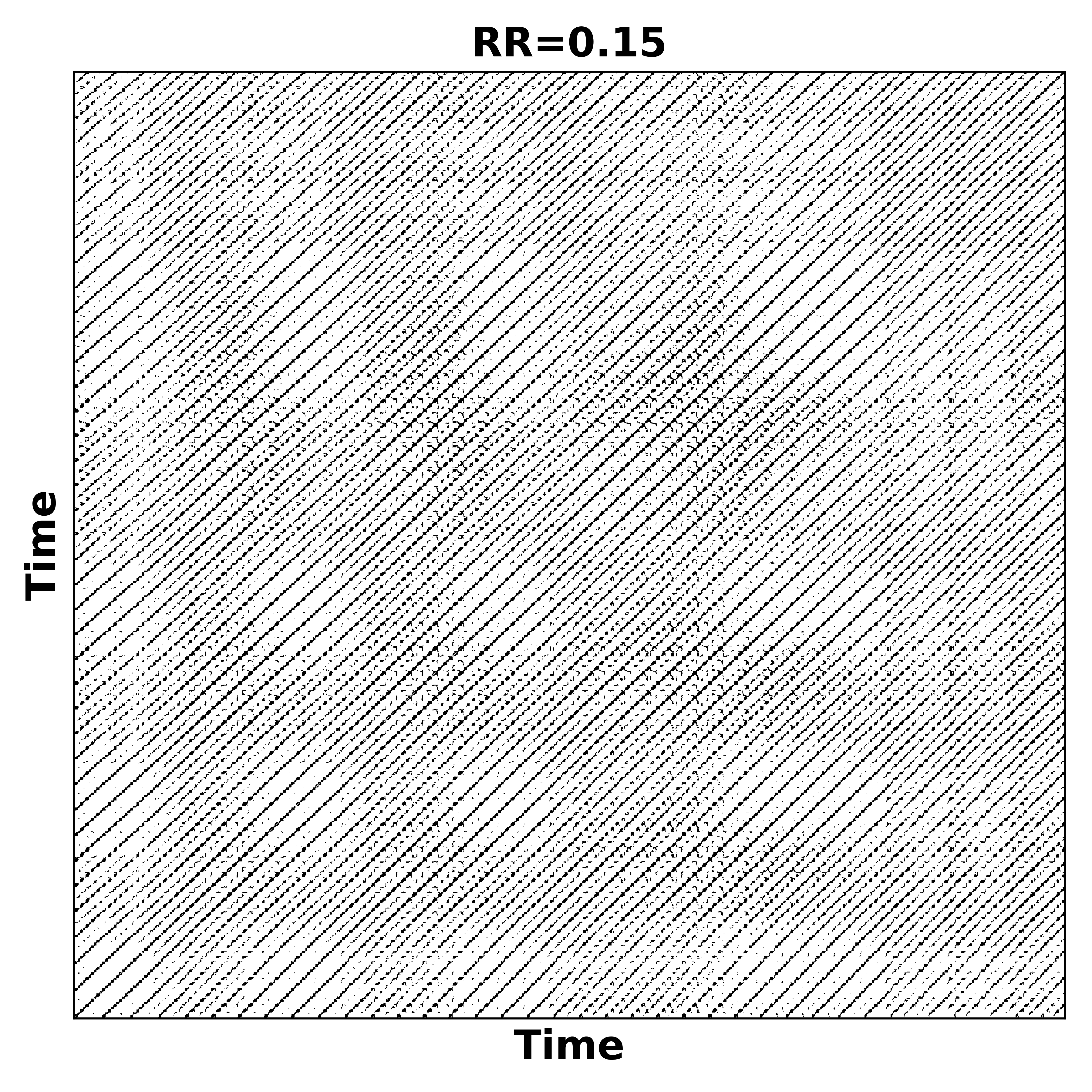}\label{frp_r_c}}
    \end{array}$
    \end{center}
    \begin{minipage}{\textwidth}
    \caption{Recurrence plots illustrating radial coordinate $r$ along three distinct orbits for Case-III, with fixed parameters $\rs = 0.15$ and $E = 90$, are presented. Fig.~\ref{frp_r}(a) displays the non-chaotic (regular) orbit corresponding to $\rhs = 0.01$. For this non-chaotic orbit, the embedding parameters are $d = 4$, $\tau = 13$, and $\epsilon = 0.6209$, with a recurrence rate $\mathcal{RR} = 0.06$.\\
    Fig.~\ref{frp_r}(b) shows the orbit at the onset of chaos for $\rhs = 0.04$. Here, the embedding dimension is $d = 3$, the delay time $\tau = 17$, the threshold $\epsilon = 0.7461$, and the recurrence rate $\mathcal{RR} = 0.08$.\\
    Finally, Fig.~\ref{frp_r}(c) depicts the fully chaotic orbit for $\rhs = 0.05$. The parameters for this chaotic orbit are $d = 5$, $\tau = 41$, $\epsilon = 1.8515$, and $\mathcal{RR} = 0.15$. We observe a distinct diagonal pattern, which is characteristic of regular or non-chaotic orbit in Fig.~\ref{frp_r}(a), slowly becoming obscured by the growing of central density $\rhs$ in Fig.~\ref{frp_r}(c).}\label{frp_r}
    \hrulefill
    \end{minipage}
	\begin{center} 
	$\begin{array}{ccc}
    \subfigure[]
    {\includegraphics[width=0.68\linewidth,height=0.65\linewidth]
    {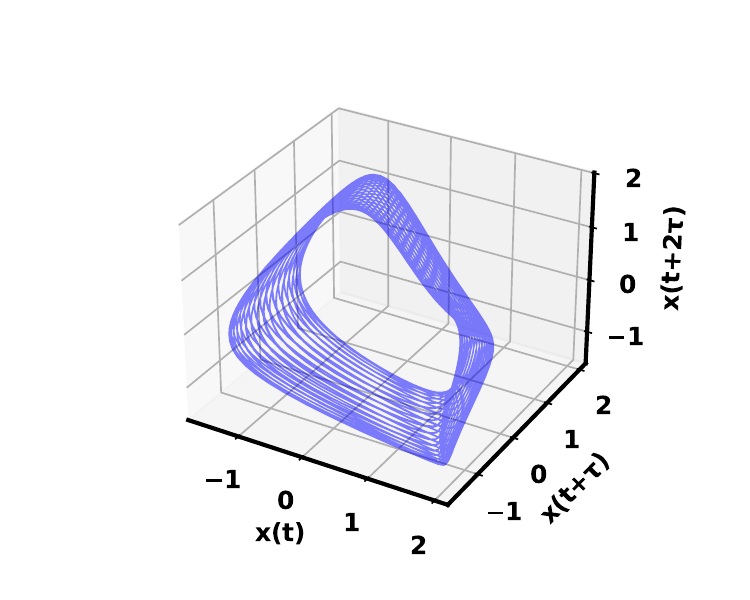}\label{fps_p}}
    \subfigure[]
    {\includegraphics[width=0.68\linewidth,height=0.65\linewidth]
    {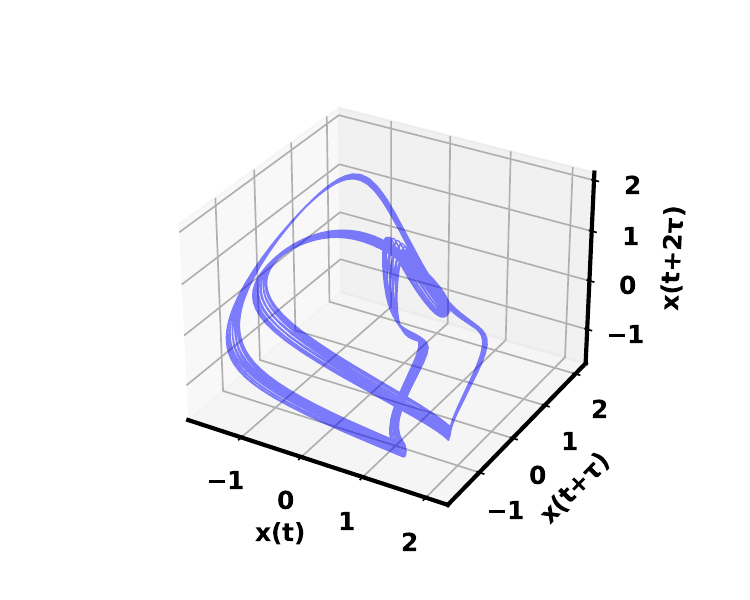}\label{fps_oc}}
	\subfigure[] 
    {\includegraphics[width=0.68\linewidth,height=0.65\linewidth]
    {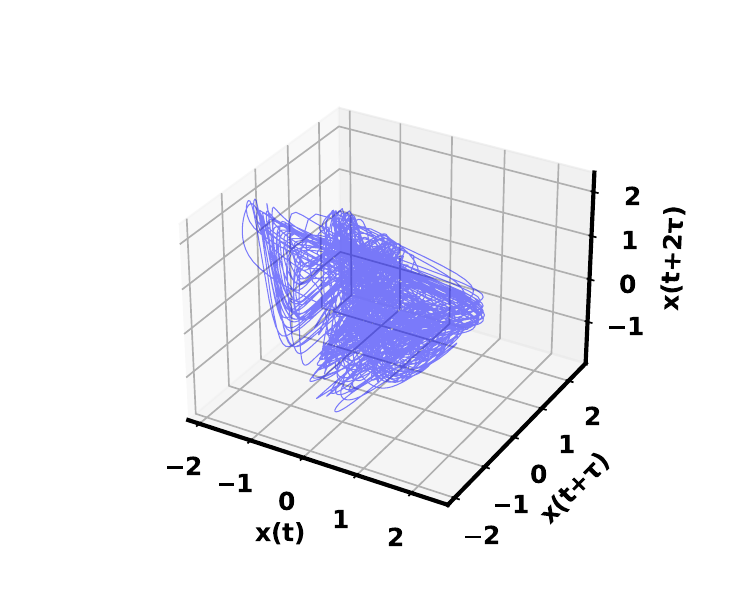}\label{fps_c}}
    \end{array}$
    \end{center}
    \begin{minipage}{\textwidth}
    \caption{The reconstructed phase space trajectories of radial coordinate $r$ in three dimensions depict the transition from the non-chaotic to chaotic behavior for fixed parameters $\rs = 0.15$ and $E = 90$ in Case-III. These trajectories correspond to three distinct regimes: regular (non-chaotic), the onset of chaos, and fully developed chaotic orbits. Figures~\ref{fps}(c) clearly demonstrate how the system evolves from a non-chaotic state (Fig.~\ref{fps}(a)) to a chaotic regime, passing through an intermediate stage where chaos first begins to emerge (Fig.~\ref{fps}(b)).}\label{fps}
    \end{minipage}
    \end{figure}
\end{widetext}

\begin{widetext}
    \begin{figure}[H]
	\centering
	\begin{center} 
	$\begin{array}{ccc}
	\subfigure[]              
    {\includegraphics[width=0.68\linewidth,height=0.65\linewidth]
    {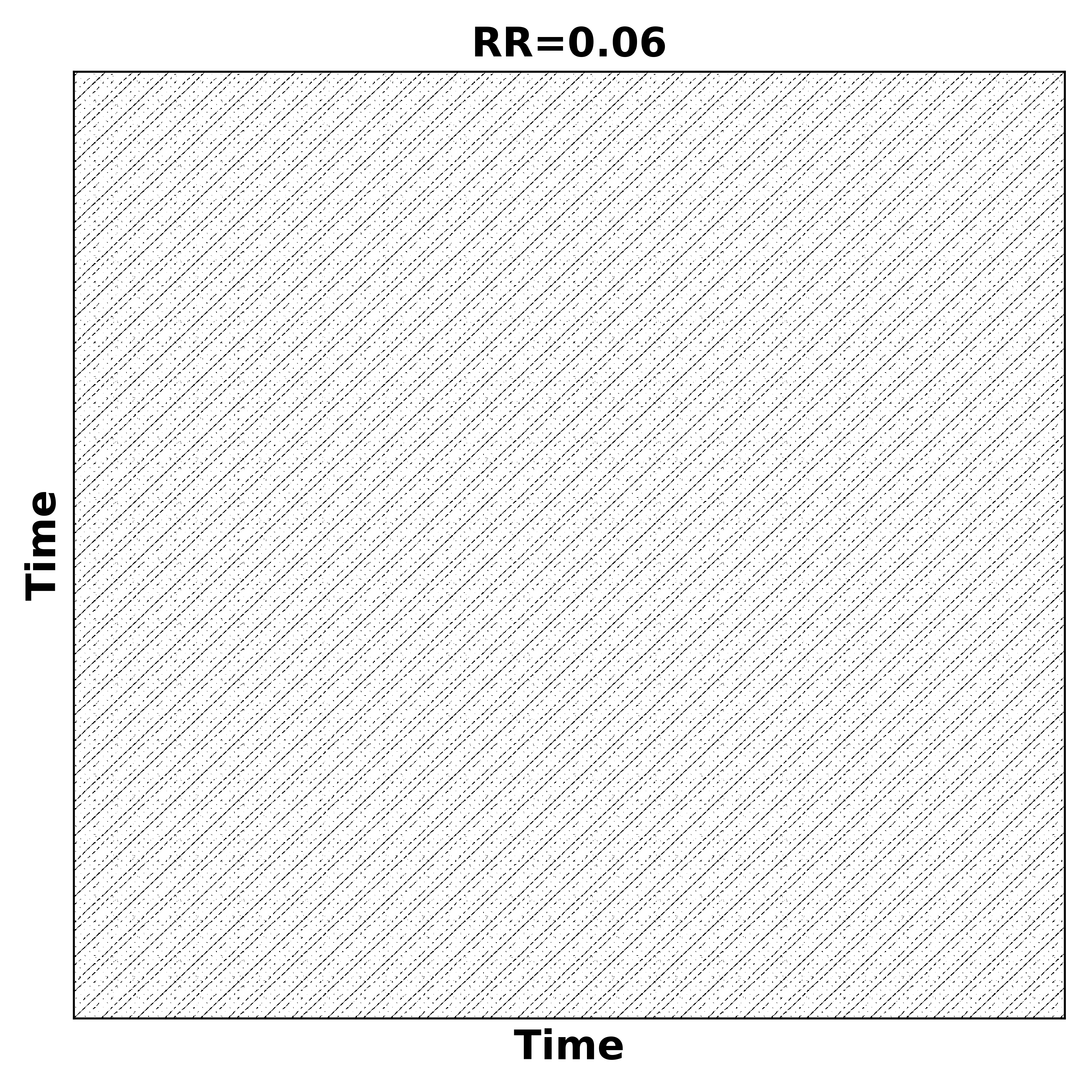}\label{frp_r_p1}}
	\subfigure[]           
    {\includegraphics[width=0.68\linewidth,height=0.65\linewidth]
    {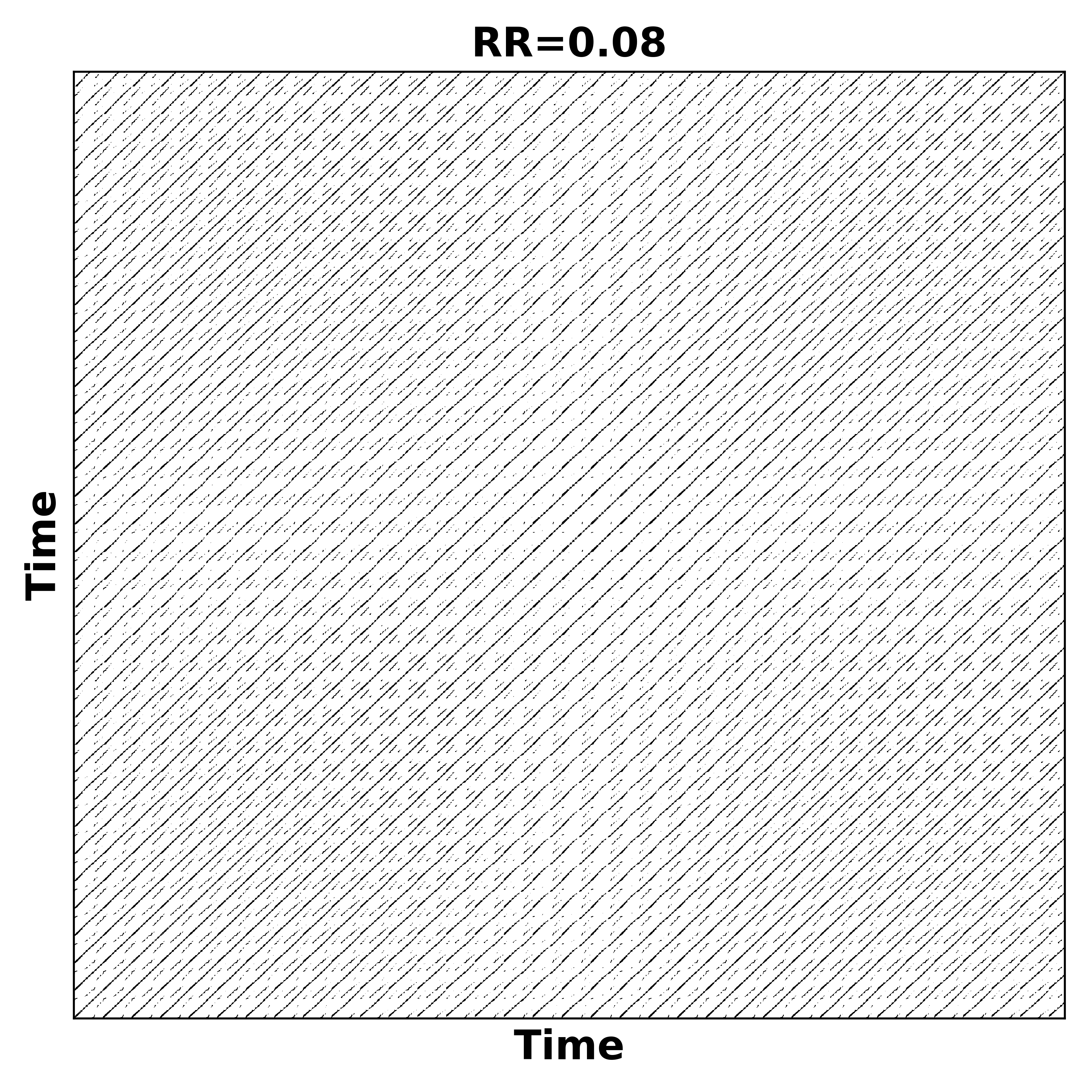}\label{frp_r_oc1}}
	\subfigure[] 
    {\includegraphics[width=0.68\linewidth,height=0.65\linewidth]
    {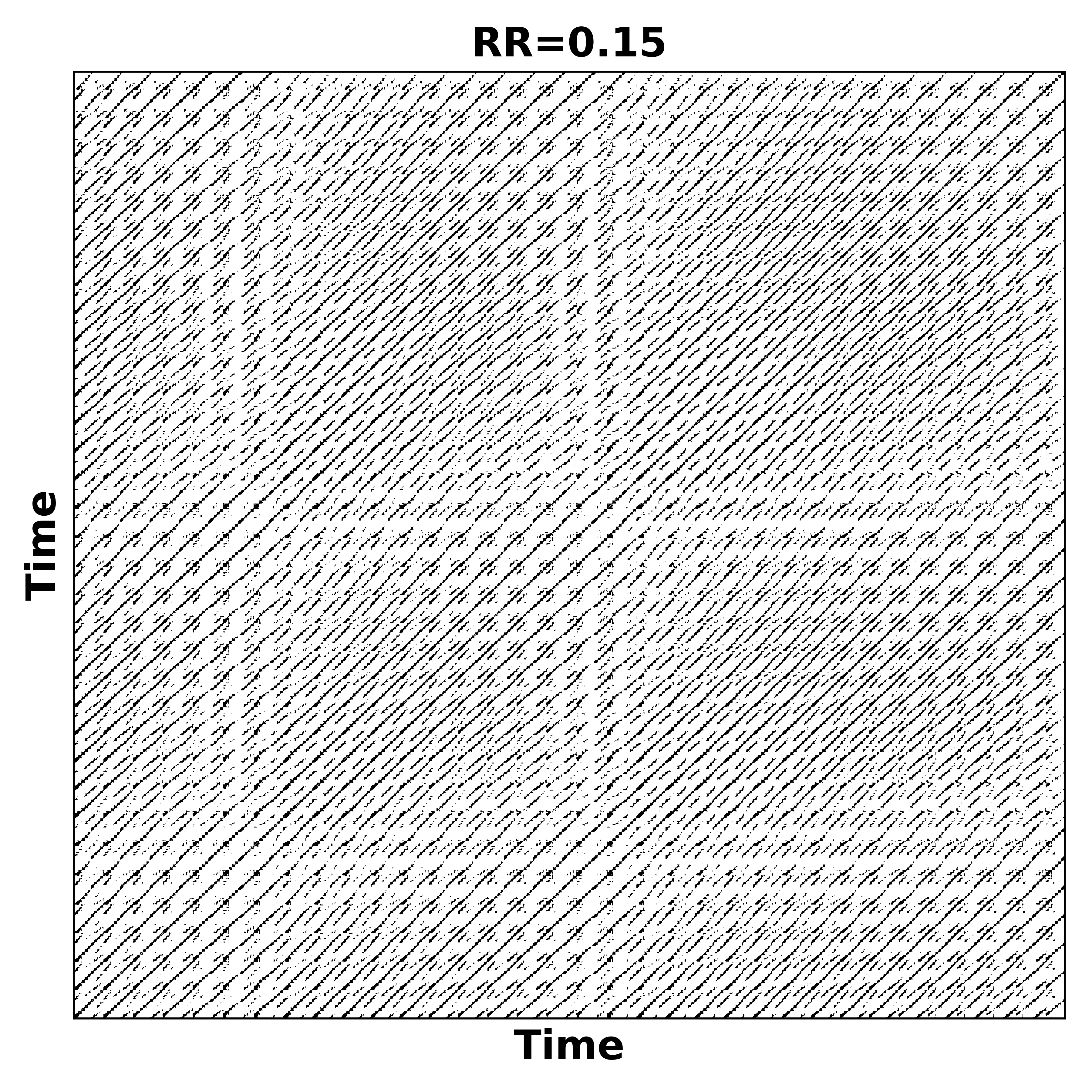}\label{frp_r_c1}}
    \end{array}$
    \end{center}
    \begin{minipage}{\textwidth}
    \caption{Recurrence plots illustrating radial coordinate $r$ along three distinct orbits for Case-IV, with fixed parameters $\rhs = 0.01$ and $E = 115$, are presented. Fig.~\ref{frp_r1}(a) displays the non-chaotic (regular) orbit corresponding to $\rs = 0.10$. For this non-chaotic orbit, the embedding parameters are $d = 4$, $\tau = 34$, and $\epsilon = 0.6938$, with a recurrence rate $\mathcal{RR} = 0.06$.\\
    Fig.~\ref{frp_r1}(b) shows the orbit at the onset of chaos for $\rs = 0.25$. Here, the embedding dimension is $d = 4$, the delay time $\tau = 20$, the threshold $\epsilon = 1.0140$, and the recurrence rate $\mathcal{RR} = 0.08$.\\
    Finally, Fig.~\ref{frp_r1}(c) depicts the fully chaotic orbit for $\rs = 0.27$. The parameters for this chaotic orbit are $d = 4$, $\tau = 23$, $\epsilon = 1.5288$, and $\mathcal{RR} = 0.15$. We observe a distinct diagonal pattern, which is characteristic of regular or non-chaotic orbit in Fig.~\ref{frp_r1}(a), slowly becoming obscured by the growing of central density $\rhs$ in Fig.~\ref{frp_r1}(c).}\label{frp_r1}
    \hrulefill
    \end{minipage}
	\begin{center} 
	$\begin{array}{ccc}
    \subfigure[]
    {\includegraphics[width=0.68\linewidth,height=0.65\linewidth]
    {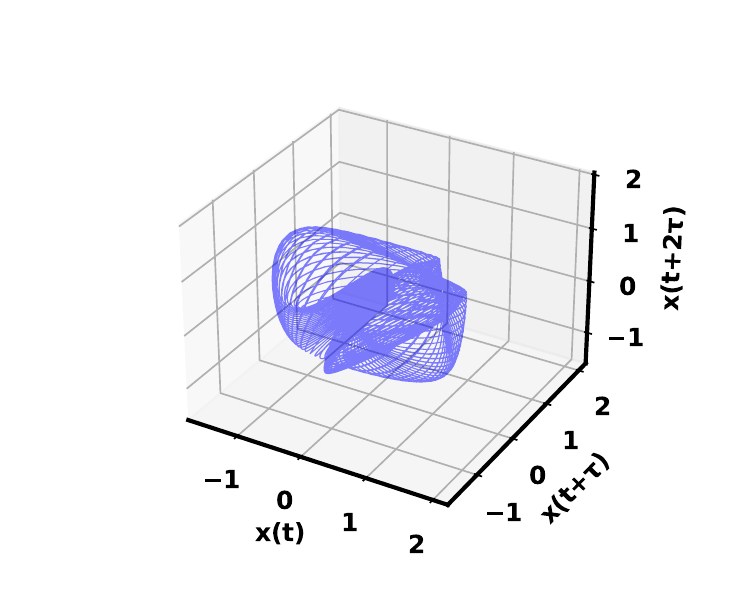}\label{fps_p1}}
    \subfigure[]
    {\includegraphics[width=0.68\linewidth,height=0.65\linewidth]
    {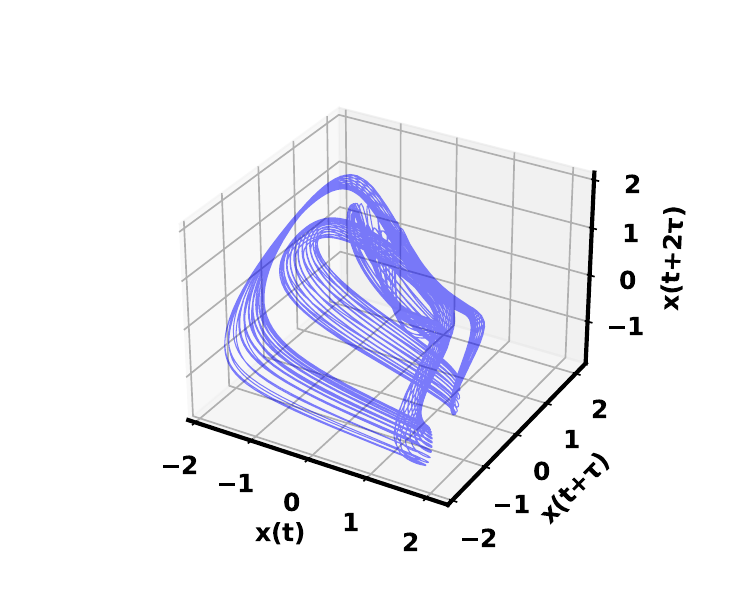}\label{fps_oc1}}
	\subfigure[] 
    {\includegraphics[width=0.68\linewidth,height=0.65\linewidth]
    {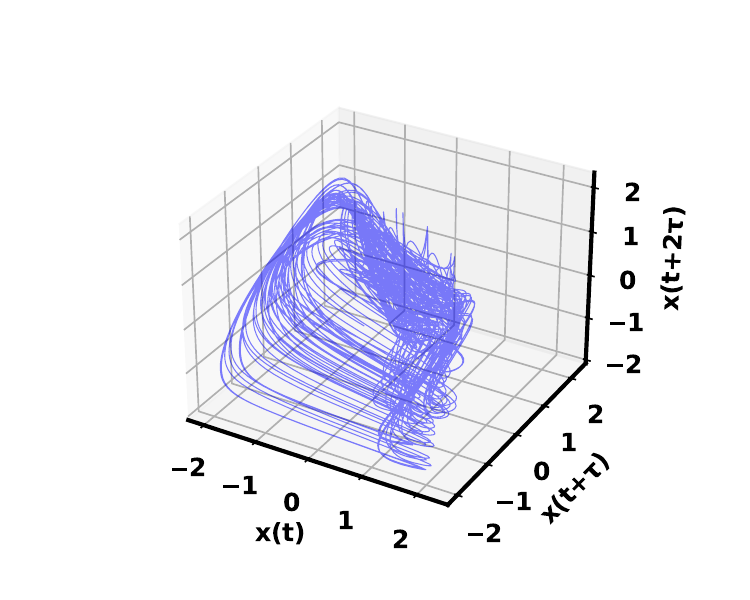}\label{fps_c1}}
    \end{array}$
    \end{center}
    \begin{minipage}{\textwidth}
    \caption{The reconstructed phase space trajectories of radial coordinate $r$ in three dimensions depict the transition from the non-chaotic to chaotic behavior for fixed parameters $\rhs = 0.01$ and $E = 115$. These trajectories correspond to three distinct regimes: regular (non-chaotic), the onset of chaos, and fully developed chaotic orbits. Figures~\ref{fps1}(c) clearly demonstrate how the system evolves from a non-chaotic state (Fig.~\ref{fps1}(a)) to a chaotic regime, passing through an intermediate stage where chaos first begins to emerge (Fig.~\ref{fps1}(b)).}\label{fps1}
    \end{minipage}
    \end{figure}
\end{widetext}

\noindent
The computed recurrence plots of orbit along radial direction are displayed in Figs.~\ref{frp_r} and \ref{frp_r1} for Case-III and IV, respectively. A visual examination of these plots reveal distinct patterns corresponding to non-chaotic and chaotic dynamical trajectories. In the non-chaotic case (Figs.~\ref{frp_r}(a), \ref{frp_r1}(a)), the recurrence plots exhibit a simple diagonal structure, which is a characteristic signature of a regular (or periodic) motion. However, this clear pattern becomes little bit disrupted when weak chaos introduced via the onset-of-chaos orbits (Figs.~\ref{frp_r}(b), \ref{frp_r1}(b)). Finally for $\rhs=0.05$ and $\rs=0.25$, we typically find a recurrence plots of strongly chaotic dynamics (Figs.~\ref{frp_r}(c), \ref{frp_r1}(c)). \\
On the other hand, the reconstructed phase space trajectories are also shown in Figs.~\ref{fps} and \ref{fps1} corresponding to different orbits for both of the considered cases. It is clearly visible in Figs.~\ref{fps}(c), \ref{fps1}(c) that chaos actually appears for DM halo parameters $\rhs=0.05$ and $\rs=0.27$, respectively, compared to the non-chaotic and onset-of-chaos orbits (Figs.~\ref{fps}(a), \ref{fps}(b) for Case-III and Figs.~\ref{fps1}(a), \ref{fps1}(b) for Case-IV).

\section*{Recurrence analysis of gravitational waveforms for Case-IV}
\begin{widetext}
    \begin{figure}[H]
	\centering
	\begin{center} 
	$\begin{array}{ccc}
	\subfigure[]              
    {\includegraphics[width=0.68\linewidth,height=0.7\linewidth]
    {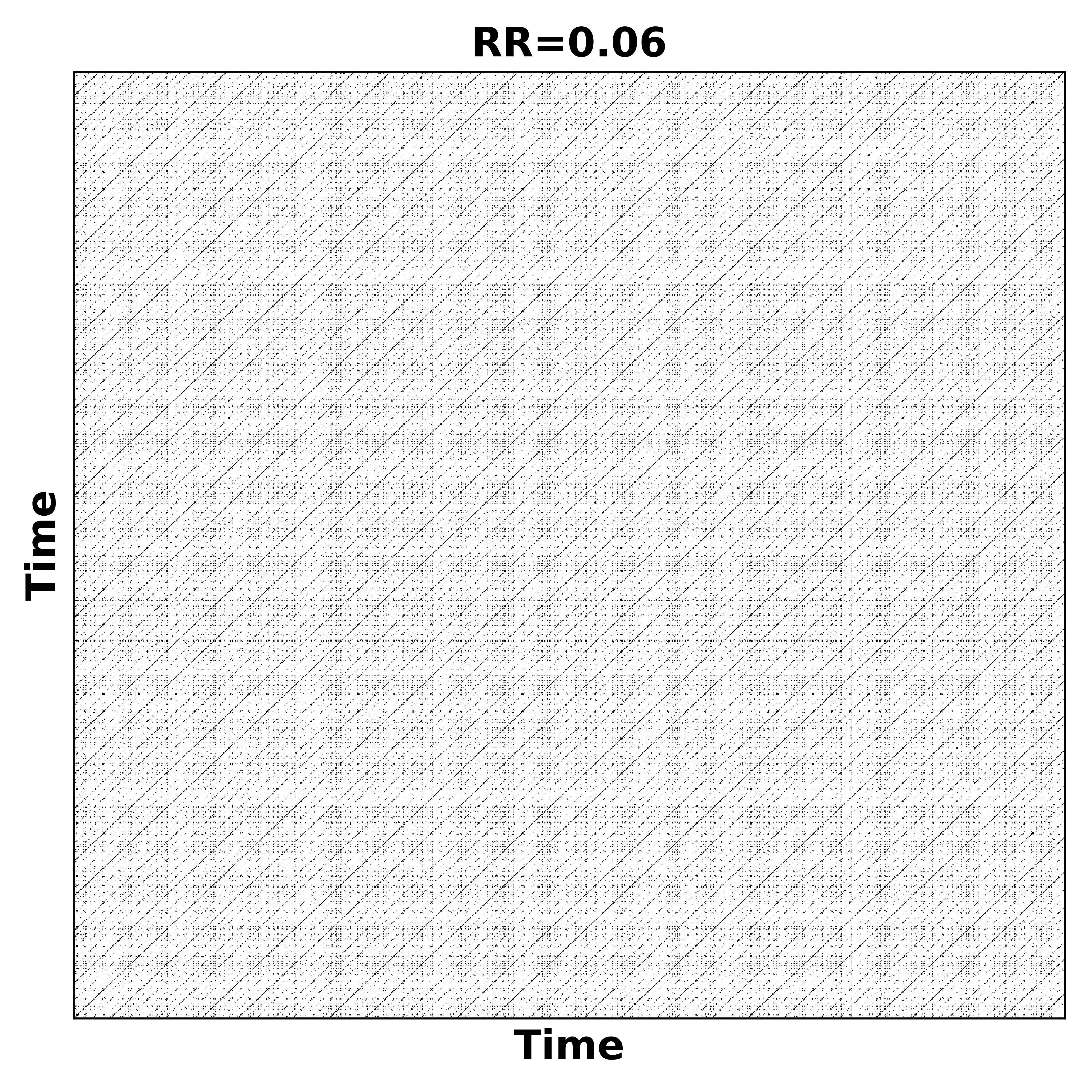}\label{frp_plus_p1}}
	\subfigure[]           
    {\includegraphics[width=0.68\linewidth,height=0.7\linewidth]
    {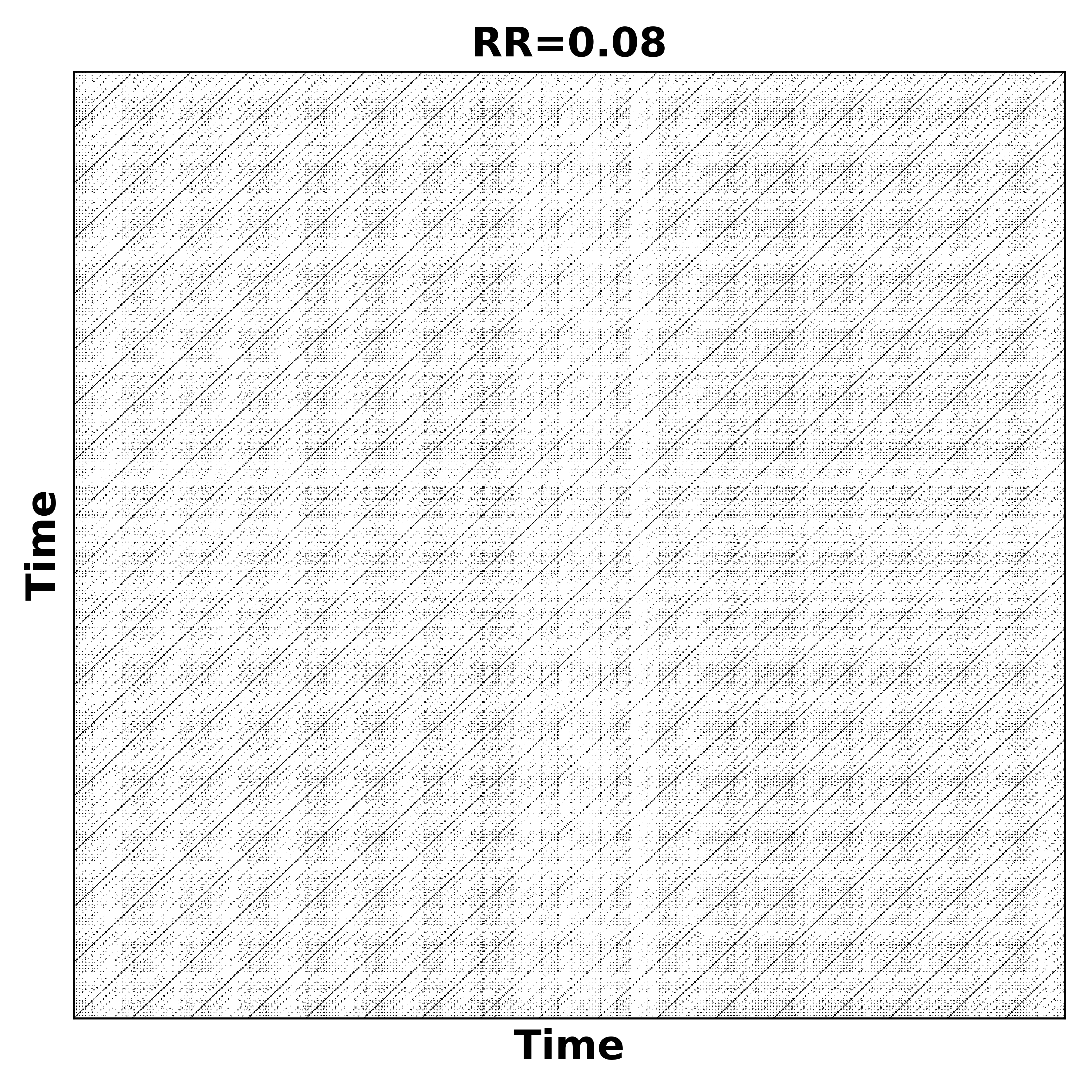}\label{frp_plus_oc1}}
	\subfigure[] 
    {\includegraphics[width=0.68\linewidth,height=0.7\linewidth]
    {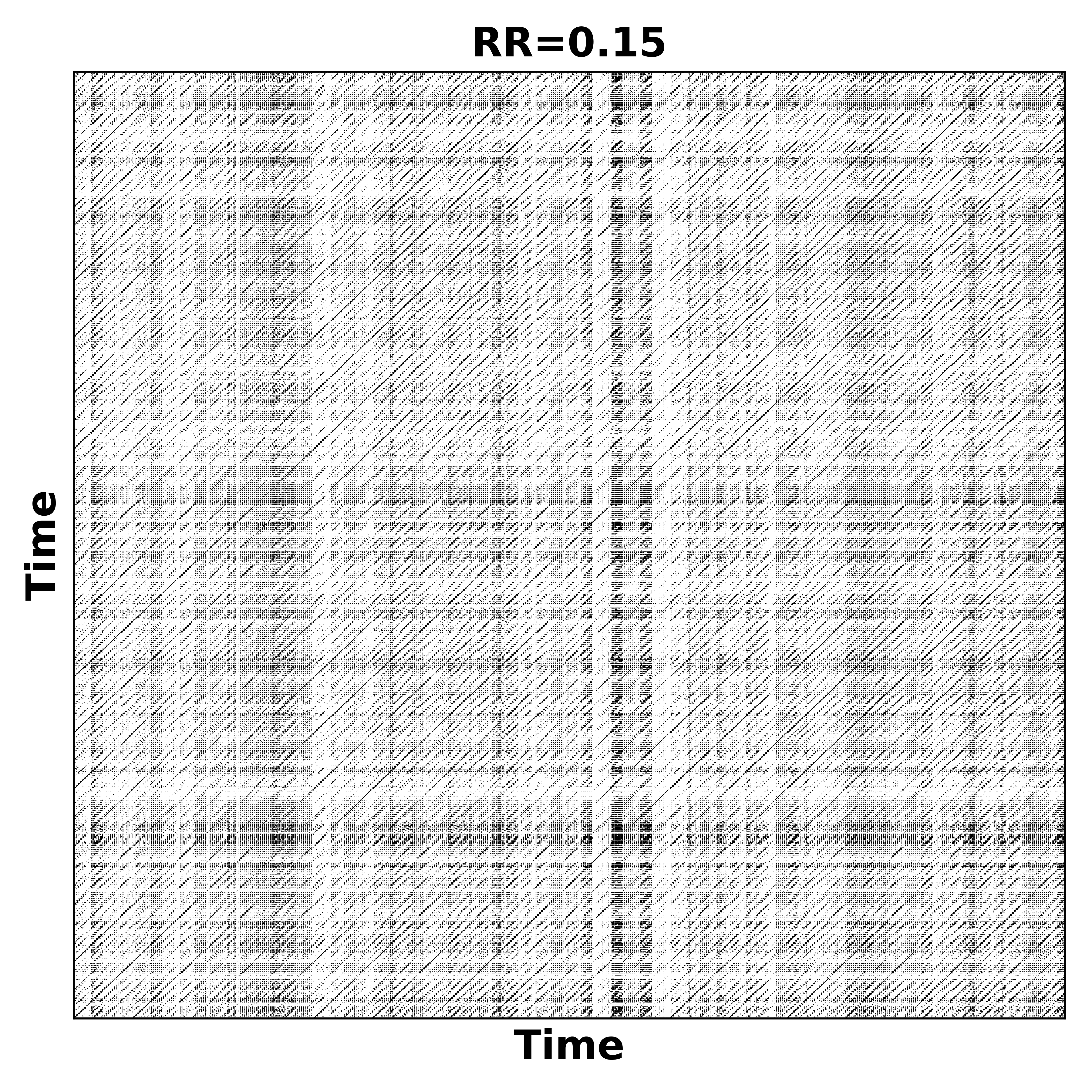}\label{frp_plus_c1}}
    \end{array}$
    \end{center}
    \begin{minipage}{\textwidth}
    \caption{Recurrence plots of the gravitational waveforms $h_+$, generated using the kludge scheme illustrating three distinct orbits for Case-IV, with fixed parameters $\rhs = 0.01$ and $E = 115$, are presented. Fig.~\ref{frp_plus1}(a) displays $h_+$ mode for the non-chaotic (regular) orbit corresponding to $\rs = 0.10$. For this regular motion, the embedding parameters are $d = 9$, $\tau = 19$, and $\epsilon = 1.8549$ with a recurrence rate $\mathcal{RR} = 0.06$.\\
    Fig.~\ref{frp_plus1}(b) shows the $h_+$ mode of orbit at the onset of chaos for $\rs = 0.25$. Here, the embedding dimension is $d = 9$, the delay time $\tau = 22$, the threshold $\epsilon = 2.2036$, and the recurrence rate $\mathcal{RR} = 0.08$.\\
    Finally, Fig.~\ref{frp_plus1}(c) depicts the $h_+$ waveform of fully chaotic orbit for $\rs = 0.27$. The parameters for this chaotic motion are $d = 9$, $\tau = 15$, $\epsilon = 2.4058$, and $\mathcal{RR} = 0.15$. We note that the plus polarization associated with non-chaotic orbits generally produces extended diagonal lines. In contrast, chaotic orbits generate a more intricate, square-shaped pattern in the corresponding plus polarization of the waveforms as the central density $\rs$ increases.}\label{frp_plus1}
    \hrulefill
    \end{minipage}
	\begin{center} 
	$\begin{array}{ccc}
    \subfigure[]
    {\includegraphics[width=0.68\linewidth,height=0.7\linewidth]
    {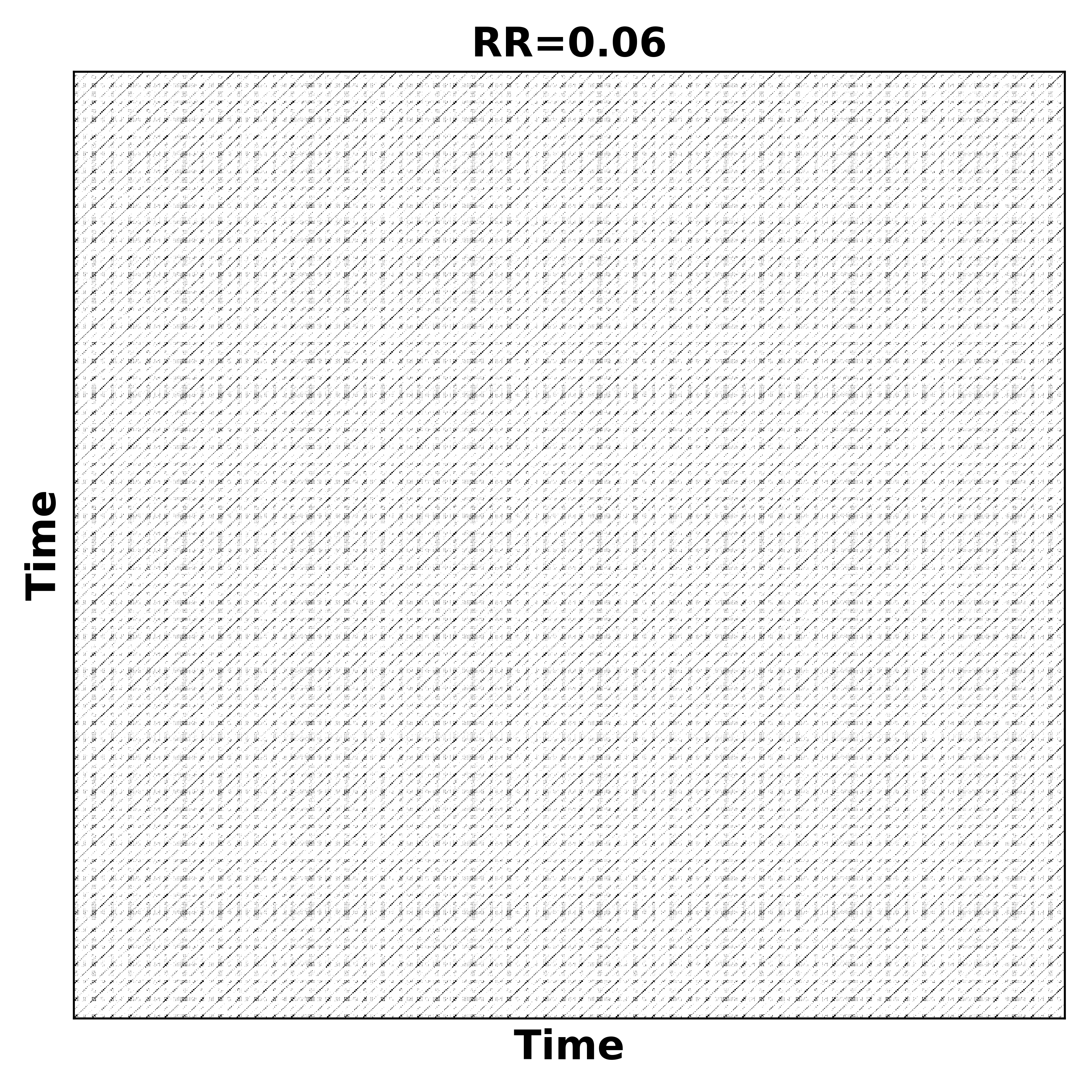}\label{frp_cross_p1}}
    \subfigure[]
    {\includegraphics[width=0.68\linewidth,height=0.7\linewidth]
    {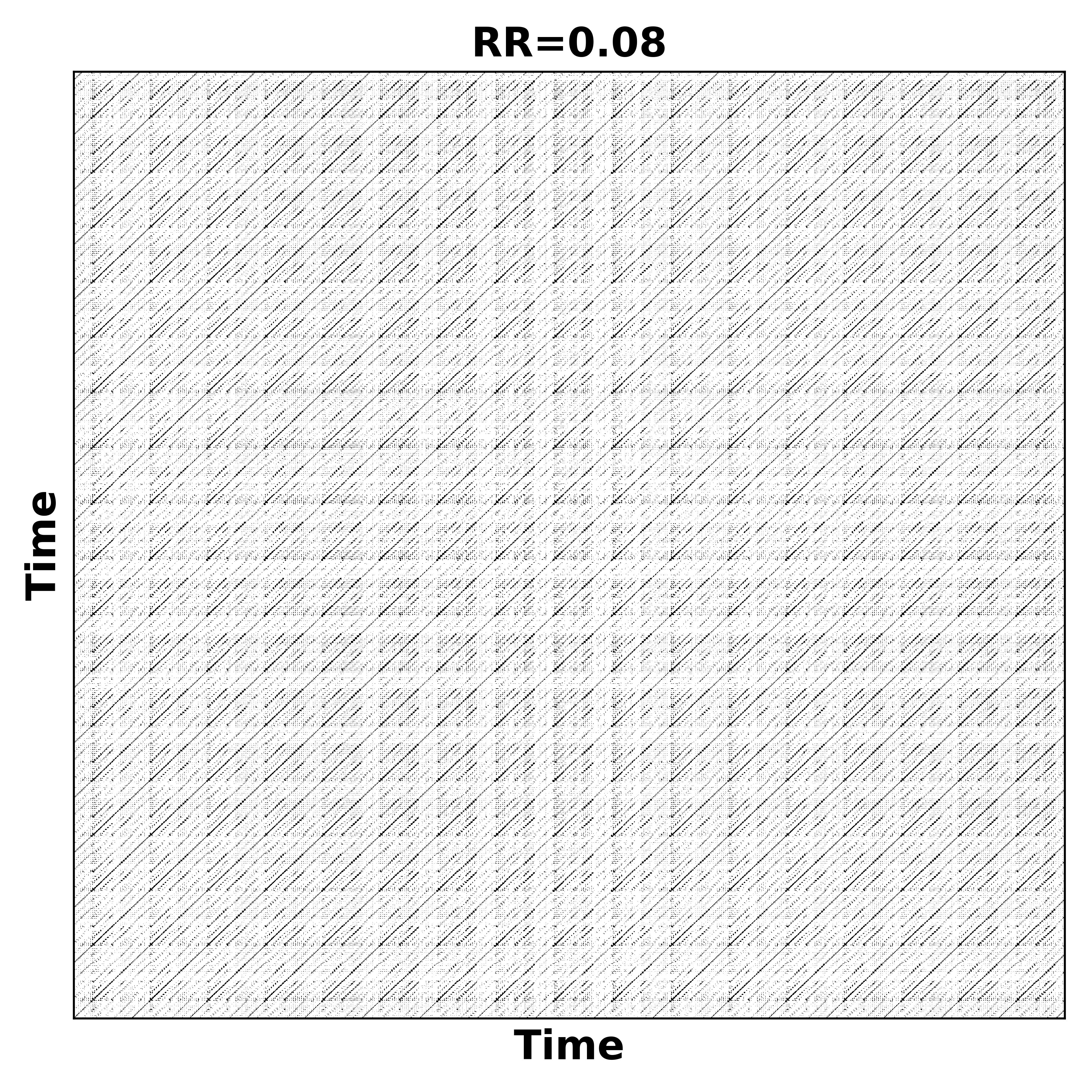}\label{frp_cross_oc1}}
	\subfigure[] 
    {\includegraphics[width=0.68\linewidth,height=0.7\linewidth]
    {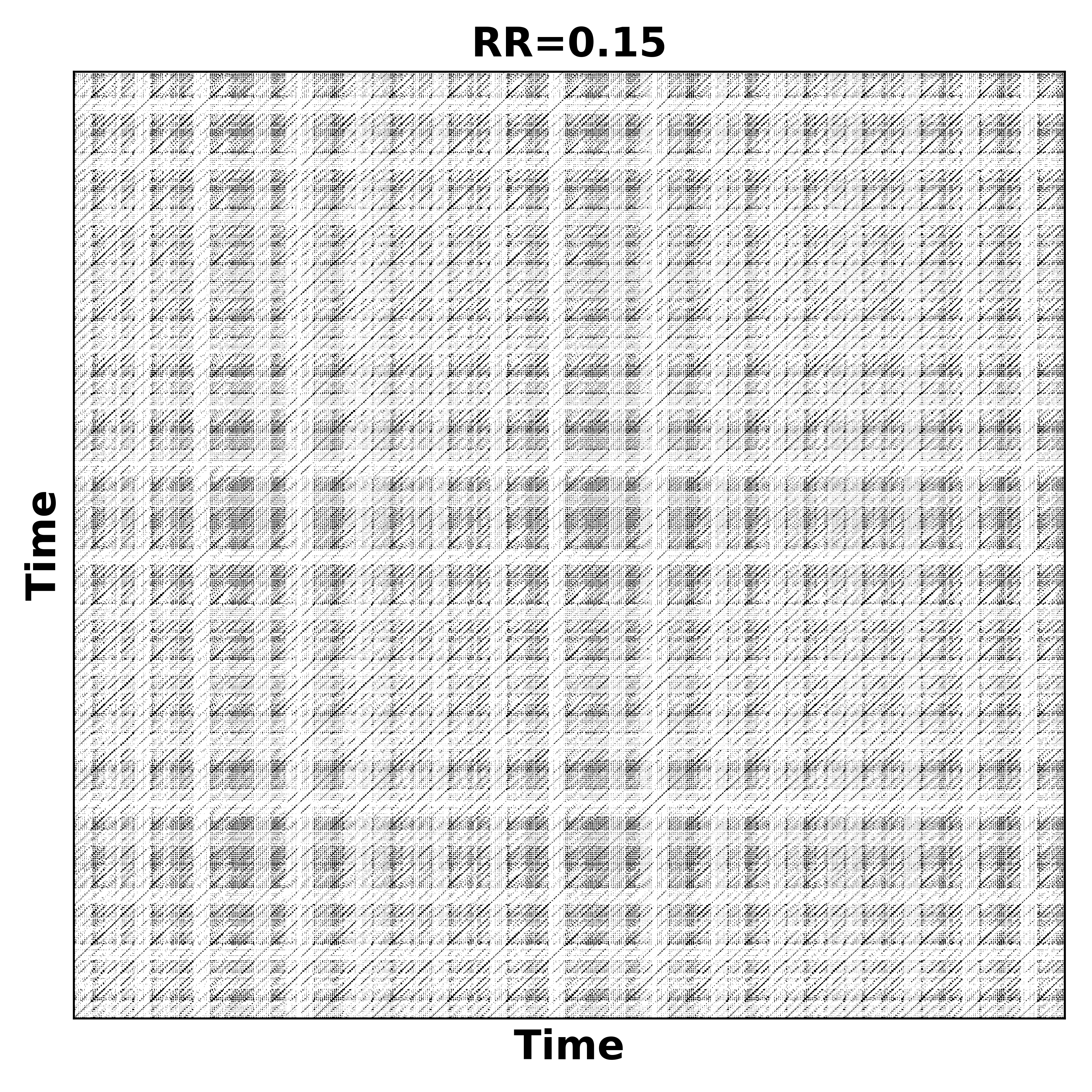}\label{frp_cross_c1}}
    \end{array}$
    \end{center}
    \begin{minipage}{\textwidth}
    \caption{Recurrence plots of the gravitational waveforms $h_\times$, generated using the kludge scheme illustrating three distinct orbits for Case-IV, with fixed parameters $\rhs = 0.01$ and $E = 115$, are presented. Fig.~\ref{frp_cross1}(a) displays $h_\times$ mode for the non-chaotic (regular) orbit corresponding to $\rs = 0.10$. For this regular motion, the embedding parameters are $d = 8$, $\tau = 12$, and $\epsilon = 1.4702$ with a recurrence rate $\mathcal{RR} = 0.06$.\\
    Fig.~\ref{frp_cross1}(b) shows the $h_\times$ mode of orbit at the onset of chaos for $\rs = 0.25$. Here, the embedding dimension is $d = 9$, the delay time $\tau = 15$, the threshold $\epsilon = 2.1360$, and the recurrence rate $\mathcal{RR} = 0.08$.\\
    Finally, Fig.~\ref{frp_cross1}(c) depicts the $h_\times$ waveform of fully chaotic orbit for $\rs = 0.27$. The parameters for this orbit are $d = 8$, $\tau = 15$, $\epsilon = 1.7991$, and $\mathcal{RR} = 0.15$. We note that similar to the plus mode, the cross polarization associated with non-chaotic orbits generally produces extended diagonal lines. In contrast, chaotic orbits generate a more intricate, square-shaped pattern in the corresponding cross polarization of the GWs as the central density $\rs$ increases.}\label{frp_cross1}
    \end{minipage}
    \end{figure}
\end{widetext}

\noindent
The resulting recurrence plots for the two modes of gravitational waveforms for Case-IV are presented in Figs.~\ref{frp_plus1}, \ref{frp_cross1}, respectively. It is clearly visible that Figs.~\ref{frp_plus1}(a), \ref{frp_cross1}(a) exhibit diagonal lines, a signature of ordered motion, while in Figs.~\ref{frp_plus1}(c), \ref{frp_cross1}(c), they are broken for the chaotic polarizations of GWs and characterized by square-shaped patterns.

\section*{Characteristic strain sensitivities of space-based gravitational wave detectors}

\subsection*{A. LISA}\label{sec:LISA}
\newpage
The LISA sensitivity curve is derived using the formulation presented in the Ref.~\cite{Robson:2018ifk}. The effective strain spectral density characterizing the noise can be expressed as \cite{Robson:2018ifk}
    \begin{equation}
        S_{\rm n}(f) \equiv \frac{P_n(f)}{\mathscr{R}(f)} + S_c(f)\nonumber~,
    \end{equation}
where $P_{\rm n}(f)$ represents the power spectral density of the detector noise and $\mathscr{R}(f)$ denotes the instrument's signal response function averaged over sky positions and polarizations. 

One can approximate the LISA sensitivity curve by expanding $P_n(f)$ and simplifying $\mathscr{R}(f)$ in the following equation \cite{Robson:2018ifk}:
    \begin{eqnarray}
        &S_{n}(f)=\frac{10}{3 L^2} \Big(P_{\rm OMS}(f) + \frac{4P_{\rm acc}(f)}{(2 \pi f)^4}\Big)\times \Big(1 + 0.6 \Big(\frac{f}{f_*}\Big)^2\Big) \non\\
        &+ S_c(f)~,\label{eq:LISA_Sn}
    \end{eqnarray}
where $L = 2.5\times 10^9~\mathrm{m}$ specifies the arm length for LISA, and $f_* = c / 2 \pi L\sim19.09~\mathrm{mHz}$ defines the transfer frequency. The single-link optical metrology noise component $(P_{OMS}(f))$ is given by \cite{Robson:2018ifk}
    \begin{equation}
    \begin{split}
        P_{\rm OMS}(f) = \left(1.5 \times 10^{-11} \mathrm{m}\right)^{2} \times \left(1+\left(\frac{2 \mathrm{mHz}}{f}\right)^{4}\right) \mathrm{Hz}^{-1}\nonumber~,
    \end{split}
    \end{equation}
The test mass acceleration noise $(P_{acc}(f))$ takes the following form \cite{Robson:2018ifk}: 
\begin{equation}
    \begin{split}
        P_{\rm acc}(f) &= \left(3 \times 10^{-15} \mathrm{ms}^{-2}\right)^{2}\left(1+\left[\frac{0.4 \mathrm{mHz}}{f}\right]^{2}\right) \times \non\\
        &\left(1+\left[\frac{f}{8 \mathrm{mHz}}\right]^{4}\right) \mathrm{Hz}^{-1}\nonumber~,
    \end{split}
    \end{equation}
Besides these instrumental noises, unresolved galactic binaries will contribute as an effective noise source, although this component is non-stationary. The galactic confusion noise $(S_c(f))$ decreases over the course of the mission as more foreground sources are identified and subtracted. In our analysis, we will not consider the confusion noise and therefore following the Refs.~\cite{Robson:2018ifk}, a well-designed analytical model for the sensitivity curve, which is sufficient for our purpose can be expressed as the following.
    \begin{eqnarray}
        &S_{n}^{L}(f)=\frac{10}{3 L^2} \Bigg(P_{\rm OMS}(f) + \frac{2P_{\rm acc}(f)}{(2 \pi f)^4} \Big(1+\cos^2\big(\frac{f}{f_*}\big)\Big)\Bigg) \times \non\\
        &\Bigg(1 + 0.6 \Big(\frac{f}{f_*}\Big)^2\Bigg)~,\label{eq:LISA_S}
    \end{eqnarray}
Here it is important to note that $\cos^2(f/f_*)$ factor, that scales the acceleration noise tends to unity in the frequency regime where acceleration noise is dominant. This simplification enables us to express the instrumental contribution to the actual sensitivity curve as presented in Eq.~\eqref{eq:LISA_Sn}.

\subsection*{B. TianQin}\label{sec:TianQin}
\noindent
The TianQin sensitivity curve follows the power spectral density formulation as \cite{TianQin:2020hid,Li:2024rnk}
    \begin{equation}
    \begin{split}
        S_{N}(f)= \frac{10}{3 L^{2}}\left[\frac{4 S_{a}}{(2 \pi f)^{4}}\left(1+\frac{10^{-4}~Hz}{f}\right)+S_{x}\right] \\
        \times\left[1+0.6\left(\frac{f}{f_{*}}\right)^{2}\right]\label{eq:tianqin}~,
    \end{split}
    \end{equation}
where $L = 1.7 \times 10^8~\mathrm{m}$ represents the arm length, $S_a = 1 \times 10^{-30}~\mathrm{m^2~s^{-4}~{Hz^{-1}}}$ quantifies the acceleration noise, $S_x = 1 \times 10^{-24}~\mathrm{m^2~{Hz^{-1}}}$ describes the displacement measurement noise, and $f_* = c / 2 \pi L\sim 280.66~mHz$ is the transfer frequency of TianQin. Notably, Eq.~\eqref{eq:tianqin} incorporates an additional $10 / 3$ factor compared to the original formulation in the Ref.~\cite{TianQin:2020hid,Li:2024rnk}, as they incorporate this factor within their waveform model rather than the power spectral density. We maintain this factor to ensure consistency with the LISA convention established by Robson {\it{et al.}} \cite{Robson:2018ifk}, enabling direct comparison between the sensitivity curves.


\subsection*{C. Taiji}\label{sec:Taiji}
\noindent
We consider the same form of the analytical model of sensitivity curve for Taiji as mentioned in Eq.~\eqref{eq:LISA_S}. Therefore for our analysis, we proceed with the following form of the sensitivity curve for Taiji \cite{Hu:2017mde,Liu:2023qap}.
    \begin{eqnarray}
        &S_{n}^{T}(f)=\frac{10}{3 L^2} \Bigg(P_{\rm dp}(f) + \frac{2P_{\rm acc}(f)}{(2 \pi f)^4} \Big(1+\cos^2\big(\frac{f}{f_*}\big)\Big)\Bigg)\non\\
        &\times \Bigg(1 + 0.6 \Big(\frac{f}{f_*}\Big)^2\Bigg)~,\label{eq:taiji}
    \end{eqnarray}
where $f_*=c/2\pi L\sim 15.90~\mathrm{mHz}$ is the transfer frequency and the arm length is $L=3\times 10^9~\mathrm{m}$ for Taiji, a little bit longer than LISA. Here $P_{\rm dp}(f),~P_{ac}(f)$ represent the power spectral density of the displacement noise and the acceleration noise, respectively, having the following forms \cite{Hu:2017mde,Liu:2023qap}.
    \begin{equation}
    \begin{split}
        P_{\rm dp}(f)&= \left(8.0 \times 10^{-12} \mathrm{m}\right)^{2} \times \left(1+\left(\frac{2 \mathrm{mHz}}{f}\right)^{4}\right) \mathrm{Hz}^{-1}\nonumber~,\\
        P_{\rm acc}(f)&= \left(3 \times 10^{-15} \mathrm{ms}^{-2}\right)^{2}\left(1+\left[\frac{0.4 \mathrm{mHz}}{f}\right]^{2}\right)\\
        &\times \left(1+\left[\frac{f}{8 \mathrm{mHz}}\right]^{4}\right) \mathrm{Hz}^{-1}\nonumber~.
    \end{split}
    \end{equation} 
It is notable to mention that both, LISA and Taiji has their same test mass acceleration noise $(P_{acc}(f)~\text{and}~P_{ac}(f))$.

\bibliography{references}
\end{document}